\documentclass[11pt,a4paper]{article}

\usepackage{amsmath,amsfonts,amssymb,amsthm,cite}

\InputIfFileExists{whatnot.tex}{}{}

\evensidemargin=\oddsidemargin
\advance\topmargin by -\headheight
\advance\topmargin by -\headsep

\newtheorem{thm}{Theorem}[section]
\newtheorem{lem}[thm]{Lemma}
\newtheorem{cly}[thm]{Corollary}
\newtheorem{prop}[thm]{Proposition}
\newtheorem{conj}[thm]{Conjecture}

\theoremstyle{definition}                    
\newtheorem{defn}[thm]{Definition}
\theoremstyle{remark}
\newtheorem{rem}[thm]{Remark}

\numberwithin{equation}{section}             

\newcommand{\A}{\mathcal{A}}                 
\renewcommand{\a}{\alpha}                    
\DeclareMathOperator{\ad}{ad}                
\newcommand{\as}{\quad\mbox{as}\enspace}     
\newcommand{\Aun}{\widetilde{\mathcal{A}}}   
\newcommand{\B}{\mathcal{B}}                 
\renewcommand{\b}{\beta}                     
\newcommand{\braCket}[3]{\langle#1\mathbin|#2\mathbin|#3\rangle}
\newcommand{\braket}[2]{\langle#1\mathbin|#2\rangle} 
\newcommand{\C}{\mathbb{C}}                  
\newcommand{\CC}{\mathcal{C}}                
\newcommand{\cc}{\mathbf{c}}                 
\DeclareMathOperator{\Cl}{C\ell}             
\newcommand{\Coo}{C^\infty}                  
\newcommand{\D}{\mathcal{D}}                 
\newcommand{\DD}{\underline{D}}              
\newcommand{\del}{\partial}                  
\DeclareMathOperator{\Dom}{Dom}              
\newcommand{\Dslash}{{D\mkern-11.5mu/\,}}    
\newcommand{\E}{\mathcal{E}}                 
\newcommand{\eps}{\varepsilon}               
\newcommand{\F}{\mathcal{F}}                 
\newcommand{\G}{\mathcal{G}}                 
\newcommand{\Ga}{\Gamma}                     
\newcommand{\ga}{\gamma}                     
\renewcommand{\H}{\mathcal{H}}               
\newcommand{\half}{\tfrac{1}{2}}             
\newcommand{\hookto}{\hookrightarrow}        
\newcommand{\Ht}{{\widetilde{\mathcal{H}}}}  
\newcommand{\I}{\mathcal{I}}                 
\DeclareMathOperator{\Junk}{Junk}            
\newcommand{\K}{\mathcal{K}}                 
\newcommand{\ket}[1]{|#1\rangle}             
\renewcommand{\L}{\mathcal{L}}               
\newcommand{\La}{\Lambda}                    
\newcommand{\la}{\lambda}                    
\newcommand{\M}{\mathcal{M}}                 

\newcommand{\mop}{{\mathchoice{\mathbin{\star_{_\theta}}}
            {\mathbin{\star_{_\theta}}}           
            {{\star_\theta}}{{\star_\theta}}}}    
\newcommand{\N}{\mathbb{N}}                  
\newcommand{\NN}{\mathcal{N}}                
\newcommand{\nb}{\nabla}                     
\newcommand{\Oh}{\mathcal{O}}                
\newcommand{\opp}{{\mathrm{op}}}             
\newcommand{\ox}{\otimes}                    
\newcommand{\oxyox}{\otimes\cdots\otimes}    
\newcommand{\pd}[2]{\frac{\partial#1}{\partial#2}}
\newcommand{\piso}[1]{\lfloor#1\rfloor}      
\newcommand{\PsiDO}{\Psi\mathrm{DO}}         
\newcommand{\R}{\mathbb{R}}                  
\newcommand{\roundbraket}[2]{(#1\mathbin|#2)} 
\newcommand{\sepword}[1]{\quad\mbox{#1}\quad} 
\newcommand{\set}[1]{\{\,#1\,\}}             
\newcommand{\Sf}{\mathbb{S}}                 
\DeclareMathOperator{\spec}{sp}              
\renewcommand{\SS}{\mathcal{S}}              
\DeclareMathOperator{\supp}{supp}            
\newcommand{\T}{\mathbb{T}}                  
\renewcommand{\th}{\theta}                   
\newcommand{\thalf}{\tfrac{1}{2}}            
\newcommand{\tihalf}{\tfrac{i}{2}}           
\newcommand{\tpi}{{\tilde\pi}}               
\DeclareMathOperator{\Tr}{Tr}                
\newcommand{\V}{\mathcal{V}}                 
\newcommand{\vac}{\ket{0}}                   
\newcommand{\vf}{\varphi}                    
\DeclareMathOperator{\wres}{wres}            
\newcommand{\x}{\times}                      
\newcommand{\Z}{\mathbb{Z}}                  
\newcommand{\7}{\dagger}                     
\newcommand{\8}{\bullet}                     
\renewcommand{\.}{\cdot}                     
\renewcommand{\:}{\colon}                    

\def\<#1,#2>{\langle#1,#2\rangle}            
\def\ee_#1{e_{{\scriptscriptstyle#1}}}       
\def\wick:#1:{\mathopen:#1\mathclose:}       

\newcommand{\opname}[1]{\mathop{\mathrm{#1}}\nolimits}

\newcommand{\hideqed}{\renewcommand{\qed}{}} 

\begin{document}

\thispagestyle{empty}

\begin{center}

CENTRE DE PHYSIQUE TH\'EORIQUE$\,^1$\\
CNRS--Luminy, Case 907\\
13288 Marseille Cedex 9\\
FRANCE\\

\vspace{2cm}

{\Large\textbf{Moyal Planes are Spectral Triples}} \\

\vspace{1.5cm}

{\large V. Gayral,$^2$ J.~M. Gracia-Bond\'{\i}a,$^3$ B. Iochum,$^2$
T. Sch\"ucker$^2$ and J.~C. V\'arilly$^{4,5}$} \\

\vspace{1.5cm}

{\large\textbf{Abstract}}
\end{center}

Axioms for nonunital spectral triples, extending those introduced in
the unital case by Connes, are proposed. As a guide, and for the sake
of their importance in noncommutative quantum field theory, the spaces
$\R^{2N}$ endowed with Moyal products are intensively investigated.
Some physical applications, such as the construction of noncommutative
Wick monomials and the computation of the Connes--Lott functional
action, are given for these noncommutative hyperplanes.

\vspace{1.5cm}

July 2003\\
CPT--03/P.4546

\vspace{5cm}

\noindent $^1$ Unit\'e Propre de Recherche 7061\\
$^2$ Also at Universit\'e de Provence, gayral@cpt.univ-mrs.fr,
iochum@cpt.univ-mrs.fr,\\ schucker@cpt.univ-mrs.fr\\
$^3$ Departamento de F\'{\i}sica, Universidad de Costa Rica,
2060 San Pedro, Costa Rica\\
$^4$ Departamento de Matem\'aticas, Universidad de Costa Rica,
2060 San Pedro, Costa Rica\\
$^5$ Regular Associate of the Abdus Salam ICTP, 34014 Trieste;
varilly@ictp.trieste.it

\newpage

\tableofcontents

\newpage

\section{Introduction}

Since Seiberg and Witten conclusively confirmed~\cite{SeibergW} that
the endpoints of open strings in a magnetic field background
effectively live on a noncommutative space, string theory has given
much impetus to noncommutative field theory (NCFT). This noncommutative
space turns out to be of the Moyal type, for which there already
existed a respectable body of mathematical knowledge, in connection
with the phase-space formulation of quantum mechanics~\cite{Moyal}.

However, NCFT is a problematic realm. Its bane is the trouble with both
unitarity and causality~\cite{GomisM,SeibergST}. Feynman rules for NCFT
can be derived either using the canonical operator formalism for
quantized fields, working with the scattering matrix in the Heisenberg
picture by means of Yang--Feldman--K\"all\'en equations; or from the
functional integral formalism. These two approaches
clash~\cite{BahnsDFPnice}, and there is the distinct possibility that
both fail to make sense. The difficulties vanish if we look instead at
NCFT in the Euclidean signature. Also, in spite of the tremendous
influence on NCFT, direct and indirect, of the work by Connes, it is
surprising that NCFT based on the Moyal product as currently practised
does not appeal to the spectral triple formalism.

So we may, and should, raise a basic question: namely, whether the
Euclidean version of Moyal noncommutative field theory is compatible
with the full strength of Connes' formulation of noncommutative
geometry, or not.

The prospective benefits of such an endeavour are mutual. Those
interested in applications may win a new toolkit, and Connes' paradigm
stands to gain from careful consideration of new examples.

In order to speak of noncommutative spaces endowed with topological,
differential and metric structures, Connes has put forward an
axiomatic scheme for ``noncommutative spin manifolds'', which in fact
is the end product of a long process of learning how to express the
concept of an ordinary spin manifold in algebraic and operatorial
terms.

A \textit{compact} noncommutative spin manifold consists of a
\textit{spectral triple} $(\A,\H,D)$, subject to the six or seven
special conditions laid out in~\cite{ConnesCollege} ---and reviewed
below in due course. Here $\A$ is a unital algebra, represented on a
Hilbert space~$\H$, together with a distinguished selfadjoint operator,
the abstract Dirac operator $D$, whose resolvent is completely
continuous, such that each operator $[D,a]$ for $a\in\A$ is bounded. A
spectral triple is even if it possesses a $\Z_2$-grading operator
$\chi$ commuting with $\A$ and anticommuting with $D$.

The key result is the reconstruction theorem
\cite{ConnesCollege,ConnesGrav} which recovers the classical geometry
of a compact spin manifold $M$ from the noncommutative setup, once the
algebra of coordinates is assumed to be isomorphic to the space of
smooth functions $\Coo(M)$. Details of this reconstruction are given
in~\cite[Chapters 10 and 11]{Polaris} and in a different vein
in~\cite{RennieSpin}.

Thus, for compact noncommutative spaces, the answer to our question is
clearly in the affirmative. Indeed the first worked examples of
noncommutative differential geometries are the noncommutative tori
(NC~tori), as introduced already in 1980~\cite{ConnesTorus,RieffelRot}.
It is a simple observation that the NC~torus can be obtained as an
ordinary torus endowed with a periodic version of the Moyal product.
The NC tori have been thoroughly exploited in
NCFT~\cite{ConnesDS,Atlas}.

The restriction to compact noncommutative spaces (``compactness'' being
a metaphor for the unitality of the coordinate algebra $\A$) is
essentially a technical one, and no fundamental obstacle to extending
the theory of spectral triples to nonunital algebras was foreseen.
However, it is fair to say that so far a complete treatment of the
nonunital case has not been written down. (There have been, of course,
some noteworthy partial treatments: one can
mention~\cite{Selene,RennieLocal}, which identify some of the
outstanding issues.) The time has come to add a new twist to the tale.

\newpage

In this article we show in detail how to build noncompact
noncommutative spin geometries. The indispensable commutative example
of noncompact manifolds is considered first. Then the geometry
associated to the Moyal product is laid out. One of the difficulties
for doing this is to pin down a ``natural'' compactification or
unitization (embedding of the coordinate algebra as an essential ideal
in a unital algebra), the main idea being that the chosen Dirac
operator must play a role in this choice.

Since the resolvent of $D$ is no longer compact, some adjustments need
to be made; for instance, we now ask for $a(D - \la)^{-1}$ to be
compact for $a \in \A$ and $\la \notin \spec D$. Then, thanks to a
variation of the famous Cwikel inequality~\cite{Cwikel,SimonTrace}
---often used for estimating bound states of Schr\"odinger operators---
we prove that the spectral triple
$$
\bigl((\SS(\R^{2N}),\star_\Theta), L^2(\R^{2N})\ox\C^{2^N},
-i\del_{\mu}\ox\ga^{\mu}\bigr),
$$
where $\SS$ denotes the space of Schwartz functions and $\star_\Theta$
a Moyal product, is $2N^+$-summable and has in fact the spectral
dimension~$2N$. The interplay between all suitable algebras containing
$(\SS(\R^{2N}), \star_\Theta)$ must be validated by the orientation and
finiteness conditions~\cite{ConnesCollege,ConnesGrav}. In so doing, we
prove that the classical background of modern-day NCFTs does fit in the
framework of the rigorous Connes formalism for geometrical
noncommutative spaces.

This accomplished, the construction of noncommutative gauge theories,
that we perform by means of the primitive form of the spectral action
functional, is straightforward. The issue of understanding the
fluctuations of the geometry, in order to develop ``noncommutative
gravity''~\cite{ChamseddineGrav} has not reached a comparable degree of
mathematical maturity, and is not examined yet. As a byproduct of our
analysis, and although we do not deal here with NCFT proper, a
mathematically satisfactory construction of the Moyal--Wick monomials
is also given.

The main results in this paper have been announced and summarized
in~\cite{GayralPtit}.

\smallskip

The first order of business is to review the Moyal product more
carefully with due attention paid to the mathematical details.

\section{The theory of distributions and Moyal analysis}
\label{sec:Moyal-basics}

In this first paragraph we fix the notations and recall basic
definitions. For any finite dimension $k$, let $\Theta$ be a real
skewsymmetric $k \x k$ matrix, let $s\.t$ denote the usual scalar
product on Euclidean $\R^k$ and let $\SS(\R^k)$ be the space of
complex Schwartz (smooth, rapidly decreasing) functions on $\R^k$. One
defines, for $f,h \in \SS(\R^k)$, the corresponding Moyal
or twisted product:
\begin{equation}
f \star_\Theta h(x) := (2\pi)^{-k} \iint
f(x - \half\Theta u) \, h(x + t) \, e^{-iu\.t} \,d^ku \,d^kt,
\label{eq:moyal-prod-slick}
\end{equation}
where $d^k x$ is the ordinary Lebesgue measure on $\R^k$. In
Euclidean field theory, the entries of~$\Theta$ have the dimensions of
an area. Because $\Theta$ is skewsymmetric, complex conjugation
reverses the product: $(f \star_\Theta h)^* = h^* \star_\Theta f^*$.

Assume $\Theta$ to be nondegenerate, that is to say,
$\sigma(s,t) := s\.\Theta t$ to be symplectic. This implies even
dimension, $k = 2N$. We note that $\Theta^{-1}$ is also skewsymmetric;
let $\th > 0$ be defined by $\th^{2N} := \det\Theta$. Then
formula~\eqref{eq:moyal-prod-slick} may be rewritten as
\begin{equation}
f \star_\Theta h(x) = (\pi\th)^{-2N} \iint
f(x + s)\, h(x + t)\, e^{-2is\.\Theta^{-1}t} \,d^{2N}s \,d^{2N}t.
\label{eq:moyal-prod-gen}
\end{equation}

The latter form is very familiar from phase-space quantum
mechanics~\cite{Iapetus}, where $\R^{2N}$ is parametrized by $N$
conjugate pairs of position and momentum variables, and the entries
of~$\Theta$ have the dimensions of an action; one then selects
$$
\Theta = \hbar S
:= \hbar \begin{pmatrix} 0 & 1_N \\ -1_N & 0 \end{pmatrix}.
$$
Indeed, the product $\star$ (or rather, its commutator) was introduced
in that context by Moyal~\cite{Moyal}, using a series development in
powers of~$\hbar$ whose first nontrivial term gives the Poisson
bracket; later, it was rewritten in the above integral form. These are
actually oscillatory integrals, of which Moyal's series development,
\begin{equation}
f\star_\hbar g(x)
= \sum_{\a\in\N^{2N}} \Bigl(\frac{i\hbar}{2}\Bigr)^{|\a|}
\frac{1}{\a!}\, \pd{f}{x^\a}(x) \, \pd{g}{(Sx)^\a}(x),
\label{eq:moyal-asymp}
\end{equation}
is an asymptotic expansion. The development~\eqref{eq:moyal-asymp}
holds ---and sometimes becomes exact--- under conditions spelled out
in~\cite{Nereid}. The first integral form~\eqref{eq:moyal-prod-slick}
of the Moyal product was exploited by Rieffel in a remarkable
monograph~\cite{RieffelDefQ}, who made it the starting point for a
more general deformation theory of $C^*$-algebras.

Since the problems we are concerned with in this paper are of
functional analytic nature, there is little point in using the most
general $\Theta$ here: we concentrate on the nondegenerate case and
adopt the form $\Theta = \th S$ with $\th$ real. Therefore, the
corresponding Moyal products are indexed by the real parameter $\th$;
we denote them by $\mop$ and usually omit explicit reference to $N$ in
the notation.

The plan of the rest of this section is roughly as follows. The
Schwartz space $\SS(\R^{2N})$ endowed with these products is an algebra
without unit and its unitization will not be unique. Below, after
extending the Moyal product to large classes of distributions, we find
and choose unitizations suitable for our construction of a noncompact
spectral triple, and show that $(\SS(\R^{2N}),\mop)$ is a
pre-$C^*$-algebra. We prove that the left Moyal product by a function
$f \in \SS(\R^{2N})$ is a regularizing operator on $\R^{2N}$. In
connection with that, we examine the matter of
Calder\'on--Vaillancourt-type theorems in Moyal analysis. We inspect as
well the relation of our compactifications with NC~tori.

\subsection{Basic facts of Moyalology}

With the choice $\Theta = \th S$ made, the Moyal product can also
be written
\begin{equation}
f \mop g(x) := (\pi\th)^{-2N} \iint f(y) g(z)\,
e^{\frac{2i}{\th}(x-y)\,\.\,S(x-z)} \,d^{2N}y \,d^{2N}z.
\label{eq:moyal-prod}
\end{equation}

Of course, our definitions make sense only under certain hypotheses on
$f$ and $g$. A good chunk of Moyal analysis can be found
in~\cite{Phobos,Deimos}, from which we extract the following lemma.

\begin{lem} {\rm\cite{Phobos}}
\label{lm:propriete}
Let $f,g\in \SS(\R^{2N})$. Then
\begin{enumerate}
\item[(i)]
$f \mop g \in \SS(\R^{2N})$.
\item[(ii)]
$\mop$ is a bilinear associative product on $\SS(\R^{2N})$. Moreover,
complex conjugation of functions $f \mapsto f^*$ is an
involution for~$\mop$.
\item[(iii)]
Let $j = 1,2,\dots,2N$. The Leibniz rule is satisfied:
\begin{equation}
\pd{}{x_j}(f\mop g) = \pd{f}{x_j} \mop g + f \mop \pd{g}{x_j}.
\label{eq:Leibniz}
\end{equation}
\item[(iv)]
Pointwise multiplication by any coordinate $x_j$ obeys
\begin{equation}
x_j (f \mop g)
= f \mop (x_j g) + \frac{i\,\th}{2} \pd{f}{(Sx)_j} \mop g
= (x_j f) \mop g - \frac{i\,\th}{2} f \mop \pd{g}{(Sx)_j}.
\label{eq:mult-rule}
\end{equation}
\item[(v)]
The product has the tracial property:
$$
\<f,g> := \frac{1}{(\pi\th)^N} \int f \mop g(x) \,d^{2N}x
= \frac{1}{(\pi\th)^N} \int g \mop f(x) \,d^{2N}x
= \frac{1}{(\pi\th)^N} \int f(x) \,g(x) \,d^{2N}x.
$$
\item[(vi)]
Let $L^\th_f \equiv L^\th(f)$ be the left multiplication
$g \mapsto f \mop g$. Then
$\lim_{\th\downarrow0}{L^\th_f\,g}(x) = f(x)\,g(x)$, for
$x \in \R^{2N}$.
\end{enumerate}
\end{lem}

Property~(vi) is a consequence of the distributional identity
$\lim_{\eps\downarrow0}\eps^{-k} e^{ia\.b/\eps} =
(2\pi)^k \delta(a) \delta(b)$, for $a,b\in\R^k$; convergence takes
place in the standard topology~\cite{Schwartz} of $\SS(\R^{2N})$. To
simplify notation, we put $\SS := \SS(\R^{2N})$ and let
$\SS' := \SS'(\R^{2N})$ be the dual space of tempered distributions.
In view of~(vi), we may denote by $L^0_f$ the pointwise product
by~$f$.

\begin{thm} {\rm\cite{Phobos}}
$\A_\th := (\SS,\mop)$ is a nonunital associative, involutive
Fr\'echet algebra with a jointly continuous product and a
distinguished faithful trace.
\end{thm}

\smallskip
Introduce the symplectic Fourier transform $F$ by
\begin{equation}
Ff(x) := (2\pi)^{-N} \int f(t) e^{ix\.St} \,d^{2N}t.
\label{eq:Fsympl}
\end{equation}
It is obviously a symmetry, i.e., an involutive selfadjoint operator.
Since $\delta \mop \delta = (\pi\th)^{-2N}$, the maps
$f \mapsto (\pi\th)^N\delta \mop f$ and
$f \mapsto f \mop (\pi\th)^N\delta$ are unitary, too; they turn out
to be
$$
[(\pi\th)^N \delta \mop f](y) = (2/\th)^N \, Ff(-2y/\th),  \qquad
[f \mop (\pi\th)^N\delta](y)  = (2/\th)^N \, Ff(2y/\th).
$$
This prompts us to consider the unitary dilation operators $E_a$ given
by
$$
E_af(x) := a^{N/2} f(a^{1/2}x),
$$
and it is immediate from~\eqref{eq:Fsympl} that $F E_a = E_{1/a} F$.
We also remark that
\begin{equation}
f \mop g =
(\th/2)^{-N/2} E_{2/\th}(E_{\th/2}f \star_2 E_{\th/2}g).
\label{eq:starscale}
\end{equation}
Nearly all formulas in this paper simplify when $\th = 2$. Thanks to
the scaling relation~\eqref{eq:starscale}, it is often enough, when
studying properties of the Moyal product, to work out the case
$\th = 2$.

\subsection{The oscillator basis}

\begin{defn}
\label{df:basis-fns}
The algebra $\A_\th$ has a natural basis of eigentransitions $f_{mn}$
of the harmonic oscillator, indexed by $m,n \in \N^N$. As usual, for
$m = (m_1,\dots,m_N) \in \N^N$, we write $|m| := m_1 +\cdots+ m_N$ and
$m! := m_1!\dots m_N!$. If
$$
H_l := \half(x_l^2 + x_{l+N}^2) \sepword{for $l=1,\dots,N$\quad and}
H := H_1 + H_2 +\cdots+ H_N,
$$
then the~$f_{mn}$ diagonalize these harmonic oscillator Hamiltonians:
\begin{align}
H_l \mop f_{mn} &= \th(m_l+\half) f_{mn},
\nonumber \\
f_{mn} \mop H_l &= \th(n_l+\half) f_{mn}.
\label{eq:moyal-haml}
\end{align}
They may be defined by
\begin{equation}
f_{mn}
:= \frac{1}{\sqrt{\th^{|m|+|n|}\,m!n!}}\,(a^*)^m \mop f_{00} \mop a^n,
\label{eq:basis}
\end{equation}
where $f_{00}$ is the Gaussian function $f_{00}(x) := 2^N e^{-2H/\th}$,
and the annihilation and creation functions respectively are
\begin{equation}
a_l := \frac{1}{\sqrt{2}} (x_l + ix_{l+N})  \sepword{and}
a_l^* := \frac{1}{\sqrt{2}} (x_l - ix_{l+N}).
\label{eq:crea-annl}
\end{equation}
One finds that $a^n := a_1^{n_1} \dots a_N^{n_N} =
a_1^{\mop n_1} \mop\cdots\mop a_N^{\mop n_N}$.
\end{defn}

These Wigner eigentransitions are already found in~\cite{Groenewold}
and also in~\cite{BartlettM}. (Incidentally, the ``first'' attributions
in~\cite{FollandPhase} are quite mistaken.) The $f_{mn}$ can be
expressed with the help of Laguerre functions in the variables $H_l$:
see subsection~\ref{sec:expl-basis} of the Appendix. The next lemma
summarizes their chief properties.

\begin{lem} {\rm\cite{Phobos}}
\label{lm:osc-basis}
Let $m,n,k,l \in \N^N$. Then $f_{mn} \mop f_{kl} = \delta_{nk}f_{ml}$
and $f_{mn}^* = f_{nm}$. Thus $f_{nn}$ is an orthogonal projector and
$f_{mn}$ is nilpotent for $m \neq n$. Moreover,
$\<f_{mn}, f_{kl}> = 2^N\,\delta_{mk}\,\delta_{nl}$. The family
$\set{f_{mn} : m,n\in\N^N} \subset \SS \subset L^2(\R^{2N})$ is an
orthogonal basis.
\end{lem}

It is clear that $\ee_K := \sum_{|n|\leq K} f_{nn}$, for $K \in \N$,
defines a (not uniformly bounded) approximate unit $\{\ee_K\}$
for~$\A_\th$.

As a consequence of Lemma~\ref{lm:osc-basis}, the Moyal product has a
matricial form.

\begin{prop} {\rm\cite{Phobos}}
Let $N = 1$. Then $\A_\th$ has a Fr\'echet algebra isomorphism with
the matrix algebra of rapidly decreasing double sequences
$c = (c_{mn})$ such that, for each $k \in \N$,
$$
r_k(c) := \biggl( \sum_{m,n=0}^\infty
\th^{2k} (m+\half)^k (n+\half)^k |c_{mn}|^2 \biggr)^{1/2}
$$
is finite, topologized by all the seminorms $(r_k)$; via the
decomposition $f = \sum_{m,n\in\N^N} c_{mn} f_{mn}$ of~$\SS(\R^2)$ in
the $\{f_{mn}\}$ basis.

For $N > 1$, $\A_\th$ is isomorphic to the (projective) tensor product
of $N$ matrix algebras of this kind.
\end{prop}

\begin{defn}
\label{df:Gst}
We may as well introduce more Hilbert spaces $\G_{st}$ (for
$s,t \in \R$) of those $f \in \SS'(\R^2)$ for which the following sum
is finite:
$$
\|f\|_{st}^2 := \sum_{m,n=0}^\infty
\th^{s+t} (m+\half)^s (n+\half)^t |c_{mn}|^2.
$$
We define $\G_{st}$, for $s,t$ now in $\R^N$, as the tensor product
of Hilbert spaces $\G_{s_1t_1} \oxyox \G_{s_Nt_N}$. In other
words, the elements $(2\pi)^{-N/2} \th^{-(N+s+t)/2}
(m+\half)^{-s/2} (n+\half)^{-t/2} f_{mn}$ (with an obvious multiindex
notation), for $m,n \in \N^N$, are declared to be an orthonormal basis
for~$\G_{st}$.
\end{defn}

If $q \leq s$ and $r \leq t$ in~$\R^N$, then
$\SS \subset \G_{st} \subseteq \G_{qr} \subset \SS'$ with continuous
dense inclusions. Moreover, $\SS = \bigcap_{s,t\in\R^N} \G_{st}$
topologically (i.e., the projective limit topology of the intersection
induces the usual Fr\'echet space topology on $\SS$) and
$\SS' = \bigcup_{s,t\in\R^N} \G_{st}$ topologically (i.e., the
inductive limit topology of the union induces the usual DF topology
on~$\SS'$). In particular, the expansion
$f=\sum_{m,n\in\N^N} c_{mn}f_{mn}$ of $f \in \SS'$ converges in the
strong dual topology.

We will use the notational convention that if $F,G$ are spaces such
that $f\mop g$ is defined whenever $f\in F$ and $g\in G$, then
$F \mop G$ is the linear span of the set
$\set{f \mop g : f\in F,\, g\in G}$; in many cases of interest, this
set is already a vector space. It is now easy to show that
$\SS \mop \SS = \SS$; more precisely, the following result holds.

\begin{prop} {\rm \cite[p.~877]{Phobos}}
\label{pr:factorization}
The algebra $(\SS,\mop)$ has the (nonunique) factorization
property: for all $h \in \SS$ there exist $f,g \in \SS$ such that
$h = f \mop g$.
\end{prop}

\subsection{Moyal multiplier algebras}

\begin{defn}
\label{df:Moyal-alg}
The Moyal product can be defined, by duality, on larger sets than
$\SS$. For $T \in \SS'$, write the evaluation on $g \in \SS$ as
$\<T, g> \in \C$; then, for $f \in \SS$ we may define $T \mop f$ and
$f \mop T$ as elements of~$\SS'$ by $\<T \mop f, g> := \<T, f \mop g>$
and $\<f \mop T, g> := \<T, g \mop f>$, using the continuity of the
star product on~$\SS$. Also, the involution is extended to $\SS'$ by
$\<T^*,g> := \overline{\<T,g^*>}$.

We shall soon argue~\cite{Phobos} that if $T \in \SS'$ and
$f \in \SS$, then $T \mop f,\,f \mop T \in C^\infty(\R^{2N})$.

Consider the left and right multiplier algebras:
\begin{align*}
\M_L^\th
&:= \set{T \in \SS'(\R^{2N}) : T \mop h \in \SS(\R^{2N})
\text{ for all } h \in \SS(\R^{2N})},
\\
\M_R^\th
&:= \set{T \in \SS'(\R^{2N}) : h \mop T \in \SS(\R^{2N})
\text{ for all } h \in \SS(\R^{2N})},
\end{align*}
and set $\M^\th := \M_L^\th \cap \M_R^\th$.
\end{defn}

It is clear from Lemma~\ref{lm:propriete}(ii) that the map
$h \mapsto T \mop h$, for $T \in \M_L^\th$, is adjointable, with
adjoint given by (left) multiplication by~$T^*$. One can then define
the Moyal products $\M_R^\th \mop \SS' = \SS'$ and
$\SS' \mop \M_L^\th = \SS'$ as well.

\begin{thm} {\rm\cite{Deimos}}
$\M^\th$ is a complete nuclear semireflexive locally convex unital
$*$-algebra with hypocontinuous multiplication and continuous
involution. Moreover, in view of the previous proposition, $\M^\th$
is the maximal compactification of $\A_\th$ defined by duality
(see {\rm\cite[Sec.~1.3]{Polaris}}).
\end{thm}

This maximal unitization $\M^\th$ of $\A_\th$ contains, beyond the
constant functions (in particular 1 is the identity), the plane waves.
By plane waves we understand all functions of the form
$x \mapsto \exp(ik\.x)$ for $k$ a $2N$-vector. They are important in
physics. Also $\M^\th$ contains the Dirac $\delta$ and all its
derivatives, and all monomials $x \mapsto x^\a$ for $\a\in\N^{2N}$.
Clearly $\M^2$ is Fourier invariant, so more generally
$F\M^\th = \M^{4/\th}$.

When $\th = 0$, the place of $\M^\th$ is taken by the space
$\Oh_M$ (``$M$'' for multiplier) of smooth functions of polynomial
growth on $\R^{2N}$ in all derivatives.

\smallskip

There is a new way of defining the Moyal product for pairs of
distributions lying in the Sobolev-like spaces
$\G_{st}$~\cite{Phobos}. If $f = \sum_{m,n} c_{mn}f_{mn}\in\G_{st}$,
$g = \sum_{m,n} d_{mn}f_{mn}\in\G_{qr}$ and if $t + q \geq 0$, then
for $a_{mn} := \sum_k c_{mk} d_{kn}$, the series
$h := \sum_{m,n} a_{mn} f_{mn}$ converges in $\G_{sr}$; $f \mop g$ is
defined, and $f \mop g = h$. Furthermore, the following useful norm
estimates hold:
$$
\|f \mop g\|_{st} \leq \|f\|_{sq} \, \|g\|_{rt}
\quad\text{whenever } q + r \geq 0.
$$
In particular, $\G_{t,-t}$ is a Banach algebra, for all $t \in \R^N$.
This is consistent with the previous definition.

We let $\G_{-\infty,t} := \bigcap_{s\in\R^N} \G_{st}$ (with the
projective limit topology) and
$\G_{s,+\infty} := \bigcup_{t\in\R^N} \G_{st}$ (with the inductive
limit topology). Then $\M_L^\th = \bigcap_{s\in\R^N}\G_{s,+\infty}$
topologically, and the strong (pre-)dual $(\M_L^\th)'$ equals
$\bigcup_{t\in\R^N} \G_{-\infty,t}$ topologically. Note in passing
that $(\M_L^\th)' \hookto \M_L^\th$ with a continuous inclusion.

\smallskip

Yet alternatively, we may work with another algebra of distributions
including $(\SS,\mop)$, to wit, the multiplier algebra of
$\G_{00} = L^2(\R^{2N})$ considered in~\cite{Kammerer,Deimos}. We
first record the analogue of Lemma~\ref{lm:propriete}.

\begin{lem} {\rm\cite{Phobos,Hansen1,Hansen2,Deimos}}
\label{lm:propriete-bis}
Let $f,g\in L^2(\R^{2N})$. Then
\begin{enumerate}
\item[(i)]
For $\th \neq 0$, $f \mop g$ lies in $L^2(\R^{2N})$. Moreover,
$f \mop g$ is uniformly continuous.
\item[(ii)]
$\mop$ is a bilinear associative product on $L^2(\R^{2N})$. The
complex conjugation of functions $f \mapsto f^*$ is an involution
for~$\mop$.
\item[(iii)]
The linear functional $f \mapsto \int f(x)\,dx$ on $\SS$ extends to
$\I_{00}(\R^{2N}) := L^2(\R^{2N}) \mop L^2(\R^{2N})$, and the product
has the tracial property:
$$
\<f,g> := (\pi\th)^{-N} \! \int f \mop g(x) \,d^{2N}x
= (\pi\th)^{-N} \! \int g \mop f(x) \,d^{2N}x
= (\pi\th)^{-N} \! \int f(x) \,g(x) \,d^{2N}x.
$$
We are not asserting that $h = f\mop g$ is absolutely integrable.
We can nevertheless find $u \in \SS'$ with $u^* \mop u = 1$ and
$|h| \in \I_{00}$ so that $h = u \mop |h|$ and $|h| = l^* \mop l$ with
$l \in \G_{00}$. Writing $\|h\|_{00,1} := \<1,|h|> = \|l\|_{00}^2$, we
obtain a Banach space norm for $\I_{00}$ such that
$\|f \mop g\|_{00,1} \leq \|f\|_{00} \|g\|_{00}$.
\item[(iv)]
$\lim_{\th\downarrow0}{L^\th_f\,g}(x) = f(x)\,g(x)$ almost
everywhere on $\R^{2N}$.
\end{enumerate}
\end{lem}

In subsection~\ref{sec:expl-basis} of the Appendix it is discussed why
$\I_{00}$ \hbox{$\subset$ \hspace{-0.37cm}/} $L^1(\R^{2N})$. Since
$f\in\I_{00}$ if and only if the Schr\"odinger representative
$\sigma^\th(f)$ is trace-class (see the proof of the next
Proposition~\ref{pr:algebra}), one can obtain sufficient conditions for
$f$ to belong in $\I_{00}$ from the treatment in~\cite{DimassiS}.

\begin{defn}
Let $A_\th := \set{T \in \SS' : T \mop g \in L^2(\R^{2N})
\text{ for all } g \in L^2(\R^{2N})}$, provided with the operator norm
$\|L^\th(T)\|_{\mathrm{op}} :=
\sup\set{\|T \mop g\|_2/\|g\|_2 : 0 \neq g \in L^2(\R^{2N})}$.

Obviously $\A_\th = \SS \hookto A_\th$. But $\A_\th$ is not dense in
$A_\th$ (see below), and we shall denote by $A^0_\th$ its closure
in~$A_\th$.
\end{defn}

Note that $\G_{00} \subset A_\th$. This is clear from the following
estimate.

\begin{lem} {\rm\cite{Phobos}}
\label{lm:norm-HS}
If $f,g \in L^2(\R^{2N})$, then $f \mop g \in L^2(\R^{2N})$ and
$\|L^\th_f\|_{\mathrm{op}} \leq (2\pi\th)^{-N/2} \|f\|_2$.
\end{lem}

\begin{proof}
Expand $f = \sum_{m,n} c_{mn} \a_{mn}$ and
$g = \sum_{m,n} d_{mn} \a_{mn}$ with respect to the orthonormal basis
$\{\a_{nm}\} := (2\pi\th)^{-N/2} \{f_{nm}\}$ of $L^2(\R^{2N})$. Then
\begin{align*}
\|f\mop g\|_2^2
&= (2\pi\th)^{-2N} \biggl\| \sum_{m,l}
\Bigl( \sum_n c_{mn}\,d_{nl} \Bigr) f_{ml} \biggr\|_2^2
= (2\pi\th)^{-N} \sum_{m,l} \Bigl|\sum_n c_{mn}\,d_{nl}\Bigr|^2
\\
&\leq (2\pi\th)^{-N} \sum_{m,j} |c_{mj}|^2 \sum_{k,l} |d_{kl}|^2
= (2\pi\th)^{-N} \|f\|_2^2 \, \|g\|_2^2,
\end{align*}
on applying the Cauchy--Schwarz inequality.
\end{proof}

The algebra $A_\th$ contains moreover $L^1(\R^{2N})$ and its Fourier
transform~\cite{Daniel}, even the bounded measures and their Fourier
transforms; the plane waves; but no nonconstant polynomials, nor
derivatives of~$\delta$. The algebra $A_\th$ is selfconjugate, and it
could have been defined using right Moyal multiplication instead.

\begin{prop} {\rm\cite{Kammerer,Deimos}}
\label{pr:algebra}
$(A_\th,\|.\|_{\mathrm{op}})$ is a unital $C^*$-algebra of
operators on $L^2(\R^{2N})$, isomorphic to $\L(L^2(\R^N))$ and
including $L^2(\R^{2N})$. Also, $(\I_{00})' = A_\th$. Moreover,
there is a continuous injection of $*$-algebras
$\A_\th \hookto A_\th$, but $\A_\th$ is not dense in~$A_\th$, namely
$A_\th^0 \subsetneq A_\th$.
\end{prop}

\begin{proof}
We prove the nondensity result. The left regular representation $L^\th$
of $A_\th$ is a denumerable direct sum of copies of the Schr\"odinger
representation $\sigma^\th$ on $L^2(\R^N)$~\cite{Johnny}. Indeed, there
is a unitary operator, the Wigner transformation
$W$~\cite{FollandPhase,Deimos}, from $L^2(\R^{2N})$ onto
$L^2(\R^N) \ox L^2(\R^N)$, such that
$$
W\,L^\th(f)\,W^{-1} = \sigma^\th(f) \ox 1.
$$
If $f\in \SS$, then $\sigma^\th(f)$ is a compact (indeed, trace-class)
operator on $L^2(\R^N)$, and so $A^0_\th$ equals
$\set{W^{-1}(T \ox 1)W : T \text{ compact}}$, while $A_\th$ itself is
$\set{W^{-1}(T \ox 1)W : T \text{ bounded}}$. Clearly the dual space
is $(A^0_\th)' = \I_{00}$. Notice as well that conjugation by~$W$
yields an explicit isomorphism between $A_\th$ and $\L(L^2(\R^N))$.
\end{proof}

Consequently, $\A_\th$ is a Fr\'echet algebra whose topology is finer
than the $\|.\|_{\mathrm{op}}$-topology. Moreover, it is stable under
holomorphic functional calculus in its $C^*$-completion $A^0_\th$, as
the next proposition shows.

\begin{prop}
\label{pr:pre-S}
$\A_\th$ is a (nonunital) Fr\'echet pre-$C^*$-algebra.
\end{prop}

\begin{proof}
We adapt the argument for the commutative case
in~\cite[p.~135]{Polaris}. To show that $\A_\th$ is stable under
the holomorphic functional calculus, we need only check that if
$f \in \A_\th$ and $1 + f$ is invertible in $A^0_\th$
with inverse $1 + g$, then the quasiinverse $g$ of $f$ must lie in
$\A_\th$. From  $f + g + f \mop g = 0$, we obtain
$f \mop f + g \mop f + f \mop g \mop f = 0$, and it is enough to show
that $f \mop g \mop f \in \A_\th$, since the previous relation then
implies $g \mop f \in \A_\th$, and then
$g = -f - g \mop f \in \A_\th$ also.

Now, $A_\th \subset \G_{-r,0}$ for any $r > N$ \cite[p.~886]{Deimos}.
Since $f \in \G_{s,p+r} \cap \G_{qt}$, for $s,t$ arbitrary and $p,q$
positive, we conclude that $f \mop g \mop f \in
\G_{s,p+r} \mop \G_{-r,0} \mop \G_{qt} \subset \G_{st}$; as
$\SS = \bigcap_{s,t\in\R} \G_{st}$, the proof is complete.
\end{proof}

The Fr\'echet algebras $\A_\th$ are automatically good (their sets of
quasiinvertible elements are open); and by an old result of
Banach~\cite{Banach}, the quasiinversion operation is continuous in a
good Fr\'echet algebra. Note that a good algebra with identity cannot
have proper (even one-sided) dense ideals. However, the nonunital
$(\M_L^\th)'$ provides an example of a good Fr\'echet algebra that
harbours $\A_\th$ as a proper dense left ideal~\cite{Charon}.

\smallskip

We noticed already that the extensions $\M^\th$ and $A_\th$ of
$\A_\th$ are quite different. Clearly $\M^\th$ is associated
with smoothness; however, even though the Sobolev-like spaces
$\G_{st}$ grow more regular with increasing $s$ and $t$~\cite{Deimos},
$\M^\th$ includes none of them; in particular,
$L^2(\R^{2N})$ \hbox{$\subset$ \hspace{-0.37cm}/} $\M^\th$
for any~$\th$.

Be that as it may, the plane waves belong both to $\M^\th$ and
$A_\th$. One obtains for the Moyal product of plane waves:
\begin{equation}
\exp(ik\,\.) \mop \exp(il\,\.)
= e^{-\frac{i}{2}\th\,k\.Sl} \, \exp(i(k+l)\.),
\label{eq:Moyal-planewave}
\end{equation}
or, reinstalling the generic Moyal product:
\begin{equation}
\exp(ik\,\.) \star_\Theta \exp(il\,\.)
= e^{-\frac{i}{2}\,k\.\Theta l} \, \exp(i(k+l)\.).
\label{eq:Moyal-planewavebis}
\end{equation}
Therefore the plane waves close to an algebra, the \textit{Weyl
algebra}. It represents the translation group of~$\R^{2N}$:
$$
\bigl(\exp(ik\,\.)\mop f \mop \exp(-ik\,\.)\bigr)(x) = f(x + \th Sk),
$$
for $f \in \SS$ or $f \in \G_{00}$, say.

\subsection{Smooth test function spaces, their duals and
the Moyal product}

Here there is a fascinating interplay. Recall that a
pseudodifferential operator $A \in \PsiDO$ on $\R^k$ is a linear
operator which can be written as
$$
A\,h(x) = (2\pi)^{-k} \iint
\sigma[A](x,\xi)\, h(y)\, e^{i\xi\.(x-y)} \,d^k\xi \,d^ky.
$$
Let $\Psi^d := \set{A\in\Psi DO : \sigma[A] \in S^d}$ be the class of
$\PsiDO$s of order $d$, with
$$
S^d := \set{\sigma \in \Coo(\R^k \times \R^k) :
|\del_x^\a \del_\xi^\b \sigma(x,\xi)|
\leq C_{K\a\b}(1 + |\xi|^{2})^{(d-|\b|)/2} \text{ for } x \in K},
$$
where $K$ is any compact subset of $\R^k$, $\a,\b\in\N^k$, and
$C_{K\a\b}$ is some constant. Also
$\Psi^{\infty} := \bigcup_{d\in\R} \Psi^d$ and
$\Psi^{-\infty} := \bigcap_{d\in\R}\Psi^d$. Recall, too, that a
$\PsiDO$ $A$ is called regularizing or smoothing if
$A \in \Psi^{-\infty}$, or equivalently~\cite{Hormander,Shubin}, if
$A$ extends to a continuous linear map from the dual of the space of
smooth functions $\Coo(\R^k)$ to itself.

\begin{lem}
\label{lm:cojoreg}
If $f \in \SS$, then $L^\th_f$ is a regularizing $\PsiDO$.
\end{lem}

\begin{proof}
{}From~\eqref{eq:moyal-prod-slick}, one at once sees that left Moyal
multiplication by~$f$ is the pseudodifferential operator on~$\R^{2N}$
with symbol $f(x - \frac{\th}{2}S\xi)$. Clearly $L^\th_f$ extends to a
continuous linear map from
$\Coo(\R^{2N})' \hookto \SS'$ to $\Coo(\R^{2N})$.
The lemma also follows from the inequality
$$
|\del_x^\a \del_\xi^\b f(x - \tfrac{\th}{2} S\xi)|
\leq C_{K\a\b} (1 + |\xi|^2)^{(d-|\b|)/2},
$$
valid for all $\a,\b \in \N^{2N}$, any compact $K \subset \R^{2N}$,
and any $d \in \R$, since $f\in \SS$.
\end{proof}

\begin{rem}
Unlike for the case of a compact manifold, regularizing $\PsiDO$s are
not necessarily compact operators. For instance, for each $n$,
$L^\th(f_{nn})$ possesses the eigenvalue~1 with infinite multiplicity,
so it cannot be compact.
\end{rem}

\begin{defn}
\label{df:distr-zoo}
For $m \in \N$, $f \in C^m(\R^k)$ ---functions with $m$ continuous
derivatives--- and $\ga,l \in \R$, let
$$
q_{\ga lm}(f) := \sup\set{(1 + |x|^2)^{(-l+\ga|\a|)/2}|\del^\a f(x)| :
x\in\R^k,\ |\a| \leq m};
$$
and then let $\underline{\V}^m_{\ga,l}$, respectively $\V^m_{\ga,l}$,
be the space of functions in $C^m(\R^k)$ for which
$$
(1 + |x|^2)^{(-l+\ga|\a|)/2} \,\del^\a f(x)
$$
vanishes at infinity for all $|\alpha|\leq m$, respectively is finite
for all $x \in \R^k$, normed by $q_{\ga lm}$. Note that
$\underline{\V}^m_{0,l}$ is Horv\'ath's space
$\SS_{-2l}^m$~\cite{Horvath}. We define
$$
\V_\ga := \bigcup_{l\in\R}\,\bigcap_{m\in\N} \V^m_{\ga,l},
\sepword{and, more generally,}
\V_{\ga,l} := \bigcap_{m\in\N} \V^m_{\ga,l},
$$
so that $\V_\ga = \bigcup_{l\in\R} \V_{\ga,l}$. Particularly
interesting cases include the space $\K := \V_1$ of
Grossmann--Loupias--Stein functions~\cite{GrossmannLS}, whose dual
$\K'$ is the space of Ces\`aro-summable distributions~\cite{Odysseus},
the space $\Oh_C := \V_0$ whose dual $\Oh'_C$ is the space of
convolution multipliers (Fourier transforms of $\Oh_M$), and the space
$\Oh_T := \V_{-1}$~\cite{Phobos}. Similarly, $\K_r := \V_{1,r}$ and
$\Oh_r := \V_{0,r}$ are defined. We see that
$$
\SS   = \bigcap_{m\in\N}\,\bigcap_{l\in\R} \underline{\V}^m_{0,l},
\qquad
\Oh_M = \bigcap_{m\in\N}\,\bigcup_{l\in\R} \underline{\V}^m_{0,l}.
$$
Following Schwartz, we denote $\B := \Oh_0$, the space of smooth
functions bounded together with all derivatives.

We shall also need
$\dot{\B} := \bigcap_{m\in\N} \underline{\V}^m_{0,0}$, the space of
smooth functions vanishing at infinity together with all derivatives,
and the weighted test space $\D_{L^2}$, the space of elements of
$L^2(\R^{2N})$ all of whose (distributional) derivatives also lie in
$L^2$~\cite{OrtnerW,Schwartz}; by Sobolev's lemma, these are in fact
smooth functions and moreover $\D_{L^2}\subset\dot{\B}$
\cite{Schwartz}: actually if $f \in \D_{L^2}$, then the (ordinary)
Fourier transform $\F(f)$ satisfies $(1+|\xi|^{2n}) \F(f) \in L^2$ for
all integer $n$, and by the Cauchy--Schwarz inequality $\F(f) \in L^1$,
thus $f$ tends to zero at infinity. In the notation of~\cite{Treves},
$\D_{L^2}$ is $H^{2,\infty}$.
\end{defn}

There are continuous inclusions
$\D \hookto \V_{\ga} \hookto \V_{\ga'} \hookto \Oh_M \hookto \D'$ for
$\ga > \ga'$; these are all normal spaces of distributions, namely,
locally convex spaces which include $\SS$ as a dense subspace and are
continuously included in $\SS'$. Also $\D_{L^2}$ (density of~$\SS$ in
this space follows from density of the Schwartz functions in $L^2$ and
invariance of~$\SS$ under derivations) and $\M^\th_L$, $\M^\th_R$ and
$\M^\th$~\cite{Deimos} are normal space of distributions.

By the way, there are suggestive Tauberian-type theorems for these
spaces, establishing when their intersections with their respective
dual spaces are included in~$\SS$. Concretely, we quote the following
result from~\cite{EstradaTauber}.

\begin{prop}
\label{pr:thin-testfns}
If $\CC$ is a space of smooth functions on~$\R^{2N}$ which is closed
under complex conjugation, and if the pointwise product space $\K\CC$
lies within $\CC$, then $\CC \cap \CC' \subseteq \SS$.
\end{prop}

In particular, $\V_\ga \cap \V'_\ga = \SS$ for $\ga \leq 1$. Also
$\Oh_M \cap \Oh'_M = \SS$ and $\Coo \cap ({\Coo})' = \D \subset \SS$.

\smallskip

Now, what can be said about the relation of all these spaces with
$\M^\th$? In~\cite{Phobos} it is established that $\Oh_T'$, and a
fortiori $\Oh_M'$, is included in~$\M^\th$, for all $\th$. Therefore
by Fourier analysis $\Oh_C$ is included in~$\M^\th$ for all $\th$, and
$g\mop f$ is defined as a tempered distribution whenever
$f,g \in \Oh_C$. Growth estimates may be obtained as follows. It is
true that $\Oh_C = \bigcup_{r\in\R} \Oh_r$ topologically. If
$g \in \Oh_r$ and $f \in \Oh_s$, the following crucial proposition
shows that the $\Oh_r$ spaces have similar behaviour under pointwise
and Moyal products.

\begin{prop}
\label{pr:hector}
The space $\Oh_C$ is an associative $*$-algebra under the Moyal
product. In fact, the Moyal product is a jointly continuous map from
$\Oh_r \x \Oh_s$ into $\Oh_{r+s}$, for all $r,s \in \R$. Moreover,
$\A_\th$ is a two sided essential ideal in $\Oh_C$.
\end{prop}

\begin{proof}
For the reader's convenience, we reproduce part of Theorem~2
of~\cite{Amalthea}. Let $f \in \Oh_r$ and $g \in \Oh_s$. By the
Leibniz rule for the Moyal product,
$\del^\a(f \mop g) =
\sum_{\b+\ga=\a} \binom{\a}{\b}\, \del^\b f \mop \del^\ga g$. Hence we
need only show that there are constants $C_{rsm}$ such that
\begin{equation}
(1 + |x|^2)^{-(r+s)/2} |(\del^\b f \mop \del^\ga g)(x)|
\leq C_{rsm} \,q_{0rm}(f) \,q_{0sm}(g)
\label{eq:hectoreq}
\end{equation}
for all $x \in \R^{2N}$, for large enough $m \geq |\b| + |\ga|$. If
$k \in \N$ (to be determined later), we can write
\begin{align*}
(\del^\b f\mop \del^\ga g)(x)
&= (\pi\th)^{-2N} \iint \frac{\del^\b f(x+y)}{(1+|y|^2)^k}\,
\frac{\del^\ga g(x+z)}{(1+|z|^2)^k}\,
(1+|y|^2)^k (1+|z|^2)^k e^{\frac{2i}{\th}y\.Sz} \,d^{2N}y \,d^{2N}z
\\
&= (\pi\th)^{-2N} \iint \frac{\del^\b f(x+y)}{(1+|y|^2)^k}\,
\frac{\del^\ga g(x+z)}{(1+|z|^2)^k}\,
P_k(\del_y,\del_z) \bigl[e^{\frac{2i}{\th}y\.Sz}\bigr]
\,d^{2N}y \,d^{2N}z
\\
&= (\pi\th)^{-2N} \iint e^{\frac{2i}{\th}y\.Sz}\,P_k(-\del_y, -\del_z)
\biggl[ \frac{\del^\b f(x+y)}{(1+|y|^2)^k}\,
\frac{\del^\ga g(x+z)}{(1+|z|^2)^k}\biggr] \,d^{2N}y\,d^{2N}z,
\end{align*}
where $P_k$ is a polynomial of degree $2k$ in both $y$ and $z$
variables. From the elementary estimates
$|\del^\a((1 + |x|^2)^{-k})| \leq c_{\a,k} (1 + |x|^2)^{-k}$ it
follows that
\begin{align*}
&|\del^\b f \mop \del^\ga g|(x)
\leq \sum_{k',k''\leq 2k} C'_{k'k''}
\iint \left|\frac{\del^{\b+k'}f(x+y)}{(1+|y|^2)^k}
\frac{\del^{\ga+k''}g(x+z)}{(1+|z|^2)^k}\right| \,d^{2N}y\,d^{2N}z
\\
&\qquad \leq C''_{rsm} \,q_{0rm}(f) \,q_{0sm}(g)\,
\iint \frac{(1 + |x+y|^2)^{r/2}}{(1+|y|^2)^k}
\frac{(1 + |x+z|^2)^{s/2}}{(1+|z|^2)^k} \,d^{2N}y\,d^{2N}z
\\
&\qquad \leq C'''_{rsm} \,q_{0rm}(f) \,q_{0sm}(g)\,
(1+|x|^2)^{(r+s)/2} \int (1+|y|^2)^{r/2-k} \,d^{2N}y
\int (1+|z|^2)^{s/2-k} \,d^{2N}z,
\end{align*}
provided $m \geq |\b| + |\ga| + 2k$; here the Cauchy inequality
$1 + |x + y|^2 \leq 2(1 + |x|^2)(1 + |y|^2)$ has been used to extract
the $x$~variables. If we now choose $k > N + \max\{r,s\}/2$ (and
therefore take $m \geq |\b| + |\ga| + 2N + \max\{r,s\}$), the
integrals will be finite. The joint continuity now follows directly
from the estimates~\eqref{eq:hectoreq}.

That $\SS$ is a two-sided ideal in $\Oh_C$ follows from the inclusion
$\Oh_C \subset \M^\th$. Essentiality for the ideal $\SS = \A_\th$ is
equivalent~\cite[Prop.~1.8]{Polaris} to $g \mop \SS \neq 0$ for any
nonzero $g \in \Oh_s$; but if $g \mop f_{mn} = 0$ for all $m,n$, then
in the expansion $g = \sum_{m,n} c_{mn} f_{mn}$ (as an element
of~$\SS'$, say) all coefficients must vanish, so that $g = 0$.
\end{proof}

Similar results hold for $\V_\ga$ when $\ga > 0$. Indeed, the Moyal
product $(f,g) \mapsto f \mop g$ is a jointly continuous map from
$\K_r \x \K_s$ into $\K_{r+s}$; moreover,
$f \mop g - fg \in \K_{r+s-2}$, which is a bonus for semiclassical
analysis (while on the contrary the similar statement for $\Oh_r \x
\Oh_s$ is in general false). For $\ga < 0$, we lose control of the
estimates; indeed, Lassner and Lassner~\cite{LassnerL} gave an example
of two functions in~$\Oh_T$ whose twisted product can be defined but is
not a smooth function, but rather a distribution (of noncompact
support). Also, in the next subsection we prove by counterexample that
$\Oh_T$ \hbox{$\subset$ \hspace{-0.37cm}/} $\M_L^\th$. The integral
estimates on the derivatives of $g\mop f$ can be refined to show that
in fact $\Oh_M \mop \Oh_C = \Oh_M$. However, since these estimates
depend on the order of the derivatives in a complicated way, it is
doubtful that the twisted product can be extended to~$\Oh_M$.

The regularizing property of~$\mop$ proved at the beginning of the
section can be vastly improved, as follows.

\begin{prop} {\rm\cite{Phobos}}
If $T \in \SS'$ and $f \in \SS$, then $T \mop f$ and $f \mop T$ lie in
$\Oh_T$. Moreover, these bilinear maps of $\SS' \x \SS$ and
$\SS \x \SS'$ into $\Oh_T$ are hypocontinuous.
\end{prop}

In fact, $\SS \mop \SS'$ equals $(\M_L^\th)'$, so the latter is made
of smooth functions. But $(\M_L^\th)' \cap (\M_L^\th)'' =
(\M_L^\th)' \cap \M_L^\th = (\M_L^\th)' \supsetneq \SS$; so
$(\M_L^\th)'$ and $(\M_R^\th)'$ do not satisfy the conclusion of
Proposition~\ref{pr:thin-testfns}. (Here $''$ of course denotes the
strong bidual space, not a bicommutant.) As distributions, the
elements of $(\M_L^\th)'$ and $(\M_R^\th)'$ belong to $\Oh'_C$, and a
fortiori they are Ces\`aro summable~\cite{Odysseus}.

\smallskip

Finally, it is important to know when smooth functions give rise to
elements of $A^0_\th$ or~$A_\th$. Sufficient conditions are the
following (quite strong) results of the Calder\'on--Vaillancourt
type \cite{FollandPhase,Howe}.

\begin{thm}
\label{th:CVai}
The inclusion $\V_{0,0}^{2N+1} \subset A_\th$ holds. In particular,
$\B \subset A_\th$. The inclusion
$\underline{\V}_{00}^{2N+1} \subset A^0_\th$ also holds. In
particular, $\dot{\B} \subset A^0_\th$. Moreover, if
$b\in\V_{0,0}^{2N+1}$ belongs to $A^0_\th$, then
$b\in\underline{\V}_{0,0}^{2N+1}$.
\end{thm}

We have also proved that the function space $\B$ is a $*$-algebra
under the Moyal product $\mop$ for any~$\th$, in which $\A_\th$ is a
two sided essential ideal. Recall that
$\D_{L^2} \subset \dot{\B} \subset \M^\th$.
We will now show that $\D_{L^2}$ is a $*$-algebra under the Moyal
product as well.

\begin{lem}
\label{lm:DL2}
$(\D_{L^2},\mop)$ is a $*$-algebra with continuous product and
involution. Moreover, it is an ideal in~$(\B,\mop)$.
\end{lem}

\begin{proof}
The closure under the twisted product follows from the Leibniz rule
and Lemma~\ref{lm:norm-HS}:
$$
\|\del^\a(f \mop g)\|_2 \leq (2\pi\th)^{-N/2} \sum_{\b\leq\a}
\binom{\a}{\b}\, \|\del^\b f\|_2 \,\|\del^{\a-\b}g\|_2.
$$
This also shows that the product is separately continuous, indeed
jointly continuous since $\D_{L^2}$ is a Fr\'echet space. The
continuity of the involution $f \mapsto f^*$ is immediate.

The fact that $\D_{L^2}$ is a two sided ideal in $\B$ comes directly
from the stability of these spaces under partial derivations and from
the inclusion $\B \subset A_\th$ given by the previous theorem,
since then $\|\del^\a f \mop \del^\b g\|_2 < \infty$ for all
$f \in \B$, $g \in \D_{L^2}$ and all $\a,\b \in \N^{2N}$.
\end{proof}

\subsection{The preferred unitization of the Schwartz Moyal algebra}

As with Stone--\v{C}ech compactifications, the algebras $\M^\th$ are too vast
to be of much practical use (in particular, to define noncommutative vector
bundles). A more suitable unitization of~$\A_\th$ is given by the algebra
$\Aun_\th := (\B,\mop)$. This algebra possesess an intrinsic characterization
as the smooth commutant of \emph{right} Moyal multiplication (see our comments
at the end of subsection~4.5). The inclusion of~$\A_\th$ in~$\B$ is not dense,
but this is not needed. $\Aun_\th$ contains the constant functions and the
plane waves, but no nonconstant polynomials and no imaginary-quadratic
exponentials, such as $e^{iax_1x_2}$ in the case $N = 1$ (we will see later
the pertinence of this).

\begin{prop}
\label{pr:pre-Aun}
$\Aun_\th$ is a unital Fr\'echet pre-$C^*$-algebra.
\end{prop}

\begin{proof}
We already know that $\B$ is a unital $*$-algebra with the Moyal
product, and that~$\mop$ is continuous in the topology of the
Fr\'echet space $\B$ defined by the seminorms $q_{00m}$, for
$m \in \N$. Its elements have all derivatives bounded, and so are
uniformly continuous functions on~$\R^{2N}$, as are their derivatives:
the group of translations $\tau_y f = f(\. - y)$, for $y \in \R^{2N}$,
acts strongly continuously on $\Aun_\th$ (i.e., $y\mapsto\tau_y f$ is
continuous for each~$f$).

This action preserves the seminorms $q_{00m}$, and it is clear that
$\B$ is a subspace of the space of smooth elements for~$\tau$, which
we provisionally call $A_\th^\infty$. The latter space has its own
Fr\'echet topology, coming from the strongly continuous action.
Rieffel~\cite[Thm.~7.1]{RieffelDefQ} proves two important properties
in this setting: firstly, based on a density theorem of Dixmier and
Malliavin~\cite{DixmierM}, that the inclusion $\B\hookto A_\th^\infty$
is continuous and dense. Secondly, using a ``$\Theta$-twisting'' of
$C^*$-algebras with an $\R^k$-action which
generalizes~\eqref{eq:moyal-prod-slick}, whereby the pointwise product
can be recovered as $(\B,\star_0) = (\Aun_\th,\star_{-\th})$, one
obtains the reverse inclusion; thus, $\B = A_\th^\infty$. (Thus, the
smooth subalgebra is independent of~$\Theta$.)

It is now easy to show that $\Aun_\th$, as a subalgebra of the
$C^*$-algebra $A_\th$, is stable under the holomorphic functional
calculus. Indeed, since $G(\tau_y(f)) = \tau_y(G(f))$ for any function
$G$ which is holomorphic in the neighbourhood of
$\spec L^\th(f) = \spec L^\th(\tau_y(f))$, it is clear that
$f \in \Aun_\th$ entails $G(f) \in \Aun_\th$.
\end{proof}

Clearly the $C^*$-algebra completion of $\Aun_\th$ properly contains
$A^0_\th$; it is not known to us whether it is equal to $A_\th$. At any
rate, $\Aun_\th \equiv \B$ is nonseparable as it stands; there is,
however, another topology on $\B$, induced by the topology of
$\Coo(\R^{2N})$ \cite[p.~203]{Schwartz}, under which this space is
separable. That latter topology is very natural in the context of
commutative and Connes--Landi spaces (see subsections~3.3 and~3.4). To
investigate its pertinence in the context of Moyal spaces would take us
too far afield.

An advantage of~$\Aun_\th$ is that the covering relation of the
noncommutative plane to the NC torus is made transparent. To wit, the
smooth noncommutative torus algebra $\Coo(\T_\Theta^{2N})$ can be
embedded in~$\B$ as periodic functions (with a fixed period
parallelogram). In that respect, it is well to recall
\cite{RieffelComQGp,Larissa} how far the algebraic structure of
$\Coo(\T_\Theta^{2N})$ can be obtained from the integral
form~\eqref{eq:moyal-prod-slick} of (a periodic version of) the Moyal
product.

Anticipating on the next section, we finally note the main reason for
suitability of $\Aun_\th$, namely, that each
$[\Dslash, L^\th(f) \ox 1_{2^N}]$ lies in $A_\th \ox M_{2^N}(\C)$, for
$f \in \Aun_\th$ and $\Dslash$ the Dirac operator on~$\R^{2N}$.

The previous proposition has another useful consequence.

\begin{cly}
\label{cr:pre-DL2}
$(\D_{L^2},\mop)$ is a (nonunital) Fr\'echet pre-$C^*$-algebra, whose
$C^*$-completion is $A^0_\th$.
\end{cly}

\begin{proof}
The argument of the proof of Proposition~\ref{pr:pre-S} applies, with
the following modifications. Firstly,
$\SS \subset \D_{L^2} \subset A^0_\th$ with continuous inclusions, so
that $A^0_\th$ is indeed the $C^*$-completion of $(\D_{L^2},\mop)$.
Indeed, for the second inclusion one can notice that if
$f \in \D_{L^2}$, then $W\,L^\th(f)\,W^{-1} = \sigma^\th(f) \ox 1$
where $\sigma^\th(f)$ is a Hilbert--Schmidt operator, hence compact.
The same conclusion follows from Theorem~\ref{th:CVai}.

Secondly, if $f \in \D_{L^2}$ has a quasiinverse $g \in A^0_\th$, then
the previous proposition shows that $g \in \Aun_\th$, too. Since
$(\D_{L^2},\mop)$ is an ideal in~$\Aun_\th$ by Lemma~\ref{lm:DL2}, we
conclude that $f \mop g \mop f \in \D_{L^2}$, which is enough to
establish that $g \in \D_{L^2}$.
\end{proof}

\smallskip

A relevant group of inner automorphisms of the ``big algebras''
$\M^\th$ or $A_\th$ is given by the \textit{metaplectic
representation}. Real symplectic $2N \x 2N$ matrices act on functions
by $Mf(x) := f(M^{-1}x)$. We can consider inhomogeneous symplectic
transformations, i.e., affine transformations leaving the symplectic
structure invariant. Let $(s,M)$ denote an element of the
inhomogeneous symplectic group $ISp(2N,\R)$, i.e., the semidirect
product of the group of translations and the symplectic group, with
group law
$$
(s_1, M_1) (s_2, M_2) = (M_2^{-1}s_1 + s_2, M_1M_2),
$$
acting by
\begin{equation}
(s,M)f(x) = f(M^{-1}x - s).
\label{eq:group-act}
\end{equation}
The equivariance of the twisted product is readily checked:
\begin{equation}
(s,M)f \,\mop\, (s,M)g = (s,M)(f \mop g).
\label{eq:lin-symplinv}
\end{equation}

We concentrate on the homogeneous $(0,M)$ transformations. The
symplectic action is realized by the adjoint $\star$-action of
unitaries $E(M,\.)$, belonging also to the multiplier Moyal
algebra~$\M^\th$. They constitute a variant of the metaplectic
representation; $E(M,\.)$ is a distribution on the space of smooth
sections of a nontrivial line bundle over $ISp(2N,\R)$, that works like
the exponential kernel of a noncommutative Fourier transform:
\begin{equation}
E(M,\.)\, \mop f \mop \, E(M,\.)^* = Mf,
\label{eq:internaut}
\end{equation}
for all $f \in \SS$ or $f \in L^2(\R^{2N})$ or even $f \in \SS'$.
Explicitly, for elements $M$ of $Sp(2N,\R)$ which are
``nonexceptional'', i.e., $\det(1 + M) \neq 0$, there is the
presentation
\begin{equation}
E(M,x) = e^{i\a}\, \frac{2^N}{\sqrt{\det(1+M)}}
\exp\biggl(-ix\,\.S\,\frac{1-M}{\th(1+M)}\,x\biggr).
\label{eq:internautbis}
\end{equation}
Thus, such $E(M,.)$ are imaginary-quadratic exponentials; the quadratic
form in the exponent is actually an important symplectic invariant,
solving a modified Hamilton--Jacobi equation, introduced by
Poincar\'e~\cite{Poincare} and nowadays all but forgotten. The phase
prefactor in~\eqref{eq:internautbis} reflects the ambiguity inherent
in~\eqref{eq:internaut}, which can be reduced to a sign, so that
$$
E(M,\.) \mop E(M',\.) = \pm E(MM',\.).
$$
The curious reader can directly check the last two formulas, aided by
the method of the stationary phase; a look
at~\cite{AmietH,Triton,Ariel} will help.

In contradistinction to the Weyl algebra, the $E(M,\.)$ do not belong
to $\B$, and so they yield \textit{outer} automorphisms
of~$\Aun_\th$ ---and of course of~$\A_\th$.

\smallskip

Note that the values
$\exp\bigl(\pm 2i\th^{-1}(x_1x_{N+1} +\cdots+ x_Nx_{2N})\bigr)$ are
never reached by $E$ in~\eqref{eq:internautbis}. For good reason:
these functions do not belong to the multiplier algebras $\M^\th$ or
$A_\th$, as the following lemma shows.

\begin{lem}
Let $h_a(x) := \exp\bigl(ia(x_1x_{N+1} +\cdots+ x_Nx_{2N})\bigr)$
for $a\ne0$. Then $h_a \in \M^\th$, or $h_a \in A_\th$, if and only if
$|a| \neq 2/\th$.
\end{lem}

\begin{proof}
We show this for $N = 1$, the general case follows immediately. In
view of~\eqref{eq:starscale}, it suffices to consider the case
$\th = 2$. We must determine whether $h_a \star_2 f_{mn} \in \SS$;
because of the multiplication rule~\eqref{eq:mult-rule}, it is enough
to check this for the Gaussian function~$f_{00}$.
{}From~\eqref{eq:moyal-prod},
$$
h_a \star_2 f_{00}(x) = \frac{1}{2\pi^2} \iint \exp\bigl(
iay_1y_2 - \half z_1^2 - \half z_2^2
+ i(x_1 - y_1)(x_2 - z_2) - i(x_2 - y_2)(x_1 - z_1)\bigr)\,d^2y\,d^2z.
$$
With $u = (y_1,y_2,z_1,z_2)$, the integral is of the type
$\int \exp(-\thalf u\.Qu - iu\.R_x) \,d^4u$, where the quadratic form
$u\.Qu$, with $\Re Q \geq 0$, is degenerate if and only if
$\det Q = a^2 - 1 = 0$. Thus if $|a| = 1$, then
$h_a \star_2 f_{00} \notin \SS$, and also
$h_a \star_2 f_{00} \notin L^2$, while if $|a| \neq 1$, an explicit
calculation shows that $h_a \star_2 f_{00} \in \SS$.
\end{proof}

This shows, by the way, that $\Oh_T$
\hbox{$\subset$ \hspace{-0.37cm}/} $\M^\th$ and that the
$\M^\th$ and $A_\th$ for different $\th$ are all distinct spaces of
tempered distributions.

\smallskip

Next we look briefly at the derivations of $\A_\th$, $\Aun_\th$. Linear
functions, not belonging to $\B\equiv\Aun_\th$ either, at the
infinitesimal level double as Hamiltonians for the translations;
quadratic functions double as Hamiltonians for linear
symplectomorphisms. For $h$ affine quadratic
$$
[h,f]_\mop = i\th \, \{h,f\},
$$
that is, the Moyal and Poisson brackets in this case essentially
coincide. (Note that the derivations of~$\Aun_\th$ corresponding to
quadratic Hamiltonians are unbounded.)

On the other hand, all derivations of $\M^\th$ are inner, as it is
easily proved using the Poincar\'e lemma in the distributional
context~\cite{DuboisVKMM}.

An important task is to compute the Hochschild cohomologies of $\A_\th$
and $\Aun_\th$, as Connes did for the NC~torus in~\cite{ConnesNCDiffG};
we have already seen that they are not entirely trivial. The ``big
algebras'' $\M^\th$ and $A_\th$, on the other hand, risk having
uninteresting cohomology.

Rennie has proposed to equip nonunital noncommutative algebras $\A$
like $\A_\th$ with a ``local ideal'' $\A_c \subset
\A$~\cite{RennieLocal}, which would be a noncommutative generalization
of the space $\Coo_c(M)$ of smooth functions with compact support. A
Fr\'echet algebra $\A$ is local in his sense if it has a dense ideal
$\A_c$ with local units; an algebra $\A_c$ has local units when, for
any finite subset of elements $\{a_1,\dots,a_k\}$ of~$\A_c$, there
exists $u \in \A_c$ such that $ua_i = a_iu = a_i$ for $i = 1,\dots,k$.

Certainly the Moyal product $\mop$ is not ``local'' in the ordinary
sense: the formulas~\eqref{eq:moyal-asymp} and~\eqref{eq:moyal-prod}
are two different definitions, as may be noticed in the simple example
of a couple $f,g$ with disjoint supports; then \eqref{eq:moyal-asymp}
gives zero outside the supports; while~\eqref{eq:moyal-prod} does not.
The algebras $\A_\th$ are not known to have bilateral ideals; it is
very likely that they are simple, and if so, they would not be local in
the sense of~\cite{RennieLocal}, either (thus, it is not clear if
Rennie's device can carry the full weight of noncommutative spin
geometry).

However, one can define a useful weaker notion of locality:

\begin{defn}
A Fr\'echet algebra $\A$ is \textit{quasilocal} if it has a dense
\textit{$*$-subalgebra} $\A_c$ with local units.
Here, we choose
$$
\A_c := \bigcup_{K\in\N} \A_{c,K},  \sepword{where}  \A_{c,K} :=
\biggl\{f\in\SS : f = \sum_{0\leq|m|,|n|\leq K} c_{mn} f_{mn}\biggr\}.
$$
That is, $\A_c$ is the algebra of finite linear combinations of the
$\set{f_{mn} : m,n \in \N^N}$; it possesses local units, and so
$\A_\th$ is quasilocal.
\end{defn}

Rennie further argues that possession of a local ideal in his sense
guarantees $H$-unitality~\cite{Wodzicki} of the original algebra.
Certainly our $\A_c$ is algebraically $H$-unital, as it possesses local
units~\cite{StrassbourgMeister}. It would be good to know whether
$\A_\th$ is topologically $H$-unital.

\section{Axioms for noncompact spin geometries}

\subsection{Generalization of the unital case conditions}
\label{sec:genld-axioms}

To define and construct noncommutative spin manifolds, one starts from
an operatorial version of ordinary spin geometry, that can be
generalized to noncommutative manifolds. Ideally, one should prove a
reconstruction theorem, allowing to recover all of the (topological,
smooth, geometrical) concrete structure from the abstract geometry
over a suitable commutative algebra; this has been performed to
satisfaction for compact manifolds without boundary
\cite{ConnesReal,ConnesCollege,Polaris,RennieSpin}. However, to rush
to that at the present stage would not do. It is better for now to
patiently listen to what the possible examples have to say.

As the first part of our task, therefore, we seek a collection of
Connes-like axioms for not necessarily compact noncommutative
manifolds. Such a list of conditions should be compatible with the
previous axiomatic framework, and be fulfilled by noncompact
commutative manifolds. We expect it also to encompass other
interesting cases. Our main task will then be to prove that
noncommutative Moyal-product algebras constitute one of the examples.
(To eventually reach this goal, we use heavy machinery wholesale; we
do not claim to have the ``best'' proofs.)

The discussion in this section will be relatively informal; a formal
proposal is made in the next one.

We set out by discussing what a \textit{real noncompact spectral
triple} might be. As mentioned in the Introduction, the basic data
$(\A,\H,D)$ ---or $(\A,\H,D,\chi)$--- for a spectral triple consist of
an algebra $\A$ represented by bounded operators on a Hilbert space
$\H$ and an unbounded selfadjoint operator $D$ on~$\H$, such that each
commutator $[D,a]$, for $a\in\A$ (densely defined as an operator
on~$\H$) extends to a bounded operator; it is understood that
$a\Dom D \subseteq \Dom D$.

To get an idea of the difficulties involved in the choice of $\A$,
consider the commutative case, say of the manifold $\R^k$. Depending
on the fall-off conditions deemed suitable, the smooth nonunital
algebras that can represent the manifold are numerous as the stars in
the sky. The problem is compounded in the noncommutative case, say
when $\A$ is a deformation of an algebra of functions. To be on the
safe side, we will take a relatively small algebra at the start of our
investigation of the Moyal examples; during its course, a larger
candidate will emerge.

Also, when $\A$ is not unital, we need choose a preferred unitization
$\Aun$. Consideration of the links to $K$-theory and $K$-homology
makes it prudent to require that $\A,\Aun$ be pre-$C^*$-algebras,
whose $K$-theories then coincide with those of their respective
$C^*$-completions~\cite{Book}.

Denote by $\K(\H)$ the compact operators on $\H$, and by $\L^p(\H)$ the
Schatten ideal in $\K(\H)$ defined by a finite norm
$\|A\|_p := \Tr(|A|^p)^{1/p}$, for $p \geq 1$. For compact or unital
spectral triples, it is further required that the operator $D$ have
compact resolvent, that is, $(D - \la)^{-1}$ must belong to $\K(\H)$
for $\la \notin \spec D$. Consequently $D$ must have discrete
spectrum of finite multiplicity. Since this is clearly not the case
for the Dirac operator $\Dslash$ on $\R^k$, in the nonunital case we
only demand~\cite{ConnesReal} that $a(D - \la)^{-1}$ be compact for
$a \in \A$. This condition ensures that the spectral triple
$(\A,\H,D)$ corresponds to a well-defined $K$-homology
class~\cite[Chap.~10]{HigsonR}, and could be termed `Axiom~0' for an
---in general noncompact--- noncommutative geometry.

\smallskip

We turn to the several conditions which spectral triples must satisfy
to yield noncommutative spin geometries. To formulate the
generalization to the noncompact case, we focus first on commutative
geometries. First in line there is a \textit{summability} condition,
namely that the operators $a(D^2 + \eps^2)^{-1}$ be not merely
compact, but belong to the generalized Schatten class called
$\L^{k+}$, with $k$ an integer; this is a kind of $k$th~root in the
sense of operator products of the Dixmier trace class
$\L^{1+}$~\cite{Book,Polaris}. More concretely, a compact operator $T$
belongs to $\L^{k+}$ if its singular values satisfy
$\mu_m(T) = O(m^{1/k})$ as $m \to \infty$.

In the compact commutative case of a $k$-dimensional spin manifold,
choosing $D$ to be the ordinary Dirac operator $\Dslash$ on a spinor
space $\H$, one finds that
$$
\Tr^+(a|\Dslash|^{-k}) = C_k\int_M a(x)\,d^kx,
$$
where $\Tr^+$ denotes any Dixmier trace, for a universal
constant~$C_k$. For the noncompact commutative case, we expect
$\Tr^+(\.|\Dslash|^{-k})$ still to exist for a suitable algebra of
integrable functions, and we regard $\Tr^+(\.|\Dslash|^{-k})$ as a
noncommutative integral. These two summability conditions together
constitute Axiom~1.

A further necessary condition was \textit{regularity} or
\textit{smoothness} of the spectral triple. If $\delta(T) := [|D|,T]$
for an operator $T$ on~$\H$, regularity means that each $a \in \A$ and
each $[D,a]$ lies in the domain of $\delta^n$ for all $n \in \N$. In
the commutative case, $|\Dslash|$ is a first-order pseudodifferential
operator, and computing $\delta^n(a)$ is onerous; it is somewhat
easier to handle the (commuting) operations
\begin{equation}
L(T) := |\Dslash|^{-1}\,[\Dslash^2, T],  \qquad
R(T) := [\Dslash^2, T]\,|\Dslash|^{-1},
\label{eq:lr-smooth}
\end{equation}
and one can show that the smooth domain of~$\delta$ equals the common
smooth domain of $L$ and~$R$~\cite{ConnesMIndex}. If $f$ is a Schwartz
function acting on $L^2(\R^k) \ox \C^{2^{\piso{k/2}}}$ as an (ordinary)
multiplication operator, we can regard it as a pseudodifferential
operator with symbol $f(x) \ox 1_{2^{\piso{k/2}}}$, and one checks that
$L^n R^m f$ is a bounded pseudodifferential operator of order (at most)
zero. On the subject of regularity, the reader is advised to look at
the discussions in~\cite[Sec.~10.3]{Polaris} and also
in~\cite{HigsonRes,RennieLocal}.

There is no obvious need to modify this axiom in the noncompact case.
However, at the technical level, $|\Dslash|^{-1}$ is a somewhat more
problematic object than in the compact case, and one must find a
substitute for it.

The condition of \textit{finiteness}, in the unital case, is that the
smooth domain $\H^\infty$ of~$D$ in~$\H$ be a finitely generated
projective (left) module over the unital algebra~$\A$, that is,
$\H^\infty \simeq \A^mp$ for some projector $p = p^* = p^2$
in~$M_m(\A)$ with a suitable~$m$.

In the case of $M = \R^k$, under either the pointwise or the Moyal
product, the module of smooth spinors is free since the spinor bundle
is trivial. However, when $\A$ is nonunital, to get a projective
$\A$-module one should select the projector~$p$ in a matrix algebra
over the preferred compactification~$\Aun$. See
Rennie~\cite{RennieProj} for a discussion both of this point and of
``pullback modules''. Concretely, if $\A_1$ is an ideal of $\Aun$ and
if $\E$ is a left $\Aun$-module, its \textit{pullback} to~$\A_1$ is the
left $\A_1$-module $\E_1 := \A_1\E$.

The finiteness condition for the nonunital case should then demand
that $\H^\infty$ densely contains a pullback of a finite projective
$\Aun$-module to $\A$; or, better still, that it can be identified with
$\A_1^m p$, with $\A$ an ideal in $\A_1$ (thus $\Aun$ is also a
unitization of $\A_1$), for some $m$ and some projector
$p \in M_m(\Aun)$. Moreover, a hermitian structure should be defined
on the module $\H^\infty$ through the noncommutative integral; we
shall see the details of this further on.

\smallskip

We bring up next the axioms having an algebraic flavour. The
\textit{reality} condition is the existence of an antilinear
conjugation operator~$J$ on~$\H$ such that $a \mapsto J a^* J^{-1}$
gives a second representation of~$\A$ on~$\H$ commuting with the
original one, and with certain algebraic properties listed
in~\cite{ConnesReal,ConnesGrav} and reviewed later: for the
commutative case of spin manifolds, $J$ is just the charge conjugation
operator on spinors. There is no need to modify this axiom in the
noncompact case.

The \textit{first order} condition is that
$$
[[D, a], J b^* J^{-1}] = 0,  \sepword{for all}  a,b \in \A.
$$
For the commutative case this is a simple check, since $D = \Dslash$
is a first-order differential operator. There is no need to modify
this axiom in the noncompact case.

The \textit{orientability} condition is that the spectral triple
$(\A,\H,D)$ carry an algebraic version of a ``volume $k$-form'', where
$k$ is the integer summability exponent ($k = 2N$ in the nondegenerate
Moyal case). Let $b_0^\opp$ denote $b_0 \in \A$ as an element of the
opposite algebra $\A^\opp$, with the product reversed; this algebraic
version consists of a Hochschild $k$-cycle $\cc$, that is, a sum of
terms of the form $(a_0 \ox b_0^\opp) \ox a_1 \oxyox a_k$ satisfying
$b\,\cc = 0$ (cycle property), that we represent by bounded operators
\begin{equation}
\pi_D((a_0 \ox b_0^\opp) \ox a_1 \oxyox a_k)
:= a_0\,J b_0^* J^{-1} \,[D,a_1] \dots [D,a_k],
\label{eq:cycle-rep}
\end{equation}
and on which we impose $\pi_D(\cc) = \chi$ (orientation), where $\chi$
is the given $\Z_2$-grading operator on~$\H$. We just use $\chi = 1$
if $k$ is odd; and, in the even case for ordinary spinors, one uses
$\chi := (-i)^m\, \ga^1\ga^2 \dots\ga^{2m}$.

For the commutative or noncommutative torus $\Coo(\T_\Theta^k)$, with
unitary generators $u_1,\dots,u_k$ satisfying
\begin{equation}
u_k u_j = e^{i\Theta_{jk}} \,u_j u_k,
\label{eq:tori-relns}
\end{equation}
the good Hochschild cycle is known~\cite{ConnesGrav,Polaris} to be
\begin{equation}
\cc = \frac{(-i)^{\piso{k/2}}}{k!} \sum_\sigma (-1)^\sigma
(u_{\sigma(1)} u_{\sigma(2)} \dots u_{\sigma(k)})^{-1}
\ox u_{\sigma(1)} \ox u_{\sigma(2)} \oxyox u_{\sigma(k)},
\label{eq:vol-tori}
\end{equation}
where the sum is over all permutations of $1,2,\dots,k$.

For nonunital algebras, we might expect something similar. However,
the fact that the plane waves belong to $\B$ suggests, in the light of
the NC torus example, taking the cycle over the unitization $\Aun$
rather than $\A$ itself. This has the happy consequence of bypassing
the many difficulties of Hochschild cohomology for nonunital algebras.

\smallskip

\textit{Poincar\'e duality} for a noncompact orientable manifold $M$ is
usually expressed as the isomorphism between the compactly supported de
Rham cohomology and the homology of $M$, mediated by the fundamental
class $[M]$. In noncommutative geometry a $K$-theoretic version is in
order. One would expect that some kind of compactly supported
$K$-homology of the initial nonunital algebra $\A$ be isomorphic to its
$K$-theory, through a fundamental $K$-homology class of $\A\ox\A^\opp$
given by the spectral triple itself. We shall actually leave aside the
final condition of Poincar\'e duality in $K$-theory, since it is not
central to the present form of the reconstruction theorem in the
compact case~\cite{Polaris}, and the details of its reformulation in
the nonunital noncommutative case are still somewhat clouded.

\subsection{Modified conditions for nonunital spectral triples}
\label{sec:axioms-of-evil}

\begin{defn}
By a \textit{real noncompact spectral triple} of dimension~$k$, we
mean the data
$$
(\A, \Aun, \H, D, J, \chi),
$$
where $\A$ is an (a priori nonunital) algebra acting faithfully (via a
representation sometimes denoted by~$\pi$) on the Hilbert space $\H$,
$\Aun$ is a preferred unitization of~$\A$, acting the same Hilbert
space, and $D$ is an unbounded selfadjoint operator on $\H$ such that
$[D,a]$, for each $a$ in $\Aun$, extends to a bounded operator on~$\H$.

Furthermore, $J$ and $\chi$ are respectively an antiunitary and
a selfadjoint operator, such that $\chi = 1$ when $k$ is odd, and
otherwise $\chi^2 = 1$, $\chi a = a \chi$ for $a \in \A$, and
$D \chi = -\chi D$, satisfying the conditions which follow.
\end{defn}

\begin{enumerate}
\addtocounter{enumi}{-1}

\item\textit{Compactness}:

The operator $a(D - \la)^{-1}$ is compact for $a \in \A$ and
$\la \notin \spec D$.

\item\textit{Spectral dimension}:

There is a unique nonnegative integer $k$, the spectral or ``classical"
dimension of the geometry, for which $a(D^2 + \eps^2)^{-1/2}$
belongs to the generalized Schatten class $\L^{k+}$ for each
$a \in \A$ and moreover $\Tr^+(a(|D| + \eps)^{-k})$ is finite and not
identically zero, for any $\eps > 0$. This $k$ is even if and only if
the spectral triple is even.

\item\textit{Regularity}:

The bounded operators $a$ and $[D,a]$, for each $a \in \Aun$, lie in
the smooth domain of the derivation $\delta : T \mapsto [|D|,T]$.

\item\textit{Finiteness}:

The algebra $\A$ and its preferred unitization $\Aun$ are
pre-$C^*$-algebras. There exists an ideal $\A_1$ of $\Aun$, including
$\A$, which is also a pre-$C^*$-algebra with the same $C^*$-completion
as $\A$, such that the space of smooth vectors
$$
C^\infty(D) \equiv \H^\infty := \bigcap_{k \in \N} \Dom(D^k)
$$
is an $\A_1$-pullback of a finite projective $\Aun$-module.
Moreover, an $\A_1$-valued hermitian structure $\roundbraket{\.}{\.}$
is implicitly defined on $\H^\infty$ with the noncommutative integral,
as follows:
\begin{equation}
\Tr^+\bigl( \roundbraket{a\xi}{\eta}(|D| + \eps)^{-k} \bigr)
= \braket{\eta}{a\xi},
\label{eq:abs-cont}
\end{equation}
where $a \in \Aun$ and $\braket{\.}{\.}$ denotes the standard inner
product on~$\H$. This is an absolute continuity condition, since
$\roundbraket{\.}{\.}$ is a kind of Radon--Nikod\'ym derivative with
respect to the functional $\Tr^+(\.\,(|D| + \eps)^{-k})$.

\item\textit{Reality}:

There is an antiunitary operator $J$ on~$\H$, such that
$[a, Jb^*J^{-1}] = 0$ for all $a,b\in\Aun$ (thus $b \mapsto Jb^*J^{-1}$
is a commuting representation on $\H$ of the opposite algebra
$\A^\opp$). Moreover, $J^2 = \pm 1$ and $JD = \pm DJ$, and also
$J\chi = \pm \chi J$ in the even case, where the signs depend only on
$k\bmod 8$. Here is the table for the even case; see the full table in
\cite[p.~405]{Polaris}.
\begin{equation}
\begin{array}{|c|cccc|}
\hline
                 N \bmod 4     & 0 & 1 & 2 & 3 \rule[-5pt]{0pt}{17pt} \\
\hline
               J^2 = \pm\,1    & + & - & - & + \rule[-5pt]{0pt}{17pt} \\
                JD = \pm\,DJ   & + & + & + & + \rule[-5pt]{0pt}{17pt} \\
J\chi = \pm\,\chi J    & + & - & + & - \rule[-5pt]{0pt}{17pt} \\
\hline
\end{array}
\label{eq:real-signs}
\end{equation}

\item\textit{First order}:

The bounded operators $[D,a]$ also commute with the opposite algebra
representation: $[[D,a], Jb^*J^{-1}] = 0$ for all $a,b \in \Aun$.

\item\textit{Orientation}:

There is a \textit{Hochschild $k$-cycle} $\cc$ on~$\Aun$, with values
in $\Aun \ox \Aun^\opp$. Such a $k$-cycle is a finite sum of terms
like $(a \ox b^\opp) \ox a_1 \oxyox a_k$, whose natural representative
by operators on~$\H$ is given by $\pi_D(\cc)$ in
formula~\eqref{eq:cycle-rep}; the ``volume form'' $\pi_D(\cc)$ must
solve the equation
\begin{equation}
\pi_D(\cc) = \chi \sepword{(even case),\quad or}
\pi_D(\cc) = 1    \quad\text{(odd case)}.
\label{eq:vol-form}
\end{equation}

\end{enumerate}

Finally, a geometry is called \textit{connected} or irreducible if the
only operators commuting with $\A$ and $D$ are the scalars. We are
mainly interested in connected noncompact noncommutative geometries.

\smallskip

The discussion in the previous subsection, and this proposal, are very
much in the vein of~\cite{Selene}. We may also keep the concept in that
article of ``star triples'', a specialization of the spectral triple
to deformations of the algebra of functions on a noncompact manifold,
wherein the Dirac operator (is possibly deformed, but) remains an
ordinary (pseudo-)differential operator on that original manifold.
However, the authors of~\cite{Selene} got carried away in that they
confused properties of~$L^\th_f$ with properties of the Weyl
pseudodifferential operator associated (by the Schr\"odinger
representation) to the ``symbol''~$f$. And thus, the dreaded
``dimension drop'', apparent there, does not actually take place. But
before going to the Moyal case, we need to reexamine the commutative
case.

\subsection{The commutative case}

The outcome of the discussion in subsection~\ref{sec:genld-axioms} is
that the main outstanding issues, in order to obtain noncompact
noncommutative spin geometries, are the analytical ones.

Let $\A$ be some appropriate subalgebra of $C^\infty(M)$ and $\Dslash$
be the Dirac operator, with $k$ equal to the ordinary dimension of the
spin manifold~$M$. Let $\H$ be the space of square-integrable spinors.
Then $[\Dslash, f] = \Dslash(f)$, just as in the unital case, and so
the boundedness of $[D,\A]$ is unproblematic. In order to check
whether $(\A,\H,\Dslash,\chi)$ is a spectral triple in our sense, one
first needs to determine whether products of the form
$f(|\Dslash| + \eps)^{-k}$ are compact operators of Dixmier trace
class, whose Dixmier trace is (a standard multiple of)
$\int f(x) \,d^kx$. This compactness condition is guaranteed in the
flat space case (taking $\A = \SS(\R^k)$, say) by celebrated estimates
in scattering theory~\cite{SimonTrace}, that we review in
subsection~\ref{sec:axiom-comp}.

The summability condition is a bit tougher. The Ces\`aro summability
theory of~\cite{Odysseus} establishes that, for a positive
pseudodifferential operator $H$ of order~$d$, acting on spinors, the
spectral density asymptotically behaves as
$$
d_H(x,x;\la') \sim \frac{2^{\piso{k/2}}}{d\,(2\pi)^k}
\bigl(\wres H^{-k/d}\,(\la')^{(k-d)/d} +\cdots \bigr),
$$
in the Ces\`aro sense. Here $\wres$ denotes the Wodzicki residue
density~\cite{Polaris}. (If the operator is not positive, one uses the
``four parts'' argument.) In our case, $H = a(|\Dslash| + \eps)^{-k}$
is pseudodifferential of order~$-k$, so
$$
d_H(x,x;\la')
\sim -\frac{2^{\piso{k/2}}\,\Omega_k\,a(x)}{k\,(2\pi)^k} \,
({\la'}^{-2} + \cdots),
$$
as $\la' \to \infty$ in the Ces\`aro sense; here $\Omega_k$ is the
hyperarea of the unit sphere in $\R^k$. We independently know that $H$
is compact, so on integrating the spectral density over~$x$ and over
$0 \leq \la' \leq \la$, we get
$$
N_H(\la)
\sim \frac{2^{\piso{k/2}}\,\Omega_k\int a(x)\,d^kx}{k\,(2\pi)^k}
\,\frac{1}{\la} \as \la \to \infty.
$$
This holds in the ordinary asymptotic sense, and not merely the
Ces\`aro sense, by the ``sandwich'' argument used in the proof of
\cite[Cor.~4.1]{Odysseus}. So finally,
\begin{equation}
\la_m(H)
\sim \frac{2^{\piso{k/2}}\,\Omega_k\int a(x)\,d^kx}{k\,(2\pi)^k}\,
\frac{1}{m} \as m \to \infty,
\label{eq:eastindian-discovery}
\end{equation}
and the Dixmier traceability of $a(|\Dslash| + \eps)^{-k}$, plus the
value of its trace, follow at once.

The rest is a long but almost trivial verification. For instance, $J$
is the charge conjugation operator on spinors; the algebra
$(\B,\star_0)$ is a suitable compactification; the domain $\H^\infty$
consists of the smooth spinors; and so on. See below the parallel
discussion for the Moyal case.

The following theorem sums it up.

\begin{thm}
\label{th:commuters}
The triple $(\SS(\R^k), L^2(\R^k)\ox\C^{2^{\piso{k/2}}}, \Dslash)$ on
$\R^k$ defines a noncompact commutative geometry of spectral
dimension~$k$.
\end{thm}

What about the nonflat case (of a spin manifold such that $\Dslash$
is selfadjoint)? Mainly because the previous Ces\`aro summability
argument is purely local, everything carries over, if we choose
for~$\A$ the algebra of smooth and compactly supported functions. Of
course, in some contexts it may be useful to demand that $M$ also has
conic exits.

We want to remark that formula~\eqref{eq:eastindian-discovery} for the
flat case has been proved by Chakraborty et~al in~\cite{ChakrabortyGS},
using an ingenious reasoning involving two Laplacians on $\R^k$. Theirs
is a kind of ``poor man's argument'' for ours, because what it is
really used is that the spectral density has the same asymptotic
behaviour for the two Laplacians. Also, our inference is not confined
to flat manifolds, rather it is directly valid on any decent noncompact
manifold (without recourse to ``lifting'' devices).

\subsection{On the Connes--Landi spaces example}

An interesting family of compact spectral triples was constructed by
Connes and Landi~\cite{ConnesLa}, by isospectral deformation of a
commutative spectral triple wherein the Dirac operator is kept fixed
(just as for our Moyal-product example) but the algebra is
``twisted''. One starts with a smooth boundaryless manifold $M$
carrying a smooth effective action of a torus $\T^k$ of dimension
$k \geq 2$. The orbits on which $\T^k$ acts freely determine maps
$\Coo(M) \to \Coo(\T^k)$, and with these maps one can pull back the NC
torus structure on $\Coo(\T^k_\Theta) := (\Coo(\T^k),*_\Theta)$ to
get an algebra $\Coo(M_\Theta) := (\Coo(M),*_\Theta)$. This algebra is
given in fact by a periodic Moyal product just
like~\eqref{eq:moyal-prod-slick}, with the translations replaced by
the $\T^k$-action. See \cite{ConnesDV,Sitarz,Larissa,Calypso} for
several equivalent formulations of this construction.

Now, as pointed out in \cite{ConnesDV}, there is no need to assume
that the manifold $M$ be compact: we only need that the group action
on~$M$ be periodic. Taking $M = \R^k$, we get a noncompact spectral
triple which is not isomorphic to the Moyal product examples
considered in this article; one can regard it as intermediate between
the commutative case and the full Moyal cases (with nonperiodic
action).

Concretely, the sphere $\Sf^{2N-1} = SO(2N)/SO(2N-1)$ carries an
effective action of $\T^N$, namely the rotations by elements of a
maximal torus of $SO(2N)$; and this extends to a $\T^N$-action by
rotations of $\R^{2N}$ preserving the radial coordinate~$r$. Each
$f \in \SS$ is a function of coordinates
$f(r,\a_1,\dots,\a_{N-1},\phi_1,\dots,\phi_N)$ where
$\phi = (\phi_1,\dots,\phi_N) \in \T^N$. If the
equation~\eqref{eq:moyal-prod-slick} is interpreted as involving
integration over the $\phi_j$ coordinates only, it defines a new
twisted product on~$\SS$ (for each real skewsymmetric $N \x N$ matrix
$\Theta$).

To define a spectral triple over this algebra, we need an operator $D$
which is also $\T^N$-invariant. For instance, one can construct $D$ by
extending radially the Dirac operator for (say) the round metric
on~$\Sf^{2N-1}$, with its spinor bundle; it will be necessary to lift
the torus action to a doubly covering action of $\T^n$ on
spinors~\cite{ConnesDV}. It remains to check that $\B$ is still a
suitable unitization of~$\SS$ (note that abstract smoothness of $\B$
is proved like in Section~2 here~\cite{Larissa}) in the case of the
Connes--Landi twisted $2N$-planes, in order to conclude that these fit
into the framework developed in this paper.

\section{The Moyal $2N$-plane as a spectral triple}

There is a natural star triple associated to the Moyal plane and we
will see that it is part of the data for an even spectral triple
fulfilling all required conditions.

Let $\A = (\SS(\R^{2N}),\mop)$, with preferred unitization
$\Aun := (\B(\R^{2N}),\mop)$. The Hilbert space will be
$\H := L^2(\R^{2N}) \ox \C^{2^N}$ of ordinary square-integrable
spinors. The representation of $\A$ is given by
$\pi^\th \: \A \to \L(\H) : f \mapsto L^\th_f \ox 1_{2^N}$, where
$L^\th_f$ acts on the ``reduced'' Hilbert space
$\H_r := L^2(\R^{2N})$. In other words, if $a \in \A$ and
$\Psi \in \H$, to obtain $\pi^\th(a)\Psi$ we just left Moyal multiply
$\Psi$ by~$a$ componentwise.

This operator $\pi^\th(f)$ is bounded, since it acts diagonally on~$\H$
and $\|L^\th_f\| \leq (2\pi\th)^{-N/2} \|f\|_2$ was proved in
Lemma~\ref{lm:norm-HS}. Under this action, the elements of $\H$ get the
lofty name of \textit{Moyal spinors}.

The selfadjoint Dirac operator is not ``deformed'': it will be the
ordinary Euclidean Dirac operator $\Dslash := -i\,\ga^\mu \del_\mu$,
where the hermitian Dirac matrices $\ga^1,\dots,\ga^{2N}$ satisfying
$\{\ga^\mu, \ga^\nu\} = +2\,\delta^{\mu\nu}$ irreducibly represent the
Clifford algebra $\Cl(\R^{2N})$ associated to $(\R^{2N},\eta)$, with
$\eta$ the standard Euclidean metric.

As a grading operator $\chi$ we take the usual chirality associated to
the Clifford algebra:
$$
\chi := \ga_{2N+1} := 1_{\H_r} \ox (-i)^N \ga^1 \ga^2 \dots \ga^{2N}.
$$
The notation $\ga_{2N+1}$ is a nod to physicists' $\ga_5$. Thus
$\chi^2 = (-1)^N (\ga^1\dots\ga^{2N})^2 = (-1)^{2N} = 1$ and
$\chi \ga^\mu = -\ga^\mu \chi$.

The real structure $J$ is chosen to be the usual charge conjugation
operator for spinors on $\R^{2N}$ endowed with an Euclidean metric.
Here, we only assume that $J^2 = \pm 1$ according to the ``sign table''
\eqref{eq:real-signs} and that
$$
J (1_{\H_r} \ox \ga^\mu) J^{-1} = - 1_{\H_r} \ox \ga^\mu
$$
which guarantees the other requirements of \eqref{eq:real-signs}. In
general, in a given representation, it can be written as
\begin{equation}
J := CK,
\label{eq:charge-conjn}
\end{equation}
where $C$ denotes a suitable $2^N \x 2^N$ unitary matrix and $K$ means
complex conjugation. An explicit form for $J$ in a particular
representation can be found in~\cite{ZinnJ} where all $\ga^\mu$ are
hermitian matrices with purely imaginary (respectively real) entries
when $\mu$ is even (respectively odd).

An important property of~$J$ is
\begin{equation}
J(L^\th(f^*) \ox 1_{2^N}) J^{-1} = R^\th(f) \ox 1_{2^N},
\label{eq:right-Moyal}
\end{equation}
where $R^\th(f) \equiv R^\th_f$ is the right Moyal multiplication by
$f$; this follows from the antilinearity of~$J$ and the reversal of the
twisted product under complex conjugation.

\medskip

Lemma~\ref{lm:propriete}(iii) implies that $[\Dslash,\pi^\th(f)] =
-i L^\th(\del_\mu f)\ox \gamma^\mu =: \pi^\th(\Dslash(f))$; by
Theorem~\ref{th:CVai} this is bounded for
$f \in \Aun_\th = \B(\R^{2N})$ ---just as in the commutative case.

\subsection{The compactness condition}
\label{sec:axiom-comp}

In this subsection and the next, the main tools are techniques
developed some time ago for scattering theory problems, as summarized
in Simon's booklet~\cite[Chap.~4]{SimonTrace}. We adopt the convention
that $\L^\infty(\H) := \K(\H)$, with
$\|A\|_\infty := \|A\|_{\mathrm{op}}$.

Let $g\in L^\infty(\R^{2N})$. We define the operator $g(-i\nb)$ on
$\H_r$ as
$$
g(-i\nb)\psi := \F^{-1}(g\,\F\psi),
$$
where $\F$ is the ordinary Fourier transform. More in detail, for
$\psi$ in the correct domain,
$$
g(-i\nb)\psi(x)
= (2\pi)^{-2N}\iint e^{i\xi\.(x-y)}\,g(\xi)\psi(y)\,d^{2N}\xi\,d^{2N}y.
$$
The inequality
$\|g(-i\nb)\psi\|_2 = \|\F^{-1}g\F\psi\|_2\leq \|g\|_\infty\|\psi\|_2$
entails that $\|g(-i\nb)\|_\infty \leq \|g\|_\infty$.

\begin{thm}
\label{th:compacite}
Let $f\in \A$ and $\la \notin \spec \Dslash$. Then, if
$R_\Dslash(\la)$ is the resolvent operator of $\Dslash$, then
$\pi^\th(f)\, R_\Dslash(\la)$ is compact.
\end{thm}

Thanks to the first resolvent equation, $R_\Dslash(\la) =
R_\Dslash(\la') + (\la' - \la) R_\Dslash(\la) R_\Dslash(\la')$, we may
assume that $\la = i\mu$ with $\mu \in \R^*$. The theorem will follow
from a series of lemmas interesting in themselves.

\begin{lem}
\label{lm:comp}
If $f \in \SS$ and $0 \neq \mu \in \R$, then
$$
\pi^\th(f)  R_\Dslash(i\mu)    \in \K(\H)
\iff
\pi^\th(f) |R_\Dslash(i\mu)|^2 \in \K(\H).
$$
\end{lem}

\begin{proof}
We know that $L^\th(f)^* = L^\th(f^*)$. The ``only if'' part is
obvious since $R_\Dslash(i\mu)$ is a bounded normal operator.
Conversely, if $\pi^\th(f)|R_\Dslash(i\mu)|^2$ is compact, then
$\pi^\th(f)|R_\Dslash(i\mu)|^2 \pi^\th(f^*)$ is compact. Since
an operator $T$ is compact if and only if $TT^*$ is compact, the proof
is complete.
\end{proof}

The usefulness of this lemma stems from the diagonal nature of the
action of $\pi^\th(f)|R_\Dslash(i\mu)|^2$ on $\H = \H_r \ox \C^{2^N}$;
so in our arguments it is feasible to replace $\H$ by $\H_r$,
$\pi^\th(f)$ by $L^\th_f$, and to use the scalar Laplacian
$-\Delta := -\sum_{\mu=1}^{2N} \del_\mu^2$ instead of the square of
the Dirac operator $\Dslash^2$.

\begin{lem}
\label{lm:HSO}
When $f,g \in \H_r$, $L^\th_f \,g(-i\nb)$ is a Hilbert--Schmidt
operator such that, for all real~$\th$,
$$
\|L^\th_f \,g(-i\nb)\|_2 = \|L_f^0 \,g(-i\nb)\|_2
= (2\pi)^{-N} \|f\|_2\,\|g\|_2.
$$
\end{lem}

\begin{proof}
To prove that an operator $A$ with integral kernel $K_A$ is
Hilbert--Schmidt, it suffices to check that
$\int |K_A(x,y)|^2 \,dx\,dy$ is finite, and this will be equal to
$\|A\|_2^2$~\cite[Thm.~2.11]{SimonTrace}. So we compute
$K_{L^\th(f)\,g(-i\nb)}$. In view of Lemma~\ref{lm:cojoreg},
$$
[L^\th(f)\,g(-i\nb)\psi](x) = \frac{1}{(2\pi)^{2N}} \iint
f(x - \tfrac{\th}2 S\xi) \,g(\xi) \psi(y)\, e^{i\xi\.(x-y)}
\,d^{2N}\xi \,d^{2N}y.
$$
Thus
$$
K_{L^\th(f)\,g(-i\nb)}(x,y) = \frac{1}{(2\pi)^{2N}} \int
f(x - \tfrac{\th}2 S\xi)\, g(\xi) \,e^{i\xi\.(x-y)} \,d^{2N}\xi,
$$
and $\int |K_{L^\th(f)\,g(-i\nb)}(x,y)|^2 \,dx\,dy$ is given by
\begin{align*}
\frac{1}{(2\pi)^{4N}}
&\int\cdots\int \bar{f}(x - \tfrac{\th}2 S\xi)\, \bar{g}(\xi)\,
f(x - \tfrac{\th}2 S\zeta)\, g(\zeta)\, e^{i(x-y)\.(\zeta-\xi)}
\,d^{2N}x \,d^{2N}y \,d^{2N}\zeta \,d^{2N}\xi
\\
&\quad = \frac{1}{(2\pi)^{2N}} \iint
|f(x - \tfrac{\th}2 S\xi)|^2\, |g(\xi)|^2 \,d^{2N}x \,d^{2N}\xi
= (2\pi)^{-2N} \|f\|_2^2 \,\|g\|_2^2 < \infty.
\tag*\qed
\end{align*}
\hideqed
\end{proof}

\begin{rem}
As a consequence, we get
$$
\|.\|_2\mbox{-}\lim_{\th\to 0} L^\th_f \,g(-i\nb) = L_f^0 \,g(-i\nb).
$$
\end{rem}

\begin{lem}
\label{lm:schatten}
If $f \in \H_r$ and $g \in L^p(\R^{2N})$ with $2 \leq p < \infty$,
then $L^\th_f\,g(-i\nb) \in \L^p(\H_r)$ and
$$
\|L^\th_f\,g(-i\nb)\|_p
\leq (2\pi)^{-N(1/2+1/p)} \th^{-N(1/2-1/p)} \,\|f\|_2\,\|g\|_p.
$$
\end{lem}

\begin{proof}
The case $p = 2$ (with equality) is just the previous lemma. For
$p = \infty$, we estimate $\|L^\th_f\,g(-i\nb)\|_\infty \leq
(2\pi\th)^{-N/2} \|f\|_2 \,\|g\|_\infty$: since
$\|L^\th_f\,g(-i\nb)\|_\infty \leq
\|L^\th_f\|_\infty \,\|g(-i\nb)\|_\infty$, this follows from
Lemma~\ref{lm:norm-HS} and a previous remark.

Now use complex interpolation for $2 < p < \infty$. For that, we first
note that we may suppose $g \geq 0$: defining the function $a$ with
$|a| = 1$ and $g = a|g|$, we see that
\begin{align*}
\|L^\th_f\,g(-i\nb)\|_2^2
&= \Tr(|L^\th_f\,g(-i\nb)|^2)
= \Tr(\bar{g}(-i\nb)\,L^\th_{f^*}\,L^\th_f\,g(-i\nb))
\\
&= \Tr(|g|(-i\nb)\,\bar{a}(-i\nb)\,L^\th_{f^*}
\,L^\th_f\,a(-i\nb)\,|g|(-i\nb))
\\
&= \Tr(\bar{a}(-i\nb)\,|g|(-i\nb)\,L^\th_{f^*}
\,L^\th_f\,|g|(-i\nb)\,a(-i\nb))
\\
&= \Tr(|L^\th_f\,|g|(-i\nb)|^2) = \|L^\th_f\,|g|(-i\nb)\|_2^2,
\end{align*}
and
\begin{align*}
\|L^\th_f\,g(-i\nb)\|_\infty
&= \|L^\th_f\,a(-i\nb)\,|g|(-i\nb)\|_\infty
=\|L^\th_f\,|g|(-i\nb)\,a(-i\nb)\|_\infty
\\
&\leq \|L^\th_f\,|g|(-i\nb)\|_\infty \,\|a(-i\nb)\|_\infty
= \|L^\th_f\,|g|(-i\nb)\|_\infty.
\end{align*}
Secondly, for any positive, bounded function $g$ with compact support,
we define the maps:
$$
F_p : z \mapsto  L^\th_f\, g^{zp}(-i\nb)
: S = \set{z\in\C : 0 \leq \Re z \leq \half} \to \L(\H_r).
$$
For all $y\in \R$,
$F_p(iy) = L^\th_f \, g^{iyp}(-i\nb) \in \L^\infty(\H_r)$ by
Lemma~\ref{lm:HSO} since $g$, being compactly supported, lies in
$\H_r$. Moreover, $\|F_p(iy)\|_\infty \leq(2\pi\th)^{-N/2} \|f\|_2$.

Also, by Lemma~\ref{lm:HSO}, $F_p(\half+iy) \in \L^2(\H_r)$ and
$\|F_p(\half+iy)\|_2 = (2\pi)^{-N} \|f\|_2 \,\|g^{p/2}\|_2$. Then
complex interpolation (see \cite[Chap.~9]{ReedSII}
and~\cite{SimonTrace}) yields $F(z) \in \L^{1/\Re z}(\H_r)$, for all
$z$ in the strip $S$. Moreover,
$$
\|F_p(z)\|_{1/\Re z}
\leq \|F(0)\|_\infty^{1-2\Re z}\,\|F(\half)\|_2^{2\Re z}
= \|f\|_2 (2\pi\th)^{-\frac{N}{2}(1-2\Re z)}(2\pi)^{-2N\Re z}\,
\|g^{p/2}\|_2^{2\Re z},
$$
and applying this result at $z = 1/p$, we get for such $g$:
$$
\|L^\th_f\,g(-i\nb)\|_p =  \|F(1/p)\|_p
\leq(2\pi)^{-N(1/2+1/p)}\th^{-N(1/2-1/p)}\|f\|_2\,\|g\|_p .
$$
We finish by using the density of compactly supported bounded
functions in $L^p(\R^{2N})$.
\end{proof}

\begin{rem}
In the commutative case, if $f$ and $g$ are bounded on $\R^k$, then
$\|f(x)\,g(-i\nb)\|_\infty \leq \|f\|_\infty\,\|g\|_\infty$. Complex
interpolation~\cite{BirmanKS,ReedSII,SimonTrace} leads then to an
estimate of the form
$$
\|f(x)\,g(-i\nb)\|_p \leq (2\pi)^{-k/p} \,\|f\|_p\,\|g\|_p
$$
when $p \geq 2$. For $f \in \SS$ and for
$g(y) := 1/\sqrt{|y|^2 + \mu^2}$, which lies in $L^p(\R^k)$ for all
$p > k$ we conclude that $f(x)\,g(-i\nb)$ is compact and in $\L^p$ for
$p > k$. This already strongly pointed to compliance with Axiom~1
(verified above using Ces\`aro summability considerations), since
$\L^{k+}$ is larger than $\L^k$, but smaller than the intersection of
the $\L^p$ for $p > k$.
\end{rem}

\begin{lem}
\label{lm:neo-schatten}
If $f \in \SS$ and $0 \neq \mu \in \R$, then
$\pi^\th(f)\,|R_\Dslash(i\mu)|^2 \in \L^p$ for $p > N$.
\end{lem}

\begin{proof} We see that
$$
\pi^\th(f)\,|R_\Dslash(i\mu)|^2
= (L^\th_f \ox 1_{2^N})\,(\Dslash-i\mu)^{-1} (\Dslash+i\mu)^{-1}
= L^\th_f \,(-\del^\nu\del_\nu +\mu^2)^{-1} \ox 1_{2^N}.
$$
So this operator acts diagonally on $\H_r \ox \C^{2^N}$ and
Lemma~\ref{lm:schatten} implies that
$$
\bigl\| L^\th_f \,(-\del^\nu\del_\nu +\mu^2)^{-1} \bigr\|_p
\leq (2\pi)^{-N(1/2+1/p)}\th^{-N(1/2-1/p)}\,\|f\|_2
\biggl(\int\frac{d^{2N}\xi}{(\xi^\nu\xi_\nu + \mu^2)^p}\biggr)^{1/p},
$$
which is finite for $p > N$.
\end{proof}

\begin{proof}[Proof of Theorem~\ref{th:compacite}]
By Lemma~\ref{lm:comp}, it was enough to prove that
$\pi^\th(f)\,|R_\Dslash(i\mu)|^2$ is compact for a nonzero
real~$\mu$.
\end{proof}

The conclusion is that $(\A,\Aun,\H,\Dslash,\chi,J)$ defines a
noncompact spectral triple; recall that we proved in
Section~\ref{sec:Moyal-basics} that both $\A$ and its preferred
compactification $\Aun$ are pre-$C^*$-algebras.

\subsection{Spectral dimension of the Moyal planes}

\begin{thm}
\label{th:dim-spectrale}
The spectral dimension of the Moyal $2N$-plane spectral triple is~$2N$.
\end{thm}

We shall first establish existence properties. Thanks to
Lemma~\ref{lm:schatten} and because
$[\Dslash,\pi^\th(f)] = -iL^\th(\del_\mu f) \ox \ga^\mu$, we see that
$\pi^\th(f)(\Dslash^2 + \eps^2)^{-l}$ and
$[\Dslash,\pi^\th(f)]\,(\Dslash^2 + \eps^2)^{-l}$ lie in $\L^p(\H)$
whenever $p > N/l$ (we always assume $\eps > 0$). In the next lemma, we
show that $[|\Dslash|,\pi^\th(f)]\, (\Dslash^2 + \eps^2)^{-l}$ has the
same property of summability; this will become our main technical
instrument for the subsection.

\begin{lem}
\label{lm:commutateur}
If $f \in \SS$ and $\half \leq l \leq N$, then
$[|\Dslash|,\pi^\th(f)]\,(\Dslash^2 + \eps^2)^{-l} \in \L^p(\H)$
for $p > N/l$.
\end{lem}

\begin{proof}
We use the following spectral identity for a positive operator $A$:
$$
A = \frac{1}{\pi} \int_0^\infty \frac{A^2}{A^2 + \mu}
               \,\frac{d\mu}{\sqrt{\mu}},
$$
and another identity for any operators $A$, $B$ and
$\la \notin \spec A$:
\begin{equation}
[B, (A-\la)^{-1}] = (A - \la)^{-1} [A,B] (A - \la)^{-1}.
\label{eq:deriv-inv}
\end{equation}

Hence, for any $\rho > 0$,
\begin{align}
\label{eq:spectid}
[|\Dslash|, \pi^\th(f)]
&= [|\Dslash|+\rho, \pi^\th(f)]
= \frac{1}{\pi} \int_0^\infty \biggl[
\frac{(|\Dslash|+\rho)^2}{(|\Dslash|+\rho)^2+\mu},
\pi^\th(f) \biggr] \, \frac{d\mu}{\sqrt{\mu}}
\nonumber\\
&= \frac{1}{\pi} \int_0^\infty
\biggl(1 - \frac{(|\Dslash|+\rho)^2}{(|\Dslash|+\rho)^2+\mu}\biggr)
\bigl[(|\Dslash|+\rho)^2, \pi^\th(f)\bigr]
\frac{1}{(|\Dslash|+\rho)^2+\mu} \, \frac{d\mu}{\sqrt{\mu}}
\nonumber\\
&= \frac{1}{\pi} \int_0^\infty
\frac{1}{(|\Dslash|+\rho)^2+\mu}
\bigl[(|\Dslash|+\rho)^2,\pi^\th(f)\bigr]
\frac{1}{(|\Dslash|+\rho)^2+\mu} \,\sqrt{\mu} \,d\mu
\\
&= \frac{1}{\pi} \int_0^\infty
\frac{1}{(|\Dslash|+\rho)^2+\mu}
\biggl(-\pi^\th(\del^\mu \del_\mu f)
- 2i(L^\th(\del_\mu f) \ox \ga^\mu) \Dslash
+ 2\rho \bigl[ |\Dslash|, \pi^\th(f) \bigr] \biggr)
\nonumber\\
&\hspace{13em} \x
\frac{1}{(|\Dslash|+\rho)^2+\mu} \,\sqrt{\mu} \,d\mu.
\nonumber
\end{align}
This implies that
\begin{align*}
\bigl\| [|\Dslash|,\pi^\th(f)]\,(\Dslash^2 + \eps^2)^{-l} \bigr\|_p
&\leq \frac{1}{\pi} \int_0^\infty
\biggl\| \frac{1}{(|\Dslash|+\rho)^2+\mu}
\Bigl( -\pi^\th(\del^\mu \del_\mu f)
- 2i(L^\th(\del_\mu f) \ox \ga^\mu) \Dslash
\\
&\hspace{4em} + 2\rho \bigl[ |\Dslash|, \pi^\th(f) \bigr] \Bigr)
\frac{1}{(|\Dslash|+\rho)^2+\mu}\,(\Dslash^2 + \eps^2)^{-l}\biggr\|_p
\,\sqrt{\mu} \,d\mu.
\end{align*}

Thus, the proof reduces to show that for any $f \in \SS$,
\begin{equation}
\label{eq:integrale}
\frac{1}{\pi} \int_0^\infty
\biggl\| \frac{1}{(|\Dslash|+\rho)^2+\mu} \,\pi^\th(f)\Dslash
\,\frac{1}{(|\Dslash|+\rho)^2+\mu} \,(\Dslash^2 + \eps^2)^{-l}
\biggr\|_p \,\sqrt{\mu} \,d\mu < \infty.
\end{equation}

Since the Schatten $p$-norm is a symmetric norm, and since, as in the
proof of Theorem~\ref{th:compacite}, only the reduced Hilbert space is
affected, expression~\eqref{eq:integrale} is majorized by
\begin{align*}
\frac{1}{\pi} \int_0^\infty
&\biggl\| \frac{1}{(|\Dslash|+\rho)^2+\mu} \biggr\|^{3/2}
\biggl\| \frac{\Dslash}{(\Dslash^2 + \eps^2)^{1/2}} \biggr\|
\biggl\| \pi^\th(f) \frac{1}{(\Dslash^2 + \eps^2)^{l-1/2}}
\frac{1}{((|\Dslash|+\rho)^2+\mu)^{1/2}} \biggr\|_p
\,\sqrt{\mu} \,d\mu
\\
&\leq \frac{1}{\pi} \int_0^\infty
\bigl\| \pi^\th(f)\, (\Dslash^2 + \eps^2)^{-l+1/2}
((|\Dslash|+\rho)^2 + \mu)^{-1/2} \bigr\|_p
\, \frac{\sqrt{\mu} \,d\mu}{(\mu+\rho^2)^{3/2}}.
\end{align*}

Thanks to Lemma~\ref{lm:schatten}, we can estimate the
$\mu$-dependence of the last $p$-norm:
\begin{align*}
\bigl\| \pi^\th(f) &((|\Dslash|+\rho)^2+\mu)^{-1/2}
(\Dslash^2+\eps^2)^{-l+1/2} \bigr\|_p
\\
&\leq (2\pi)^{-N(1/2+1/p)}\th^{-N(1/2-1/p)} \|f\|_2\,
\bigl\| ((|\xi|+\rho)^2 +\mu)^{-1/2} (|\xi|^2 + \eps^2)^{-l+1/2}
\bigr\|_p
\\
&\leq C(p,\th) \bigl\| ((|\xi|+\rho)^2 +\mu)^{-1/2} \bigr\|_q \,
\bigl\| (|\xi|^2+\eps^2)^{-l+1/2} \bigr\|_r;
\end{align*}
with $p^{-1} = q^{-1} + r^{-1}$ appropriately chosen, these integrals
are finite for $q > 2N$ and $r > 2N/(2l-1)$; for $l = \half$, take
$r = \infty$ and $q = p$. For such values,
\begin{align*}
&\bigl\| \pi^\th(f) ((|\Dslash|+\rho)^2+\mu)^{-1/2}
(\Dslash^2+\eps^2)^{-l+1/2} \bigr\|_p
\\
&\leq C(p,\th,N;f) \|(|\xi|^2+\eps^2)^{-l+1/2}\|_r \, \Omega_{2N}^{1/q}
\biggl( \int_0^\infty \frac{R^{2N-1}}{((R+\rho)^2+\mu)^{q/2}}\,dR
\biggr)^{1/q}
\\
&= C(p,\th,N;f) \|(|\xi|^2+\eps^2)^{-l+1/2}\|_r \, \pi^{N/q} \,
\frac{\Ga^{1/q}(\tfrac{q}{2} - N)}{\Ga^{1/q}(\tfrac{q}{2})}\,
\mu^{-1/2 + N/q} =: C'(p,q,\th,N;f) \,\mu^{-1/2 + N/q}.
\end{align*}
Finally, the integral~\eqref{eq:integrale} is less than
$$
C'(p,q,\th,N;f) \int_0^\infty
\frac{\mu^{N/q}}{(\mu+\rho^2)^{3/2}} \,d\mu,
$$
which is finite for $q > 2N$ and $p > N/l$. This concludes the proof.
\end{proof}

\begin{lem}
\label{lm:Cwikel}
If  $f \in \SS$, then
$\pi^\th(f) \,(|\Dslash|+\eps)^{-1}\, \pi^\th(f^*) \in \L^{2N+}(\H)$.
\end{lem}

\begin{proof}
This is an extension to the Moyal context of the renowned inequality by
Cwikel~\cite{Cwikel,SimonTrace,Weidl}. As remarked before, it is
possible to replace $\Dslash^2$ by $-\Delta$, $\pi^\th(f)$ by $L^\th_f$
and $\H$ by $\H_r$. Consider
$g(-i\nb) := (\sqrt{-\Delta} + \eps)^{-1}$. Since $g$ is positive, it
can be decomposed as $g = \sum_{n\in\Z} g_n$ where
$$
g_n(x) := \begin{cases} g(x) &\text{if $2^{n-1} < g(x) \leq 2^n$}, \\
0 &\text{otherwise}. \end{cases}
$$

For each $n \in \Z$, let $A_n$ and $B_n$ be the two operators
$$
A_n := \sum_{k\leq n} L^\th_f \,g_k(-i\nb) \,L^\th_{f^*},  \quad
B_n := \sum_{k>n}     L^\th_f \,g_k(-i\nb) \,L^\th_{f^*}.
$$
We estimate the uniform norm of the first part:
\begin{align*}
\|A_n\|_\infty
&\leq \|L^\th_f\|^2\, \biggl\|\sum_{k\leq n} g_k(-i\nb)\biggr\|_\infty
\leq (2\pi\th)^{-N} \|f\|_2^2
\biggl\| \sum_{k\leq n} g_k \biggr\|_\infty\\
&\leq (2\pi\th)^{-N} \|f\|_2^2 \,2^n =: 2^n\,c_1(\th,N;f).
\end{align*}
The trace norm of $B_n$ can be computed using Lemma~\ref{lm:HSO}:
\begin{align*}
\|B_n\|_1
&= \biggl\|\Bigl(\smash[b]{\sum_{k>n}g_k(-i\nb)}\Bigr)^{1/2}
L^\th_{f^*} \biggr\|_2^2
= \biggl\| L^\th_f \Bigl(\smash[b]{\sum_{k>n}g_k(-i\nb)}\Bigr)^{1/2}
\biggr\|_2^2
= (2\pi)^{-2N} \|f\|_2^2 \,
\biggl\| \Bigl(\smash[b]{\sum_{k>n} g_k}\Bigr)^{1/2} \biggr\|_2^2
\\
&= (2\pi)^{-2N} \|f\|_2^2\, \biggl\| \sum_{k>n} g_k \biggr\|_1
= (2\pi)^{-2N} \|f\|_2^2 \,\sum_{k>n} \|g_k\|_1
\\
&\leq (2\pi)^{-2N} \|f\|_2^2 \,
\sum_{k>n} \|g_k\|_\infty \,\nu\{\supp(g_k)\},
\end{align*}
where $\nu$ is the Lebesgue measure on $\R^{2N}$. By definition,
$\|g_k\|_\infty \leq 2^k$ and
\begin{align*}
\nu\{\supp(g_k)\}
&= \nu\set{x \in \R^{2N} : 2^{k-1} < g(x) \leq 2^k}
\leq \nu\set{x \in \R^{2N} : (|x|+\eps)^{-1} \geq 2^{k-1}}
\\
&\leq 2^{2N(1-k)} \,c_2.
\end{align*}
Therefore
\begin{align*}
\|B_n\|_1
&\leq (2\pi)^{-2N} \|f\|_2^2 \, 2^{2N} c_2\, \sum_{k>n} 2^{k(1-2N)}
\\
&< \pi^{-2N} \,c_2 \,\|f\|_2^2 \, 2^{n(1-2N)}
=: 2^{n(1-2N)} \,c_3(N;f),
\end{align*}
where the second inequality follows because $N > \half$.

We can now estimate the $m$th singular value $\mu_m$ of $B_n$
(arranged in decreasing order with multiplicity):
$\|B_n\|_1 = \sum_{k=0}^\infty \mu_k(B_n)$. Note that, for
$m = 1,2,3,\dots$,
$\|B_n\|_1 \geq \sum_{k=0}^{m-1} \mu_k(B_n) \geq m\,\mu_m(B_n)$. Thus,
$\mu_m(B_n) \leq  \|B_n\|_1\, m^{-1}\leq 2^{n(1-2N)} \,c_3 \,
m^{-1}$. Now Fan's
inequality~\cite[Thm.~1.7]{SimonTrace} yields
\begin{align*}
\mu_m(L^\th_f \,g(-i\nb) \,L^\th_{f^*})
&= \mu_m(A_n + B_n) \leq \mu_1(A_n) + \mu_m(B_n)
\\
&\leq \|A_n\| +  \|B_n\|_1 \,m^{-1}
\leq 2^n\,c_1 + 2^{n(1-2N)} \,c_3 \,m^{-1}.
\end{align*}
Given $m$, choose $n \in \Z$ so that $2^n \leq m^{-1/2N} < 2^{n+1}$.
Then
$$
\mu_m(L^\th_f \,g(-i\nb) \,L^\th_{f^*})
\leq  c_1 \, m^{-1/2N} + c_3 \, m^{-(1-2N)/2N} m^{-1}
=: c_4(\th,N;f) \, m^{-1/2N}.
$$
Therefore
$L^\th_f\,(\sqrt{-\Delta}+\eps)^{-1}\,L^\th_{f^*} \in \L^{2N+}(\H_r)$,
and the statement of the lemma follows.
\end{proof}

\begin{cly}
\label{cr:fragmentation}
If  $f,g \in \SS$, then
$\pi^\th(f)\, (|\Dslash|+\eps)^{-1} \,\pi^\th(g) \in \L^{2N+}(\H)$.
\end{cly}

\begin{proof}
Consider
$\pi^\th(f \pm g^*)\,(|\Dslash|+\eps)^{-1}\,\pi^\th(f^* \pm g)$ and
$\pi^\th(f \pm ig^*)\,(|\Dslash|+\eps)^{-1}\,\pi^\th(f^* \mp ig)$.
\end{proof}

\begin{cly}
\label{cr:spec-dim-one}
If $h \in \SS$, then
$\pi^\th(h)\,(|\Dslash|+\eps)^{-1} \in \L^{2N+}(\H)$.
\end{cly}

\begin{proof}
Let $h = f \mop g$. Then
$$
\pi^\th(h)\, (|\Dslash|+\eps)^{-1}
= \pi^\th(f)\, (|\Dslash|+\eps)^{-1}\,\pi^\th(g)
+ \pi^\th(f)\, [\pi^\th(g), (|\Dslash|+\eps)^{-1}],
$$
and we obtain from the identity \eqref{eq:deriv-inv} that
$$
\pi^\th(h) \,(|\Dslash|+\eps)^{-1}
= \pi^\th(f) \,(|\Dslash|+\eps)^{-1} \,\pi^\th(g)
+ \pi^\th(f) \,(|\Dslash|+\eps)^{-1} \,[|\Dslash|, \pi^\th(g)]\,
(|\Dslash|+\eps)^{-1}.
$$
By arguments similar to those of lemmata \ref{lm:schatten} and
\ref{lm:commutateur}, the last term belongs to $\L^p$ for $p > N$, and
thus to~$\L^{2N+}$.
\end{proof}

Boundedness of $(|\Dslash|+\eps)(\Dslash^2+\eps^2)^{-1/2}$ follows
from elementary Fourier analysis. And so the last corollary means that
the spectral triple is ``$2N^+$-summable''. We have taken care of the
first assertion of the theorem. The next lemma is the last property of
existence that we need.

\begin{lem}
\label{lm:existence}
If $f\in \SS$, then $\pi^\th(f) (|\Dslash|+\eps)^{-2N}$ and
$\pi^\th(f) (\Dslash^2+\eps^2)^{-N}$ are in $\L^{1+}(\H)$.
\end{lem}

\begin{proof}
It suffices to prove that
$\pi^\th(f) (|\Dslash|+\eps)^{-2N} \in \L^{1+}(\H)$. We factorize
$f \in \SS$ according to Proposition~\ref{pr:factorization}, with the
following notation:
\begin{align*}
f &= f_1 \mop f_2 = f_1 \mop f_{21} \mop f_{22}
= f_1 \mop f_{21} \mop f_{221} \mop f_{222}
\\
&= \cdots = f_1 \mop f_{21} \mop f_{221} \mop\cdots\mop
f_{22\cdots 21} \mop f_{22\cdots 22}.
\end{align*}
Therefore,
\begin{align}
\pi^\th(f) \,(|\Dslash|+\eps)^{-2N}
&= \pi^\th(f_1) \,(|\Dslash|+\eps)^{-1} \,\pi^\th(f_2)
\,(|\Dslash|+\eps)^{-2N+1}
\nonumber \\
&\qquad + \pi^\th(f_1) \,(|\Dslash|+\eps)^{-1}
\,[|\Dslash|,\pi^\th(f_2)] \,(|\Dslash|+\eps)^{-2N}.
\label{eq:modtrace}
\end{align}

By Lemma~\ref{lm:schatten},
$\pi^\th(f_1)(|\Dslash|+\eps)^{-1} \in \L^p(\H)$ whenever $p > 2N$;
and by Lemma~\ref{lm:commutateur}, the term
$[|\Dslash|,\pi^\th(f_2)] (|\Dslash|+\eps)^{-2N}$ lies in $\L^q(\H)$
for $q > 1$. Hence, the last term on the right hand side of
equation~\eqref{eq:modtrace} lies in $\L^1(\H)$. We may write the
following equivalence relation:
$$
\pi^\th(f) (|\Dslash|+\eps)^{-2N}
\sim \pi^\th(f_1) (|\Dslash|+\eps)^{-1} \pi^\th(f_2)
(|\Dslash|+\eps)^{-2N+1},
$$
where $A \sim B$ for $A,B \in \K(\H)$ means that $A - B$ is
trace-class. Thus,
\begin{align*}
\pi^\th(f) &(|\Dslash|+\eps)^{-2N}
\sim \pi^\th(f_1) (|\Dslash|+\eps)^{-1} \pi^\th(f_2)
(|\Dslash|+\eps)^{-2N+1}
\\
&= \pi^\th(f_1) (|\Dslash|+\eps)^{-1} \pi^\th(f_{21})
(|\Dslash|+\eps)^{-1} \pi^\th(f_{22}) (|\Dslash|+\eps)^{-2N+2}
\\
&\qquad + \pi^\th(f_1) (|\Dslash|+\eps)^{-1} \pi^\th(f_{21})
(|\Dslash|+\eps)^{-1} \,[|\Dslash|, \pi^\th(f_{22})]\,
(|\Dslash|+\eps)^{-2N+1}
\\
&\sim \pi^\th(f_1) (|\Dslash|+\eps)^{-1} \pi^\th(f_{21})
(|\Dslash|+\eps)^{-1} \pi^\th(f_{22}) (|\Dslash|+\eps)^{-2N+2}
\sim \cdots \\
&\sim \pi^\th(f_1) (|\Dslash|+\eps)^{-1} \pi^\th(f_{21})
(|\Dslash|+\eps)^{-1} \pi^\th(f_{221}) (|\Dslash|+\eps)^{-1}
\dots \pi^\th(f_{22\cdots 22}) (|\Dslash|+\eps)^{-1}.
\end{align*}

The second equivalence relation holds because
$\pi^\th(f_1)(|\Dslash|+\eps)^{-1}\pi^\th(f_{21})(|\Dslash|+\eps)^{-1}
\in \L^p(\H)$ for $p > N$ by Lemma~\ref{lm:schatten}, and
$[|\Dslash|,\pi^\th(f_{22})] (|\Dslash|+\eps)^{-2N+1} \in \L^q(\H)$
for $q > 2N/(2N - 1)$ by Lemma~\ref{lm:commutateur} again. The other
equivalences come from similar arguments.
Corollary~\ref{cr:fragmentation}, the H\"older inequality (see
\cite[Prop.~7.16]{Polaris}) and the inclusion
$\L^1(\H) \subset \L^{1+}(\H)$ finally yield the result.
\end{proof}

\smallskip

Now we go for the computation of the Dixmier trace. Using the
regularized trace for a $\Psi$DO:
$$
\Tr_\La(A) := (2\pi)^{-2N} \iint_{|\xi|\leq\La} \sigma[A](x,\xi)
\,d^{2N}\xi \,d^{2N}x,
$$
the result can be conjectured because
$\lim_{\La\to\infty} \Tr_\La(\.)/\log(\La^{2N})$ is heuristically
linked with the Dixmier trace, and the following computation:
\begin{align*}
\lim_{\La\to\infty} &\frac{1}{2N\log\La}
\Tr_\La \bigl(\pi^\th(f)(\Dslash^2 + \eps^2)^{-N}\bigr)
\\
&= \lim_{\La\to\infty}\,\frac{2^N}{2N(2\pi)^{2N}\log\La}
\iint_{|\xi|\leq\La} f(x - \tfrac{\th}{2}S\xi)\,
(|\xi|^2 + \eps^2)^{-N} \,d^{2N}\xi\,d^{2N}x
\\
&= \frac{2^N\,\Omega_{2N}}{2N\,(2\pi)^{2N}} \int f(x) \,d^{2N}x.
\end{align*}
This is precisely the same result of~\eqref{eq:eastindian-discovery},
in the commutative case, for $k = 2N$. However, to establish it
rigorously in the Moyal context requires a subtler strategy. We shall
compute the Dixmier trace of $\pi^\th(f)\,(\Dslash^2 + \eps^2)^{-N}$
as the residue of the ordinary trace of a related meromorphic family
of operators. For this, recent results of Carey and
coworkers~\cite{CareyPS} extending Connes' trace theorem
(see~\cite{ConnesAction} and~\cite[Chap.~7]{Polaris}) come in handy.
In turn we are allowed to introduce the explicit symbol formula that
will establish measurability~\cite{Book,Polaris}, too.

In the language of~\cite{HigsonRes}, thus, we seek first to verify that
$\A_\th$ has \textit{analytical dimension} equal to $2N$; that is, for
$f \in \A_\th$ the operator $\pi^\th(f)\,(\Dslash^2 + \eps^2)^{-z/2}$
is trace-class if $\Re z>2N$.

\begin{lem}
\label{lm:in-extremis}
If $f \in \SS$, then $L^\th_f\,(\Dslash^2 + \eps^2)^{-z/2}$ is
trace-class for $\Re z>2N$, and
$$
\Tr[L^\th_f\,(\Dslash^2+\eps^2)^{-z/2}] = (2\pi)^{-2N} \iint
f(x)\, (|\xi|^2 + \eps^2)^{-z/2} \,d^{2N}\xi \,d^{2N}x.
$$
\end{lem}

\begin{proof}
If $a(x,\xi) \in \K_p(\R^{2k})$, for $p < -k$, is the symbol of a
pseudodifferential operator $A$, then the operator is trace-class and
moreover
$$
\Tr A = (2\pi)^{-k} \iint a(x,\xi) \,d^kx\,d^k\xi.
$$
This is easily proved by taking $a \in \SS(\R^{2k})$ first and
extending the resulting formula by continuity; have a look
at~\cite{DimassiS,Nicola,Voros} as well.

In our case, the symbol formula for a product of $\PsiDO$s yields, for
$p > N$,
\begin{align*}
\sigma \bigl[L^\th_f(-\Delta + \eps^2)^{-p}\bigr](x,\xi)
&= \sum_{\a\in\N^N} \frac{(-i)^{|\a|}}{\a!} \;
\del_\xi^\a \sigma[L^\th_f](x,\xi) \,
\del_x^\a \sigma\bigl[(-\Delta + \eps^2)^{-p}\bigr](x,\xi)
\\
&= \sigma[L^\th_f](x,\xi) \,
\sigma\bigl[(-\Delta + \eps^2)^{-p}\bigr](x,\xi)
\\
&= f(x - \tfrac{\th}{2}S\xi)\,(|\xi|^2 + \eps^2)^{-p}.
\end{align*}
Therefore, for $p > N$,
\begin{align*}
\Tr\bigl(L^\th_f(-\Delta + \eps^2)^{-p}\bigr)
&= (2\pi)^{-2N}
\iint f(x - \tfrac{\th}{2} S\xi)\,(|\xi|^2 + \eps^2)^{-p}
\,d^{2N}\xi \,d^{2N}x
\\
&= (2\pi)^{-2N}
\iint f(x)\, (|\xi|^2 + \eps^2)^{-p} \,d^{2N}\xi \,d^{2N}x.
\tag*\qed
\end{align*}
\hideqed
\end{proof}

We continue with a technical lemma, in the spirit
of~\cite{RennieLocal}. Consider the approximate unit
$\{\ee_K\}_{K\in\N} \subset \A_c$ where
$\ee_K := \sum_{0\leq|n|\leq K} f_{nn}$. These $\ee_K$ are projectors
with a natural ordering: $\ee_K \mop \ee_L = \ee_L \mop \ee_K = \ee_K$
for $K \leq L$, and they are local units for~$\A_c$.

\begin{lem}
\label{lm:traceclass}
Let $f \in \A_{c,K}$. Then
$$
\pi^\th(f) \,(\Dslash^2+\eps^2)^{-N} - \pi^\th(f) \,
\bigl( \pi^\th(\ee_K)(\Dslash^2+\eps^2)^{-1}\pi^\th(\ee_K) \bigr)^N
\in \L^1(\H).
$$
\end{lem}

\begin{proof}
For simplicity we use the notation $e := \ee_K$ and
$e_n := \ee_{K+n}$. By the boundedness of $\pi^\th(f)$, we may assume
that $f = e \in \A_{c,K}$.

Because $e_n \mop e = e \mop e_n = e$, it is clear that
\begin{equation}
\pi^\th(e) (\Dslash+\la)^{-1} \bigl( 1 - \pi^\th(e_n) \bigr)
= \pi^\th(e) \,(\Dslash+\la)^{-1} \,[\Dslash,\pi^\th(e_n)]
\,(\Dslash+\la)^{-1}.
\label{eq:difference}
\end{equation}
Also, $\pi^\th(e)\,[\Dslash, \pi^\th(e_n)] =
[\Dslash,\pi^\th(e\mop e_n)] - [\Dslash,\pi^\th(e)]\,\pi^\th(e_n) = 0$
because $[\Dslash,\pi^\th(e)]\,\pi^\th(e_n) = [\Dslash,\pi^\th(e)]$ for
$n = 1$ or bigger ---see equation \eqref{eq:plusun} of the Appendix. We
obtain
\begin{align*}
& A_n := \pi^\th(e) (\Dslash+\la)^{-1} [\Dslash,\pi^\th(e_n)]
(\Dslash+\la)^{-1}
\nonumber\\
&= \pi^\th(e) (\Dslash+\la)^{-1} [\Dslash,\pi^\th(e_1)]
(\Dslash+\la)^{-1} [\Dslash,\pi^\th(e_n)] (\Dslash+\la)^{-1}
\nonumber\\
&= \pi^\th(e) (\Dslash+\la)^{-1} [\Dslash,\pi^\th(e_1)] \pi^\th(e_2)
(\Dslash+\la)^{-1} [\Dslash,\pi^\th(e_n)] (\Dslash+\la)^{-1}
= \cdots
\nonumber\\
&= \bigl( \pi^\th(e) (\Dslash+\la)^{-1} \bigr)
\bigl( [\Dslash,\pi^\th(e_1)] (\Dslash+\la)^{-1} \bigr)
\bigl( [\Dslash,\pi^\th(e_2)] (\Dslash+\la)^{-1} \bigr)
\cdots \bigl( [\Dslash,\pi^\th(e_n)] (\Dslash+\la)^{-1} \bigr).
\end{align*}

Taking $n = 2N$ here, $A_{2N}$ appears as a product of $2N+1$ terms in
parentheses, each in $\L^{2N+1}(\H)$ by Lemma~\ref{lm:schatten}. Hence,
by H\"older's inequality, $A_{2N}$ is trace-class and therefore
$\pi^\th(e) (\Dslash+\la)^{-1} (1 - \pi^\th(\ee_{2N})) \in \L^1(\H)$.
Thus,
\begin{align}
\pi^\th(e)\, &(\Dslash^2+\eps^2)^{-1} \bigl(1-\pi^\th(\ee_{4N})\bigr)
\nonumber\\
&= \pi^\th(e) (\Dslash-i\eps)^{-1}
\bigl(1-\pi^\th(\ee_{2N}) + \pi^\th(\ee_{2N})\bigr)
(\Dslash+i\eps)^{-1} \bigl(1-\pi^\th(\ee_{4N})\bigr)
\nonumber\\
&= \pi^\th(e)(\Dslash-i\eps)^{-1} \bigl(1-\pi^\th(\ee_{2N})\bigr)
(\Dslash+i\eps)^{-1} \bigl(1-\pi^\th(\ee_{4N})\bigr)
\nonumber\\
&\qquad + \pi^\th(e) (\Dslash-i\eps)^{-1} \pi^\th(\ee_{2N})
(\Dslash+i\eps)^{-1} \bigl(1-\pi^\th(\ee_{4N})\bigr) \in \L^1(\H).
\label{eq:difference2}
\end{align}
This is to say $\pi^\th(e) (\Dslash^2+\eps^2)^{-1} \sim
\pi^\th(e) (\Dslash^2+\eps^2)^{-1} \pi^\th(\ee_{4N})$.
Shifting this property, we get
\begin{align*}
\pi^\th(e) (\Dslash^2+\eps^2)^{-N}
&\sim \pi^\th(e) (\Dslash^2+\eps^2)^{-1} \pi^\th(\ee_{4N})
(\Dslash^2+\eps^2)^{-N+1}
\\
&\sim \pi^\th(e) (\Dslash^2+\eps^2)^{-1} \pi^\th(\ee_{4N})
(\Dslash^2+\eps^2)^{-1} \pi^\th(\ee_{8N}) (\Dslash^2+\eps^2)^{-N+2}
\sim \cdots
\\
&\sim \pi^\th(e) (\Dslash^2+\eps^2)^{-1} \pi^\th(\ee_{4N})
(\Dslash^2+\eps^2)^{-1} \pi^\th(\ee_{8N}) \cdots
(\Dslash^2+\eps^2)^{-1} \pi^\th(\ee_{4N^2}).
\end{align*}
By identity \eqref{eq:deriv-inv}, the last term on the right equals
\begin{align*}
&\pi^\th(e) (\Dslash+i\eps)^{-1} \pi^\th(e) (\Dslash-i\eps)^{-1}
\pi^\th(\ee_{4N}) (\Dslash^2+\eps^2)^{-1} \pi^\th(\ee_{8N}) \cdots
(\Dslash^2+\eps^2)^{-1} \pi^\th(\ee_{4N^2})
\\
&\quad + \pi^\th(e) (\Dslash+i\eps)^{-1} [\Dslash,\pi^\th(e)]
(\Dslash^2+\eps^2)^{-1} \pi^\th(\ee_{4N}) (\Dslash^2+\eps^2)^{-1}
\pi^\th(\ee_{8N}) \cdots (\Dslash^2+\eps^2)^{-1} \pi^\th(\ee_{4N^2}).
\end{align*}

The last term is trace-class because it is a product of $N$ terms in
$\L^p(\H)$ for $p > N$ and one term in $\L^q(\H)$ for $q > 2N$, by
Lemma~\ref{lm:schatten}. Removing the second $\pi^\th(e)$ once again,
by the ordering property of the local units $\ee_K$ yields
\begin{align*}
&\pi^\th(e) (\Dslash+i\eps)^{-1} \pi^\th(e) (\Dslash-i\eps)^{-1}
\pi^\th(\ee_{4N}) (\Dslash^2+\eps^2)^{-1} \pi^\th(\ee_{8N}) \cdots
(\Dslash^2+\eps^2)^{-1} \pi^\th(\ee_{4N^2})
\\
&= \pi^\th(e) (\Dslash^2+\eps^2)^{-1} \pi^\th(e)
(\Dslash^2+\eps^2)^{-1} \pi^\th(\ee_{8N}) \cdots
(\Dslash^2+\eps^2)^{-1} \pi^\th(\ee_{4N^2})
\\
&\quad + \pi^\th(e) (\Dslash^2+\eps^2)^{-1} [\Dslash,\pi^\th(e)]
(\Dslash-i\eps)^{-1} \pi^\th(\ee_{4N}) (\Dslash^2+\eps^2)^{-1}
\pi^\th(\ee_{8N}) \cdots (\Dslash^2+\eps^2)^{-1} \pi^\th(\ee_{4N^2}).
\end{align*}
The last term is still trace-class, hence
$$
\pi^\th(e) (\Dslash^2+\eps^2)^{-N}
\sim \pi^\th(e) (\Dslash^2+\eps^2)^{-1} \pi^\th(e)
(\Dslash^2+\eps^2)^{-1} \pi^\th(\ee_{8N}) \cdots
(\Dslash^2+\eps^2)^{-1} \pi^\th(\ee_{4N^2}).
$$
This algorithm, applied another $(N-1)$ times, yields the result:
$$
\pi^\th(e) (\Dslash^2+\eps^2)^{-N}
\sim \bigl( \pi^\th(e) (\Dslash^2+\eps^2)^{-1} \pi^\th(e) \bigr)^N.
\eqno\qed
$$
\hideqed
\end{proof}

We retain the following consequence.

\begin{cly}
\label{cr:killproj}
$\Tr^+\bigl(\pi^\th(g)\,[\pi^\th(f),(\Dslash^2+\eps^2)^{-N}]\bigr)
= 0$ for any $g \in \SS$ and any projector $f \in \A_c$.
\end{cly}

\begin{proof}
This follows from Lemma~\ref{lm:traceclass} applied to
$\pi^\th(f)\,(\Dslash^2+\eps^2)^{-N}$ and its adjoint.
\end{proof}

Now we are finally ready to evaluate the Dixmier traces.

\begin{prop}
\label{pr:calcul}
For $f \in \SS$, any Dixmier trace $\Tr^+$ of
$\pi^\th(f)\,(\Dslash^2 + \eps^2)^{-N}$ is independent of~$\eps$, and
$$
\Tr^+ \bigl( \pi^\th(f)\,(\Dslash^2 + \eps^2)^{-N} \bigr)
= \frac{2^N\,\Omega_{2N}}{2N\,(2\pi)^{2N}} \int f(x) \,d^{2N}x
= \frac{1}{N!\,(2\pi)^N} \int f(x) \,d^{2N}x.
$$
\end{prop}

\begin{proof}
We will first prove it for $f \in \A_c$. Choose $e$ a unit for $f$,
that is, $e \mop f = f \mop e = f$. By Lemmata~\ref{lm:existence}
and~\ref{lm:traceclass}, and because $\L^1(\H)$ lies inside the kernel
of the Dixmier trace, we obtain
$$
\Tr^+( \pi^\th(f) \,(\Dslash^2+\eps^2)^{-N}) = \Tr^+ \bigl(
\pi^\th(f) \,(\pi^\th(e) (\Dslash^2+\eps^2)^{-1} \pi^\th(e))^N \bigr).
$$
Lemma~\ref{lm:traceclass} applied to $f = e$ implies that
$\bigl( \pi^\th(e) (\Dslash^2+\eps^2)^{-1} \pi^\th(e) \bigr)^N$ is a
positive operator in $\L^{1+}(\H)$, since it is equal to
$\pi^\th(e) (\Dslash^2+\eps^2)^{-N}$ plus a term in $\L^1(\H)$. Thus,
\cite[Thm.~5.6]{CareyPS} yields (since the limit converges, any
Dixmier trace will give the same result):
\begin{align}
\Tr^+ \bigl( \pi^\th(f) \,(\Dslash^2+\eps^2)^{-N} \bigr)
&= \lim_{s\downarrow 1} (s-1) \Tr\bigl[ \pi^\th(f)\,
(\pi^\th(e) (\Dslash^2+\eps^2)^{-1} \pi^\th(e))^{Ns} \bigr]
\nonumber \\
&= \lim_{s\downarrow 1} (s-1) \Tr\bigl( \pi^\th(f) \pi^\th(e)
(\Dslash^2+\eps^2)^{-Ns} \pi^\th(e) + E_{Ns} \bigr),
\label{eq:decompo}
\end{align}
where
$$
E_{Ns} := \pi^\th(f) \,
\bigl( \pi^\th(e) (\Dslash^2+\eps^2)^{-1} \pi^\th(e) \bigr)^{Ns}
- \pi^\th(f)\pi^\th(e) (\Dslash^2+\eps^2)^{-Ns} \pi^\th(e).
$$
Lemma~\ref{lm:traceclass} again shows that $E_N \in \L^1(\H)$.

Now for $s > 1$,  the first term $\pi^\th(f) \,\bigl(\pi^\th(e)
(\Dslash^2+\eps^2)^{-1} \pi^\th(e) \bigr)^{Ns}$ of $E_{Ns}$ is in
$\L^1(\H)$. In effect, using Lemma~\ref{lm:schatten} and since
$\pi^\th(e) (\Dslash^2+\eps^2)^{-1} \in \L^p(\H)$ for $p > N$, we have
$\pi^\th(e) (\Dslash^2+\eps^2)^{-1} \pi^\th(e) \in \L^{Ns}(\H)$. This
operator being positive, one concludes
$$
\bigl( \pi^\th(e) (\Dslash^2+\eps^2)^{-1} \pi^\th(e) \bigr)^{Ns}
\in \L^1(\H).
$$
The second term
$\pi^\th(f) \pi^\th(e) (\Dslash^2+\eps^2)^{-Ns} \pi^\th(e)$ lies in
$\L^1(\H)$ too, because
$$
\|\pi^\th(e) (\Dslash^2+\eps^2)^{-Ns} \pi^\th(e)\|_1
= \|(\Dslash^2+\eps^2)^{-Ns/2} \pi^\th(e)\|_2^2
= \|\pi^\th(e) (\Dslash^2+\eps^2)^{-Ns/2}\|_2^2
$$
is finite by Lemma~\ref{lm:HSO}. So $E_{Ns} \in \L^1(\H)$ for
$s \geq 1$, and \eqref{eq:decompo} implies
\begin{align*}
\Tr^+ \bigl( \pi^\th(f) \,(\Dslash^2+\eps^2)^{-N} \bigr)
&= \lim_{s\downarrow 1} (s-1) \Tr \bigl(
\pi^\th(f) \pi^\th(e) (\Dslash^2+\eps^2)^{-Ns} \pi^\th(e) \bigr)
\\
&= \lim_{s\downarrow 1}
(s-1) \Tr \bigl(\pi^\th(f) (\Dslash^2+\eps^2)^{-Ns} \bigr).
\end{align*}

Applying now Lemma~\ref{lm:in-extremis}, we obtain
\begin{align*}
\Tr^+ \bigl( \pi^\th(f) \,(\Dslash^2+\eps^2)^{-N} \bigr)
&= \lim_{s\downarrow 1} (s-1) \Tr(1_{2^N})
\Tr\bigl(L^\th_f (-\Delta +\eps^2)^{-Ns}\bigr)
\\
&= 2^N(2\pi)^{-2N} \lim_{s\downarrow 1} (s-1)
\iint f(x)\, (|\xi|^2+\eps^2)^{-Ns} \,d^{2N}\xi \,d^{2N}x
\\
&= \frac{1}{N!\,(2\pi)^N} \int f(x) \,d^{2N}x,
\end{align*}
where the identity
$$
\int (|\xi|^2+\eps^2)^{-Ns} \,d^{2N}\xi
= \pi^N \frac{\Ga(N(s-1))}{\Ga(Ns)\,\eps^{2N(s-1)}},
$$
and $\Ga(N\a) \sim 1/N\a \as \a \downarrow 0$ have been used. The
proposition is proved for $f \in \A_c$.

Finally, take $f$ arbitrary in $\SS$, and recall that $\{e_K\}$ is an
approximate unit for $\A_\th$. Since $f = g \mop h$ for some
$g,h \in \SS$, Corollary~\ref{cr:killproj} implies
\begin{align*}
\bigl| \Tr^+ &\bigl( (\pi^\th(f) - \pi^\th(\ee_K \mop f \mop \ee_K))
(\Dslash^2+\eps^2)^{-N} \bigr) \bigr|
\\
&= \bigl| \Tr^+ \bigl( (\pi^\th(f) - \pi^\th(\ee_K \mop f))\,
(\Dslash^2+\eps^2)^{-N} \bigr) \bigr|
\\
&= \bigl| \Tr^+ \bigl( (\pi^\th(g) - \pi^\th(\ee_K \mop g))\,
\pi^\th(h) (\Dslash^2+\eps^2)^{-N} \bigr) \bigr|
\\
&\leq \| \pi^\th(g) -\pi^\th(\ee_K \mop g) \|_\infty \,
\Tr^+ \bigl| \pi^\th(h)\,(\Dslash^2+\eps^2)^{-N} \bigr|.
\end{align*}
Since $\|\pi^\th(g) -\pi^\th(\ee_K \mop g)\|_\infty
\leq (2\pi\th)^{-N/2}\|g - \ee_K \mop g\|_2$ tends to zero when $K$
increases, the proof is complete because $\ee_K\mop f\mop \ee_K$ lies
in $\A_c$ and
\begin{align}
\int[\ee_K\mop f\mop \ee_K](x)\,d^{2N}x \to \int f(x)\,d^{2N}x\as
K\uparrow\infty.
\tag*{\qed}
\end{align}
\hideqed
\end{proof}

\begin{rem}
Similar arguments to those of this section (or a simple comparison
argument) show that for
$f\in\SS$,
$$
\Tr^+ \bigl( \pi^\th(f)\,(|\Dslash| + \eps)^{-2N} \bigr)
= \Tr^+ \bigl( \pi^\th(f)\,(\Dslash^2 + \eps^2)^{-N} \bigr).
$$
\end{rem}

In conclusion: the analytical and spectral dimension of Moyal planes
coincide. Lemma~\ref{lm:existence}, Proposition~\ref{pr:calcul} and the
previous remark have concluded the proof of
Theorem~\ref{th:dim-spectrale}. \qed

\subsection{The regularity condition}

\begin{thm}
\label{th:regular}
For $f \in \Aun_\th$, the bounded operators $\pi^\th(f)$ and
$[\Dslash,\pi^\th(f)]$ lie in the smooth domain of the derivation
$\delta(T) := [|\Dslash|,T]$.
\end{thm}

The traditional recursive proof \cite{ConnesMIndex,Polaris} does not
work in its original form because the useful transformations $L$ and
$R$ are undefined in the noncompact case (i.e., $|\Dslash|^{-1}$ is
not available). However, an analogue of this proof may exist if
instead of \eqref{eq:lr-smooth} we define $L_\la$ and $R_\la$, for
real~$\la$, as
$$
L_\la(\.) := (|\Dslash| + i\la)^{-1}\,[\Dslash^2,\.],  \quad
R_\la(\.) := [\Dslash^2,\.]\,(|\Dslash| - i\la)^{-1}.
$$

Here we prefer to prove the theorem by its north face: this approach
is still valid for the commutative case, compact or not.

\begin{proof}[Proof of Theorem~\ref{th:regular}]
As before, because
$[\Dslash,\pi^\th(f)] = -iL^\th(\del_\mu f) \ox \ga^\mu$, it is
sufficient to prove that $\pi^\th(f)$ lies in the smooth domain
of~$\delta$. For each $n \in \N$ and $\rho > 0$, we may iterate the
spectral identity \eqref{eq:spectid} $n$~times, to get for
$\delta^n(\pi^\th(f))$:
$$
\frac{1}{\pi^n} \int_0^\infty \!\!\cdots \int_0^\infty
\prod_{i=1}^n \frac{\sqrt{\la_i}}{(|\Dslash|+\rho)^2+\la_i} \,
(\ad(|\Dslash|+\rho)^2)^n \,(\pi^\th(f))
\prod_{i=1}^n \frac{1}{(|\Dslash|+\rho)^2+\la_i} \,d\la_n\dots d\la_1,
$$
with an obvious notation for the $n$-fold iterated commutators.

Because $[\Dslash^2,\pi^\th(f)] = \Dslash^2(f) + 2\Dslash(f)\,\Dslash$,
with the notation $\Dslash(f) := -iL^\th(\del_\mu f) \ox \ga^\mu$, we
can check that the term with the highest power of $\Dslash$ in the
expansion of $(\ad(|\Dslash|+\rho)^2)^n \,(\pi^\th(f))$ is
$2^n\Dslash^n(f)\,\Dslash^n$. For the rest of the proof, we consider
only such highest-power terms. As in the proof of
Lemma~\ref{lm:commutateur}, all commutators
$[|\Dslash|, \pi^\th(f)]$, which appear due to the artificial presence
of~$\rho$, will be treated as a sum of two first order operators.
Hence,
\begin{align}
\frac{1}{\pi^n} & \int_0^\infty \!\!\cdots \int_0^\infty
\prod_{i=1}^n \frac{\sqrt{\la_i}}{(|\Dslash|+\rho)^2+\la_i}
2^n \Dslash^n(f) \Dslash^n
\prod_{j=1}^n \frac{1}{(|\Dslash|+\rho)^2+\la_j} \,d\la_n\dots d\la_1
\nonumber \\
&=\frac{1}{\pi^n} \int_0^\infty \!\!\cdots \int_0^\infty
2^n \Dslash^n(f) \,\Dslash^n \prod_{i=1}^n
\frac{\sqrt{\la_i}}{((|\Dslash|+\rho)^2+\la_i)^2}
               \,d\la_n\dots d\la_1
\label{eq:chorizo} \\
&\qquad + \frac{1}{\pi^n} \int_0^\infty \!\!\cdots \int_0^\infty
\biggl[ \prod_{i=1}^n \frac{1}{(|\Dslash|+\rho)^2+\la_i},
2^n \Dslash^n(f) \biggr] \Dslash^n \prod_{i=1}^n
\frac{\sqrt{\la_i}}{(|\Dslash|+\rho)^2+\la_i} \,d\la_n\dots d\la_1.
\nonumber
\end{align}

Using $\int_0^\infty t(\la+t^2)^{-2} \sqrt{\la}\,d\la = \pi/2$, the
first term on the right hand side of \eqref{eq:chorizo} equals
$$
2^n \Dslash^n(f)  \frac{\Dslash^n}{(|\Dslash|+\rho)^n}
\biggl( \frac{1}{\pi} \int_0^\infty
\frac{|\Dslash|+\rho}{((|\Dslash|+\rho)^2+\la)^2}
\,\sqrt{\la}\,d\la \biggr)^n
=\Dslash^n(f)  \frac{\Dslash^n}{(|\Dslash|+\rho)^n},
$$
which is a bounded operator.

For the other term, notice that the commutator
$\bigl[\prod_i ((|\Dslash|+\rho)^2+\la_i)^{-1}, \Dslash^n(f)\bigr]$
can be rewritten as
$$
- \prod_{i=1}^n ((|\Dslash|+\rho)^2+\la_i)^{-1} \,
\biggl[ \prod_{j=1}^n ((|\Dslash|+\rho)^2+\la_j), \Dslash^n(f) \biggr]
\prod_{k=1}^n ((|\Dslash|+\rho)^2+\la_k)^{-1},
$$
and the highest-power term of this expression is, up to a constant:
$$
\prod_{i=1}^n ((|\Dslash|+\rho)^2+\la_i)^{-1} \,
\Dslash^{n+1}(f)\,\Dslash^{2n-1}
\prod_{k=1}^n ((|\Dslash|+\rho)^2+\la_k)^{-1}.
$$
So the proof reduces to showing the finiteness of the following norm:
\begin{align*}
&\biggl\| \int_0^\infty \!\!\cdots \int_0^\infty
\prod_{i=1}^n \frac{\sqrt{\la_i}}{(|\Dslash|+\rho)^2+\la_i}\,
\Dslash^{n+1}(f)\,\Dslash^{3n-1}
\biggl( \prod_{j=1}^n \frac{1}{(|\Dslash|+\rho)^2+\la_j} \biggr)^2
\,d\la_n\dots d\la_1 \biggr\|
\\
&\leq \|\Dslash^{n+1}(f)\| \int_0^\infty \!\!\cdots \int_0^\infty
\prod_{i=1}^n \biggl( \biggl\|
\frac{\Dslash^{3-1/n}}{((|\Dslash|+\rho)^2+\la_i)^{3/2-1/2n}} \biggr\|
\biggr\| \frac{1}{(|\Dslash|+\rho)^2+\la_i} \biggr\|^{3/2+1/2n}
\! \sqrt{\la_i} \,d\la_i \biggr)
\\
&\leq \|\Dslash^{n+1}(f)\| \biggl( \int_0^\infty
\frac{\sqrt{\la}}{(\rho^2 + \la)^{3/2+1/2n}} \,d\la \biggr)^n.
\end{align*}
This integral is finite for all $n\in\N$ and so is the norm since
$\del^\a f \in \Aun_\th \subset A_\th$ for $|\a| \leq n+1$. The
proof is complete.
\end{proof}

\subsection{The finiteness condition}

\begin{lem}
The smooth vectors for $\Dslash$ are given by
$$
C^\infty(\Dslash) \equiv \H^\infty
:= \bigcap_{k\in\N} \Dom(\Dslash^k) \simeq \D_{L^2} \ox \C^{2^N}.
$$
\end{lem}

\begin{proof}
Since $\D_{L^2}$ is the common smooth domain of the partial
derivatives $\del_\mu$, for $\mu = 1,\dots,2N$, and since
$\Dslash = -i\del_\mu \ox \ga^\mu$, the conclusion is clear.
\end{proof}

Take $\A_1 := \D_{L^2}$; by Lemma~\ref{lm:DL2}, this is an ideal
in $\Aun_\th$. Then $\H^\infty$ is an $\A_1$-pullback of a free left
$\Aun_\th$-module.

On $\H^\infty$, there is a natural $\A_1$-valued hermitian structure,
given by
$$
\roundbraket{\xi}{\eta}' := \sum_{j=1}^{2^N} \xi_j \mop \eta_j^*,
\sepword{for all} \xi,\eta \in \H^\infty.
$$
Because $\D_{L^2} \subset \M^\th$, the hermitian pairing
$\roundbraket{\pi^\th(a)\xi}{\eta}' = a \mop \roundbraket{\xi}{\eta}'$
is $\A_\th$-valued whenever $a \in \A_\th$.
Proposition~\ref{pr:calcul} and Lemma~\ref{lm:propriete}(v) now imply
\begin{align*}
\Tr^+\bigl( \pi^\th(\roundbraket{\xi}{\eta}')\,
(\Dslash^2 + \eps^2)^{-N} \bigr)
&= \frac{2^N\,\Omega_{2N}}{2N\,(2\pi)^{2N}} \sum_{j=1}^{2^N}
\int (\xi_j \mop \eta_j^*)(x) \,d^{2N}x
\\
&= \frac{1}{N!\,(2\pi)^N} \sum_{j=1}^{2^N}
\int \eta_j^*(x)\,\xi_j(x) \,d^{2N}x.
\end{align*}
Therefore,
$\roundbraket{\xi}{\eta} := N!\,(2\pi)^N \roundbraket{\xi}{\eta}'$ is
the desired hermitian structure satisfying~\eqref{eq:abs-cont}. Its
uniqueness can be checked in the same way as in~\cite[p.~501]{Polaris}.
In summary: the inner product on $\H$ is tightly linked to the natural
hermitian structure on $\H^\infty(D)$ by means of the resolvent of~$D$
and the noncommutative integral.

\begin{rem}
An obvious integral estimate makes it clear that
$\Oh_r \subset \D_{L^2}$ if and only if $r < -N$. Consider, therefore,
$\NN := \bigcup_{r<-N} \Oh_r \subset \D_{L^2}$. It follows from
Proposition~\ref{pr:hector} that $\NN$ is a $*$-algebra for the
twisted product~$\mop$, and that it is also an ideal in
$\Aun_\th = \B$. This space has already been used with physical
motivations in~\cite{LangmannM}.
\end{rem}

\subsection{The other axioms for the Moyal $2N$-plane}

\begin{description}

\item[$\quad \bullet$]
The signs for the table \eqref{eq:real-signs} are easily checked in
the representation \eqref{eq:charge-conjn}; indeed, since neither $J$
nor $\Dslash$ depend directly on~$\th$, it suffices to check these
signs in the commutative case. The reality property follows at once.

\item[$\quad \bullet$]
The first-order property comes directly from \eqref{eq:right-Moyal},
since
$$
[[\Dslash, \pi^\th(f)], J \pi^\th(g) J^{-1}]
= [\pi^\th(\Dslash(f)), J \pi^\th(g) J^{-1}]
= -i\,[L^\th(\del_\mu f) \ox \ga^\mu, R^\th(g^*) \ox 1] = 0,
$$
because left and right twisted multiplications commute.

\item[$\quad \bullet$]
The orientation property requires a suitable Hochschild $2N$-cycle
over the preferred unitization $\Aun_\th = \B$. As already
mentioned, there is a natural embedding of the noncommutative torus
$\Coo(\T_\Theta^k)$ in $\B$ as periodic functions. Indeed, the
generators $u_j$, $j = 1,\dots,2N$, of the NC torus correspond exactly
to the elementary plane waves $u_j(x):=e^{ix_j}$, which are unitary
elements for the twisted product and satisfy the same algebraic
relations~\eqref{eq:tori-relns}. Now the very same
formula~\eqref{eq:vol-tori}, rewritten with the Moyal product, yields
the desired Hochschild $2N$-cycle; and the volume-form
relation~\eqref{eq:vol-form} can be checked with the same
calculation~\cite{Polaris} as for the NC torus:
$$
\frac{(-i)^N}{(2N)!} \sum_\sigma (-1)^\sigma
(u_{\sigma(1)} \mop u_{\sigma(2)} \mop\cdots\mop
u_{\sigma(2N)})^{\star_{\Theta}-1}
[\Dslash,u_{\sigma(1)}]\cdots[\Dslash,u_{\sigma(2N)}] = \chi.
$$
Here the $\pi^\th$-representation has been understood.

The plane waves belong not to the Schwartz algebra $\SS$, but rather
to its unitization $\B$. No finite sum of tensors with entries from
the ``small'' algebra will make up a Hochschild cycle $\cc$ satisfying
$\pi_\Dslash(\cc) = \chi$, because of decay at infinity; thus, if one
wants to avoid approximation sequences both of operators and volume
forms, passage to a compactification containing at least the plane
waves is ineluctable.

\end{description}

The commutant of $\A_\th$ acting by left multiplication consists of
right multipliers. Indeed, it has been shown~\cite{MeloM} that among
those operators on $L^2(\R^{2N})$ which are smooth for the adjoint
action of the Heisenberg group, the commutant of $R^\th(\SS)$ is
exactly $L^\th(\B)$. Right multipliers do not commute with $\Dslash$
unless they are scalars. Therefore the Moyal spin geometries are
\emph{connected}. In this respect, left Moyal quantization behaves like
a prequantization~\cite{Tuynman,vanHove}: see our remark at the end of
the conclusions section.

\smallskip

Checking back our arguments and estimates, we find that we have proved
something stronger than what we set out to show: most properties hold
for $\A_1$, which is determined solely by $D$; and so, the outcome of
the tug-of-war between the operator and the algebra witnessed in the
previous pages is the triumph of the operator, which goes a very long
way to determine both the algebra and the inner product on the triple's
Hilbert space.

\begin{thm}
\label{th:main}
The Moyal planes $(\A,\Aun,\H,\Dslash,J,\chi)$ are connected real
noncompact spectral triples of spectral dimension~$2N$, for which
$\A_1$ (as introduced in subsections 3.1, 3.2), is equal to $\D_{L^2}$.
Moreover, all postulates for noncompact noncommutative geometries
except the first are fulfilled if we replace $\A$ by $\A_1$ throughout.
\qed
\end{thm}

This is the main result of the paper.

\section{Moyal--Wick monomials}
\label{sec:Moyal-Wick}

\subsection{An algebraic mould}

In this section we put the theory developed in the previous sections to
good use in clarifying some fundamentals of quantum noncommutative
field theory. In the NCFT literature, models based on spacetime
equations like $(\square + m^2)\vf(x) = g\,\vf^{\star r}(x)$,
where~$\vf$ is a quantized scalar field and $r$ a suitable integer, are
commonplace (we suppress for a while the $\th$ in the notation for the
star product). However, this is a formal equation, and in practically
all the treatments $\vf^{\star r}$ is in want of rigorous definition.
The Moyal product does not help with the ordering issue in quantum
field theory, and therefore that equation should be given in normally
ordered form
$$
(\square + m^2)\vf(x) = g\,\wick:\vf^{\star r}(x): \,.
$$
Thus we need a concept of normally-ordered Moyal products of fields, or
Moyal--Wick monomials. Such a definition should work at least for free
fields, to serve as basis for a perturbative treatment of interacting
ones.

In order to avoid excessively model-dependent casuistics in the
definition of what noncommutative Wick monomials should be, it is
imperative to employ an algebraic framework. Such a framework
fortunately exists~\cite{BaezBZ}, and it turns out to mesh very well
with Connes' formulation of noncommutative geometry in terms of
spectral triples. We contend with Baez, Segal and Zhou as well that it
is natural to regard those monomials as quadratic form-valued, rather
than operator-valued, distributions; this improves and simplifies the
usual definition \textit{\`a la} Wightman and
G{\aa}rding~\cite{WightmanG}. In turn this will be helpful with the
explicit rigorous construction of noncommutative Wick monomials in the
Moyal algebra context, that we perform next, in which $\A_1 = \D_{L^2}$
is again found to play the main role. In short: the theory of
noncompact spectral triples is born in intimate contact with quantum
field theory.

According to Segal, the boson quantization of a separable complex
Hilbert space $\H$ with inner product $\braket\cdot\cdot$ (assuming the
simplest circumstance in which a suitably unique quantum vacuum can be
chosen) consists of a quadruple $(\K,\vac,\b,\Ga)$, where
$\K \equiv \Ga(\H)$ is another separable Hilbert space; $\vac$ is a
distinguished unit vector in~$\K$; $\b$ is a strongly continuous map
from~$\H$ to the group of unitary operators on~$\K$ satisfying:
$$
\b(v) \b(v') = \b(v+v') \exp[-i\Im\braket{v}{v'}]
$$
for all $v,v' \in \H$, and such that the span of
$\{\b(v)\vac : v \in \H\}$ is dense in~$\K$; and $\Ga$ is a unitary
representation on $\K$ of the group of unitaries on~$\H$, fulfilling
the covariance condition
$$
\Ga(U) \b(U^{-1}v) \Ga(U)^{-1} = \b(v),
$$
for which $\vac$ is stationary, and such that the infinitesimal
generator $d\Ga(A)$ of the one-parameter group $\Ga(\exp(itA))$ is
positive selfadjoint on~$\K$ whenever $A$ is positive selfadjoint on
$\H$. Up to unitary equivalence, this abstract setting uniquely leads
to the standard boson (Hopf) algebra on~$\H$, with the customary
construction of the second-quantized operators.

The most important condition on a compact Connes triple, in regard to
our subject, is the finiteness prescription. For the purposes of this
paper, where the vector bundle aspect is completely trivial, we could
as well identify $\A \equiv \H^\infty(D)$ as vector spaces. In the
present nonunital case, in order to have projective modules an
auxiliary multiplier algebra $\Aun$ of $\A$ is needed; still
$\H^\infty(D)$ must be an $\A$-pullback of a finite projective
$\Aun$-module, and we can keep the identification of $\A$ and
$\H^\infty(D)$.

Assume, then, that the Hilbert space $\H$ for Segal's framework has
been identified. It is clear that some more structure is required if
one is to construct singular operators like the Wick polynomials. The
role of a distinguished operator $\DD$ ---and of its quantum
counterpart $d\Ga(\DD)$--- is precisely to determine domains of
regularity for them. For technical reasons, in the context
of~\cite{BaezBZ} the operator $\DD$ must be taken strictly positive:
$\DD \geq \eps$ for some $\eps > 0$; so in particular it is invertible.
The operator $\DD$ might not be strictly positive at the outset; but a
related operator will do. For instance, for the scalar case,
commutative or not, we may use $\DD := (\Dslash^2 + \eps^2)^{1/2}$.

Denote $\K^\infty(\DD) := \bigcap_{k\in\N}\Dom(d\Ga(\DD)^k)\subset\K$.
A typical element of $\K^\infty(\DD)$ is a symmetrized tensor power of
elements of $\H^\infty(\DD)$; in fact the algebraic span of such
vectors is dense in $\K^\infty(\DD)$. The boson field $\vf(v)$ is just
the selfadjoint generator of $\b(tv)$; the Segal field
$\vf(v) = a(v) + a^\7(v)$ (here $a(v)$ and $a^\7(v)$ are the usual
annihilation and creation operators) is essentially selfadjoint on
$\K^\infty(\DD)$, and it is easy to see that for
$v \in \H^\infty(\DD)$, it sends $\K^\infty(\DD)$ continuously into
itself.

It is advantageous to think of $\vf(v)$ as a quadratic form; we recall
how this comes about. Let $L$ be a dense subspace of $\H$, gifted with
a topology stronger than that of $\H$ (in our case, $\H^\infty(\DD)$
and $\K^\infty(\DD)$ are given the projective Fr\'echet space
topologies associated to the families of norms $\|\DD^n(\cdot)\|$ and
$\|d\Ga(\DD)^n(\cdot)\|$, respectively), and let $f$ be a continuous
sesquilinear form on~$L$. One could try to introduce a Hilbert space
operator $T^\H_f$ through $f(u,v) =: \braket{u}{T^\H_f v}$, defined on
elements $v$ of $L$ for which $f(u,v) \leq c_v \|u\|$ for all~$u$; but
that condition might only hold for, say, $v = 0$. However, if
$L^\sharp$ is the antidual of $L$, then $\H \hookto L^\sharp$ with a
continuous embedding, since $u \mapsto \braket{u}{v}$ is an antilinear
continuous functional on~$L$, and $f$ defines a map
$T_f\: L \to L^\sharp$ by $T_fv(u) := f(u,v)$. The elements of
$L^\sharp$ in a concrete representation for $\H$ are distributions; and
so quadratic forms are generalized operators. Often,
$(\H^\infty(\DD))^\sharp$ is denoted $\H^{-\infty}(\DD)$.

We refer to~\cite[Sec.~7.3]{BaezBZ} for the following estimate: for
all $v \in \H$, $\Phi \in \K^\infty(\DD)$ and $m \geq 1$,
$$
\|a(v)\Phi\| \leq C\,\|\DD^{-m}v\| \, \|d\Ga(\DD)^m\Phi\|.
$$
{}From that, and the formula
\begin{equation}
\braCket{\Psi}{\wick:\vf(w_1)\dots\vf(w_n):}{\Phi}
= \sum_{I\subseteq\{1,\dots,n\}} \biggl<
\prod_{i\in I^c} a(w_i)\Psi \biggm| \prod_{j\in I} a(w_j)\Phi \biggr>,
\label{eq:mama-de-Tarzan}
\end{equation}
with $\Psi,\Phi\in\K^\infty(\H)$ and $I^c = \{1,\dots,n\}\setminus I$,
it is immediate that one can define a \textit{Wick map} from monomials
in the free algebra over $\H^{-\infty}(\DD)$ to quadratic forms on
$\K^\infty(\DD)$, extending the similar map in the subalgebra
generated by~$\H$.

\smallskip

To fix ideas in the following, the reader can put $2N = 4$. We work on
Euclidean space rather than on Minkowski spacetime, but formal passage
to relativistic field theory (where however everything takes place
on-shell) is quite simple. The conservative approach is to have the
$\wick:\vf^{\star r}(x):$ living in the commutative context, that is,
in the boson algebra $\K = \Ga(\H)$ over the Hilbert space $\H$ of
square-summable functions on momentum space. As already indicated,
$\K \simeq \bigoplus_{n=0}^\infty \H^{\vee n}$, where $\H^{\vee n}$
is identified to the space of complex symmetric functions $\Phi$,
square-integrable with respect to the standard volume form
$d^{2N}p_1 \dots d^{2N}p_n$ in~$\R^{2Nn}$. Precisely, the norm on
$\H^{\vee n}$ is taken to be
$$
\|\Phi^{(n)}\|^2 := \idotsint n!\, |\Phi^{(n)}(p_1,\dots,p_n)|^2
\,\prod_{i=1}^n d^{2N}p_i.
$$
Then $\H^{\infty}(\DD)$ is nothing other than the space $\D_{L^2}\,$!
Furthermore, it is possible to take $w_i(x) = \delta(x - x_i)$ in the
above~\eqref{eq:mama-de-Tarzan}, as the distribution
$\delta(\cdot - x_i)$ belongs to
$\H^{-\infty}(\DD) = H^{2,-\infty} = \D'_{L^2}$.

An outcome of the previous discussion is that the Wick products
$$
\wick:\vf(x_1) \dots \vf(x_l):\,
:= \,\wick:\vf(\delta(x - x_1)) \dots \vf(\delta(x - x_l)):
$$
used by physicists make perfect sense as continuous sesquilinear forms
on the corresponding $\K^\infty(\DD)$, and \emph{a fortiori} on the
space of Fock vectors with finitely many nonvanishing components, each
one belonging to (a symmetrized tensor power of) $\D_{L^2}$. The
function from $\R^{2Nl} \x (\K^\infty(\DD))^2$ to $\C$ given by
$$
(x_1,\dots,x_n;\Psi,\Phi) \mapsto
\braCket{\Psi}{\wick:\vf(x_1)\dots\vf(x_l):}{\Phi},
$$
being continuous (indeed, smooth) in $x_1,\dots,x_l$, can be restricted
to the diagonal; and this defines the (ordinary) Wick monomials
$\wick:\vf^l(x):$ for any~$l$. That is to say,
\begin{equation}
\braCket{\Psi}{\wick:\vf^l(x):}{\Phi} =
\braCket{\Psi}{\< \wick:\vf(x_1) \dots \vf(x_l):,
  \delta(x - x_1) \dots \delta(x - x_l) >_{x_1,\dots,x_l}}{\Phi}
\label{eq:papa-de-los-tomates}
\end{equation}
is a well-defined expression. Thus, and more important still, we have
established that manipulations with Dirac delta functions ---such as
the ones we are going to use later to define Moyal--Wick monomials---
are justifiable. In this respect, the good behaviour of $\D_{L^2}$
under the Moyal product, as under the ordinary one, becomes crucial.
Also, for the same reason that the better algebra to represent the
Moyal plane is $(\D_{L^2},\mop)$ rather than $(\SS,\mop)$, the use of
Schwartz functions and tempered distributions in the classic paper by
Wightman and G{\aa}rding~\cite{WightmanG}, in which Wick products and
Wick monomials were defined as operator-valued distributions, has been
revealed as artificial.

\subsection{The noncommutative Wick monomials}
\label{sec:NC-Wick}

For ease of reference, we give here the explicit expression of the
ordinary commuting Wick products
\begin{align}
&\bigl[\wick:\vf(x_1)\dots\vf(x_l):\,\Phi\bigr]^{(n)}(p_1,\dots,p_n)
\nonumber \\
&\quad = (2\pi)^{-Nl} \sum_{j=0}^l \idotsint \sum_{|X|=l-j}
\frac{1}{j!\,(l-j)!} \sum_P
\bigl[ P\,e^{i(x_1\eta_1 +\cdots+ x_j\eta_j
-x_{j+1}\eta_{j+1} -\cdots- x_l\eta_l)} \bigr]
\nonumber\\
&\hspace{8em} \x
\Phi^{(n-l+2j)}(\eta_1,\dots,\eta_j,p_1,\dots,\widehat{\eta_{j+1}},
\dots,\widehat{\eta_l},\dots,p_n) \,\prod_{k=1}^j d^{2N}\eta_k,
\label{eq:as-de-oros}
\end{align}
where $P$ runs over all permutations of the momentum variables, and
$X = \{\eta_{j+1},\dots,\eta_l\}$ ranges over all subsets of $l - j$
distinct elements of $\{p_1,\dots,p_n\}$. Consequently, for good
measure:
\begin{align*}
\bigl[ \wick:\vf^l:(x)\, &\Phi \bigr]^{(n)}(p_1,\dots,p_n)
:= \bigl[ \wick:\vf^l(x):\,\Phi \bigr]^{(n)}(p_1,\dots,p_n)
\\
&= (2\pi)^{-Nl} \sum_{j=0}^l \idotsint \sum_{|X|=l-j}
\frac{1}{j!\,(l-j)!} \sum_P
\bigl[ P\,e^{ix(\eta_1+\dots+\eta_j-\eta_{j+1}-\dots-\eta_l)} \bigr]
\\
&\hspace{5em} \x
\Phi^{(n-l+2j)}(\eta_1,\dots,\eta_j,p_1,\dots,\widehat{\eta_{j+1}},
\dots,\widehat{\eta_l},\dots,p_n) \,\prod_{k=1}^j d^{2N}\eta_k.
\end{align*}
We have used operator rather than sesquilinear-form notation, although
$\wick:\vf^l(x): \,\Phi$ for $\Phi \in \K^\infty(\DD)$ is not in $\K$,
instead it is an (actually rather tame) vector-valued distribution. But
it is guaranteed that
$\braCket{\K^\infty(\DD)}{\wick:\vf^l(x):}{\K^\infty(\DD)}$ is finite.

Let us now reinstate the Moyal product associated to a $k \x k$
skewsymmetric matrix $\Theta$; for now, we assume $\Theta$ to be
nondegenerate. Formula~\eqref{eq:moyal-prod-gen} can be construed as
meaning
$$
\delta(x-s) \star_\Theta \delta(x-t) = (\pi\th)^{-2N}
e^{-2i(s\.\Theta^{-1}t)} e^{-2i(x\.\Theta^{-1}s+t\.\Theta^{-1}x)}.
$$
The left hand side could of course have been written, somewhat more
correctly, as $(\delta_s \star_\Theta \delta_t)(x)$. More generally, an
easy two-step induction gives
\begin{subequations}
\label{eq:delta-chain}
\begin{align}
\delta(x-x_1&) \star_\Theta\dots\star_\Theta \delta(x-x_{2m})
\nonumber \\
&= (\pi\th)^{-2Nm}\, e^{2i\sum_{i<j}(-)^{i+j}x_i\.\Theta^{-1}x_j}
\,e^{-2ix\.\Theta^{-1}(x_1-x_2+x_3-\cdots-x_{2m})},
\label{eq:delta-chain-even}
\\
\delta(x-x_1&) \star_\Theta\dots\star_\Theta \delta(x-x_{2m+1})
\nonumber \\
&= (\pi\th)^{-2Nm}\, e^{2i\sum_{i<j}(-)^{i+j}x_i\.\Theta^{-1}x_j}
\,\delta(x-x_1+x_2-x_3+\cdots-x_{2m+1}).
\label{eq:delta-chain-odd}
\end{align}
\end{subequations}
These functionals of $x_1,\dots,x_{2m}$ or $x_1,\dots,x_{2m+1}$ belong
to $(\D'_{L^2})^{\!2m}$ or respectively $(\D'_{L^2})^{\!2m+1}$
---recall that the space of rapidly decreasing distributions $\Oh'_C$
is a subspace of $\D'_{L^2}$~\cite{Schwartz}.

\smallskip

There can be no question of making $\wick:\vf(x_1)\dots\vf(x_l):$
``noncommutative''; so, how are we to define the Moyal--Wick products
$\wick:\vf^{\star_\Theta l}:(x)$?

A ``quantum Wick product'' was recently introduced
in~\cite{BahnsDFPouf}; but it is at variance with Moyal NCFT, and so is
unsuitable for our present purposes. A different course is suggested by
the older duality theory of~\cite{Phobos, Deimos} and the discussion in
the previous subsection. Our declared tactics are to construct
$\wick:\vf^{\star_\Theta l}:(x)$ on the very same Fock space of the
real scalar field. This would seem to run against the spirit of
noncommutative geometry, but is in fact demanded by our results here so
far, and the treatment in the previous subsection. We posit
\begin{equation}
\wick:\vf^{\star_\Theta l}:(x)
:= \< \wick:\vf(x_1)\dots\vf(x_l):, \delta(x-x_1)
\star_\Theta\dots\star_\Theta \delta(x-x_l)>_{x_1,\dots,x_l},
\label{eq:as-de-bastos}
\end{equation}
to be compared with \eqref{eq:papa-de-los-tomates}.

We may also define Moyal products of Moyal--Wick monomials with
suitable scalar functions or distributions on configuration space:
\begin{align*}
\wick:\vf^{\star_\Theta l}: \star_\Theta h (x)
&= \< \wick:\vf(x_1)\dots\vf(x_l):,
\delta(x-x_1) \star_\Theta\cdots\star_\Theta \delta(x-x_l)
\star_\Theta h(x) >_{x_1,\dots,x_l}
\\
h \star_\Theta \wick:\vf^{\star_\Theta l}: (x)
&= \< \wick:\vf(x_1)\dots\vf(x_l):,
h(x) \star_\Theta \delta(x-x_{1}) \star_\Theta\cdots\star_\Theta
\delta(x-x_l) >_{x_1,\dots,x_l}.
\end{align*}
What it is required is that the functional $\delta(x-x_1)
\star_\Theta\cdots\star_\Theta \delta(x-x_l) \star_\Theta h(x)$, in the
$x_1,\dots,x_l$ variables, belong to $(\H^{-\infty}(\DD))^l$. A
seemingly alternative definition is given by
\begin{align*}
\<\wick:\vf^{\star_\Theta l}: \star_\Theta h(x), g(x)>
&= \<\wick:\vf^{\star_\Theta l}:(x), h \star_\Theta g(x)>,
\\
\<h \star_\Theta \wick:\vf^{\star_\Theta l}:(x), g(x)>
&= \<\wick:\vf^{\star_\Theta l}:(x), g \star_\Theta h(x)>,
\end{align*}
in the spirit of~\cite{Phobos,Deimos}, for suitable spaces of
functions $g$ and distributions $h$. The verification that both kinds
of definition coincide is immediate.

Note that the identity
$$
\<\wick:\vf^{\star_\Theta l}:(x), h(x)>
= \<\wick:\vf(x_1)\dots\vf(x_l):,
h(x_1) \star_\Theta \delta(x_1-x_2) \star_\Theta\cdots\star_\Theta
\delta(x_1-x_l) >_{x_1,\dots,x_l}
$$
affords a definition of the Moyal--Wick monomials \textit{\`a la}
Wightman and G{\aa}rding.

Using now \eqref{eq:delta-chain} together with \eqref{eq:as-de-oros}
and~\eqref{eq:as-de-bastos}, we obtain the completely explicit formula
on the boson Fock space
\begin{align*}
(2\pi)^{Nl}\,
&\bigl[\vf^{\star_\Theta l}(x)\Phi\bigr]^{(n)}(p_1,\dots,p_n)
\\
&\quad = \sum_{j=0}^l \idotsint \sum_{|X|=l-j} \frac{1}{j!\,(l-j)!}
\sum_P \bigl[ P\,e^{ix(\eta_1+\cdots+\eta_j-\eta_{j+1}-\cdots-\eta_l)}
e^{\mp\tihalf\sum_{m<r}\eta_m\.\Theta\eta_r} \bigr]
\\
&\hspace{5em} \x
\Phi^{(n-l+2j)}(\eta_1,\dots,\eta_j,p_1,\dots,\widehat{\eta_{j+1}},
\dots,\widehat{\eta_l},\dots,p_n) \,\prod_{k=1}^j d^{2N}\eta_k.
\end{align*}
Here in the exponent quadratic in the $\eta$'s the $-$~sign applies
when $r \leq j$ or $m > j$, the $+$~sign otherwise. In the simplest
instance, we get
\begin{align*}
(2\pi)^{3N}
&\bigl[\<\vf^{\star2}(x), h(x)>\Phi\bigr]^{(n)}(k_1,\dots,k_n)
\\
&= \iint \hat{h}(\kappa_1+\kappa_2) \cos{\thalf\kappa_1\Theta\kappa_2}
\,\Phi^{(n+2)}(\kappa_1,\kappa_2,k_1,\dots,k_n)
\,d^{2N}\kappa_1 \,d^{2N}\kappa_2
\\
&\qquad + \sum_{j=1}^n \int
\bigl[ \hat{h}(\kappa - k_j)e^{\tihalf\kappa\Theta k_j}
+ \hat{h}(k_j - \kappa)e^{\tihalf k_j\Theta\kappa} \bigr]\,
\Phi^{(n)}(\kappa,k_1,\dots,\widehat{k_j},\dots,k_n)
\,d^{2N}\kappa
\\
&\qquad + \sum_{1\leq j\ne l\leq n} \hat{h}(-k_j - k_l)
\cos{\thalf k_j\Theta k_l}\,
\Phi^{(n-2)}(k_1,\dots,\widehat{k_j},\dots,\widehat{k_l},\dots,k_n).
\end{align*}

(We underline again that the $\wick:\vf(x_1)\dots\vf(x_l):$, here as
in~\eqref{eq:papa-de-los-tomates}, are the usual commutative boson
products of fields, with all creation operators to the left of the
annihilation operators ---see~\cite[Sec.~4.1]{BrouderO}--- as for
instance in
\begin{align*}
&(2\pi)^{3N} \wick:\vf(x_1)\vf(x_2)\vf(x_3):
\\
&= \idotsint \bigl[ e^{i(k_1x_1+k_2x_2+k_3x_3)} a(k_1) a(k_2) a(k_3)
+ e^{-i(k_1x_1-k_2x_2-k_3x_3)} a^\7(k_1) a(k_2) a(k_3)
\\
&\qquad
+ e^{i(k_1x_1-k_2x_2+k_3x_3)} a^\7(k_2) a(k_1) a(k_3)
+ e^{i(k_1x_1+k_2x_2-k_3x_3)} a^\7(k_3) a(k_1) a(k_2)
\\
&\qquad
+ e^{-i(k_1x_1+k_2x_2-k_3x_3)} a^\7(k_1) a^\7(k_2) a(k_3)
+ e^{-i(k_1x_1-k_2x_2+k_3x_3)} a^\7(k_1) a^\7(k_3) a(k_2)
\\
&\qquad
+ e^{i(k_1x_1-k_2x_2-k_3x_3)} a^\7(k_2) a^\7(k_3) a(k_1)
+ e^{-i(k_1x_1+k_2x_2+k_3x_3)} a^\7(k_1) a^\7(k_2) a^\7(k_3) \bigr]
\, \prod_{i=1}^3 d^{2N}k_i.
\end{align*}
In turn we are assured that $\vf^{\star_\Theta l}(x)$ is normally
ordered. Had we tried to use in~\eqref{eq:as-de-bastos} the operator
product instead of the normal product, we would have been punished by
extra divergent terms of the type
$\int \delta(k_1 - k_2) \,d^{2N}k_1 \,d^{2N}k_2$, just as in the
commutative case. Thus, as anticipated, the twisted product does not
help with the ordering problem.)

\smallskip

The previous formulae have been obtained under the assumption that
$\det\Theta > 0$. For $k = 2N$, the set of nonsingular skewsymmetric
$k \x k$ matrices is open and dense in the set of all skewsymmetric
$k \x k$ matrices, and the same formulae are valid when
$\det\Theta = 0$ by continuity. We also conclude their validity in the
case that $k$ is odd, by consideration of an extra dimension with
trivial commutation relations.

\section{The functional action}

The functional action plays a great role in the applications to
physics of noncommutative geometry, because it reproduces not only the
Yang--Mills action but also the full
Yang--Mills--Higgs~\cite{IochumS,Sirius} and even, in its more general
incarnation~\cite{AliAlain}, the Einstein--Hilbert action.

Here we choose, for the reasons indicated in the introduction, to
compute the Connes--Lott action
\cite{ConnesL1,ConnesL2,ConnesAction,Cordelia}, which notionally is
$\Tr^+(F^2\,\Dslash^{-2N})$, for $F$ the field strength or curvature
associated to a vector potential $\a$. Due to some ambiguity in the
transition to $F$ from $\a$, unimportant for the general theory but
crucial for physics, we need to deal with ``junk'': that is, to
quotient by an ideal living in the representation $\pi_D$ of the
universal differential algebra on~$\H$. Then we show that the action
coincides with the noncommutative Yang--Mills action currently used in
Moyal gauge theory. Of course, physicists have not waited for formal
developments of this kind before forging ahead (see~\cite{Camilla},
for instance); but for our purposes this is an indispensable check.

We first make some necessary remarks on the bimodule nature of the
image of~$\pi_D$.

\subsection{Connes--Terashima fermions}

That bimodule nature is completely familiar to customers of Connes'
noncommutative geometry, and basically means that $\H^\infty$ can
sustain a bimodule action of two algebras. The reconstruction of the
Standard Model Lagrangian in~\cite{ConnesReal} uses actions of this
type, exchanged by the charge conjugation operator.

Independently, in the traditional context of Lie algebras,
Terashima~\cite{Terashima} summarized to similar effect some natural
methods and restrictions, that were scattered in practice, to
introduce noncommutative gauge fields.

First of all, assume an infinitesimal gauge variation given by
$$
\delta_\la A_\mu(x) = \del_\mu \la(x) - i[A_\mu,\la]_\mop(x),
$$
or explicitly,
$$
\delta_\la A_\mu(x)
= T^a \del_\mu \la_a(x) - \tfrac{i}{2} \bigl(
[T^a,T^b]\,\{A^a_\mu,\la^b\}_\mop(x)
+ \{T^a,T^b\}\,[A^a_\mu,\la^b]_\mop(x) \bigr)
$$
where the $T^a$ denote the gauge ``group'' generators, normalized as
$$
\Tr(T^a T^b) = \half\,\delta^{ab},
$$
closing to a Lie algebra
$$
[T^a, T^b] = i\,f^{ab}_c\,T^c,
$$
and $[\.,\.]_\mop$, $\{\.,\.\}_\mop$ denote the Moyal commutator and
anticommutator brackets, respectively. Let us think of the Lie algebra
of $SU(n)$, to fix ideas. Then
$$
\{T^a, T^b\} = \frac{1}{n} \,\delta^{ab} + d^{ab}_c\,T^c,
$$
where the $d^{ab}_c$ are totally symmetric and real. This is not a
linear combination of $T^d$'s. Therefore noncommutative gauge
transformations are consistent only for unitary groups (there are some
ways round this obstacle; but they are not very appealing). But then
the gauge group of unitary transformations is identified to the unitary
endomorphism group of a module, and we are back in Connes' context.

The second remark by Terashima is that the same closure requirement
and consideration of the covariant derivative forces the
representation of $U(n)$ to be fundamental or antifundamental. It is
possible, however, for a gauge group to act from the left, say in the
fundamental representation, and (perhaps a different one) from the
right in the antifundamental one, with gauge transformations given by
$$
\Psi \mapsto U_{(1)} \mop \Psi \mop U^*_{(2)}.
$$
Again, this is completely natural in the context of algebra bimodules.
We remark that already the chiral anomaly for these fermions has been
calculated~\cite{CarmeloJuggernaut}.

We want to add that, even in the context of pure group theory, the
concept of bimodule is called for. We formalize this remark, in the
spirit of~\cite{Wildberger}. By definition, a linear space $V$ is a
$(G,H)$-bimodule if it carries a left action $\triangleright$ of the
group $G$ (with the usual continuity or smoothness conditions) and a
right action $\triangleleft$ of the group $H$ which are compatible,
that is to say,
$$
g \triangleright (v \triangleleft h)
= (g \triangleright v) \triangleleft h,
$$
for $g \in G$, $h \in H$, $v \in V$. A bimodule is irreducible if
there is no proper subspace of $V$ stable under both actions. If $V$
is a $G$ left-module and $W$ is an $H$ right-module, then $V \ox W$
is a $(G,H)$-bimodule. When $G = H$, interesting bimodules
usually have a conjugation operator that exchanges the actions. In
that case, the bimodule is a very canonical object in harmonic
analysis: the space of functions on a group $G$ is a $(G,G)$-bimodule;
and if $G$ possesses a representation on a space $W$, then
$V = \opname{End}W$ is a $(G,G)$-bimodule; but, strangely enough, it
does not seem to be in use.

\subsection{The differential algebra}

In the Connes--Lott approach one works with the tensor product of some
finite dimensional Eigenschaften algebra and a spacetime algebra (that
here is no longer commutative). We disregard the Eigenschaften algebra
in what follows; in other words, we concentrate on the analytical
details of the $U(1)$ Moyal gauge theory.

In this subsection we do not need to consider the preferred
unitization of~$\A_\th$. As in \cite{ConnesL1,ConnesL2,Book}, let
$\Omega^\8 \A_\th := \bigoplus_{p\in\N} \Omega^p\A_\th$ be the
universal differential graded algebra over $\A_\th$, where
$\Omega^p \A_\th :=
\set{f_0\,\delta f_1 \dots \delta f_p : f_i \in \A_\th}$ and the only
constraint on $\delta$ is to satisfy the Leibniz rule
$\delta(f_1\mop f_2) = \delta f_1\,f_2 + f_1\,\delta f_2$ so $\delta$
can be extended on $\Omega^\8 \A_\th$. Since $\A_\th$ has no unit, we
define \cite[III.1.$\a$]{Book} $\Omega^0\A_\th := \A_\th\oplus \C$,
which is the minimal unitization of $\A_\th$, and
$\delta(0\oplus 1) := 0$. Moreover $(\delta f)^* := \delta{f^*}$.

The representation $\pi^\th$ of $\A_\th$ by elements of $\L(\H)$
extends naturally to $\Omega^\8\A_\th$, by
$$
\tpi^\th : \Omega^p\A_\th \to \L(\H) :
f_0\,\delta f_1 \dots \delta f_p \mapsto
i^p\,\pi^\th(f_0)\,[\Dslash,\pi^\th(f_1)]\dots[\Dslash,\pi^\th(f_p)].
$$

\begin{lem}
If $f_i \in \A_\th$, then
$\tpi^\th(f_0\,\delta f_1 \dots \delta f_p) =
L^\th(f_0 \mop \del_{\mu_1}f_1 \mop\cdots\mop \del_{\mu_p}f_p)
\ox \ga^{\mu_1} \dots \ga^{\mu_p}$.
\end{lem}

\begin{proof}
This follows from
$[\Dslash, L^\th_f \ox 1_{2^N}] = -i L^\th(\del_\mu f) \ox \ga^\mu$
and $L^\th_f\, L^\th_g = L^\th(f \mop g)$.
\end{proof}

To overcome the unfaithfulness of $\tpi^\th$ (even if $\pi^\th$ is
faithful), one introduces a graded 2-sided ideal of $\Omega^\8\A_\th$,
namely $\Junk := \bigoplus_{p\in \N} J^p =
\bigoplus_{p\in\N} J_0^p + \delta J_0^{p-1}$,
$J_0^p := \set{\omega\in \Omega^p\A_\th :\tpi^\th(\omega) = 0}$, and
finally
$$
\Omega_\Dslash\,\A_\th := \tpi^\th(\Omega^\8\A_\th)/\tpi^\th(\Junk).
$$
Here, the 2-junk is particularly simple since it is isomorphic to
$\pi^\th(\A_\th)$, as we now show.

\begin{prop}
There is a natural identification
$\tpi^\th(J^2) \simeq \pi^\th(\A_\th) = L^\th(\A_\th) \ox 1_{2^N}$.
\end{prop}

\begin{proof}
Any $\omega \in \tpi^\th(J^2) \subset \tpi^\th(\Omega^2\A_\th)$ can be
written as $\omega = \sum_{j\in I}
L^\th(\del_\mu f_j) L^\th(\del_\nu g_j) \ox \ga^\mu\ga^\nu$ where $I$
is a finite set, and satisfies
$\sum_{j\in I} L^\th(f_j \mop \del_\mu g_j) \ox \ga^\mu = 0$. By the
Leibniz rule,
\begin{align*}
\omega &= \sum_{j\in I}
L^\th(\del_\mu(f_j \mop \del_\nu g_j) - f_j \mop \del_\mu\del_\nu g_j)
\ox \ga^\mu \ga^\nu
= -\sum_{j\in I} L^\th(f_j \mop \del_\mu \del_\nu g_j)
\ox \ga^\mu \ga^\nu\\
& = -\sum_{j\in I} L^\th(f_j \mop \del_\mu \del_\nu g_j)
\ox \eta^{\mu\nu}\,1_{2^N}.
\end{align*}

Hence
$\tpi^\th(J^2) \subset \pi^\th(\A_\th) = L^\th(\A_\th) \ox 1_{2^N}$.

Consider
$\omega_{mnkl} := f_{mk}\,\delta f_{kn} - f_{ml}\,\delta f_{ln}$ (no
summation) in $\Omega^1\A_\th$. In subsection~\ref{sec:more-junk} of
the Appendix, it is shown that $\tpi^\th(\omega_{mnkl}) = 0$ and
$\tpi^\th(\delta\omega_{mnkl}) = \frac2{\th} \sum_{j=1}^N (k_j - l_j)
\,L^\th(f_{mn}) \ox 1_{2^N}$, which is nonzero if $|l| \neq |k|$. Thus,
$L^\th(f_{mn}) \ox 1_{2^N}$ lies in $\tpi^\th(J^2)$ for all
$m,n \in \N^N$. Since $\{f_{mn}\}$ is a basis for $\A_\th$, we
conclude that $\pi^\th(\A_\th) \simeq
L^\th(\A_\th) \ox 1_{2^N} \subset \tpi^\th(J^2)$.
\end{proof}

It is easy to generalize the above proof, to get the next Corollary.

\begin{cly}
For $p \geq 2$, $\tpi^\th(J^p)$ is the linear span of the elements in
$\tpi^\th(\Omega^p\A_\th)$ of the form
$L^\th_f \ox \ga^{\mu_1} \dots \ga^{\mu_k}$, with $k \leq p - 2$ and
of the same parity as~$p$.
\end{cly}

\subsection{The action}

Let $\Ht_p$ be the Hilbert space obtained by completion of
$\tpi^\th(\Omega^p \A_\th)$ under the scalar product
$$
\braket{\tpi^\th(\omega)}{\tpi^\th(\omega')}_p
:= \Tr^+ \bigl( \tpi^\th(\omega)^* \,\tpi^\th(\omega')\,
(\Dslash^2 + \eps^2)^{-N} \bigr),
$$
for $\omega,\omega' \in \Omega^p \A_\th$. This defines a natural
pre-action $I(\eta)$ when $p = 2$ and
$\omega' = \omega = \delta\eta + \eta^2$:
\begin{equation}
\label{eq:pre-action}
I(\eta) := \Tr^+ \bigl( \tpi^\th(\omega)^* \,\tpi^\th(\omega)\,
(\Dslash^2 + \eps^2)^{-N} \bigr).
\end{equation}
Let $P$ be the orthogonal projector on $\Ht_p$ whose range is the
orthogonal complement of $\tpi^\th(\delta J_0^{p-1})$, and define
$\H_p := P\Ht_p$. Then $P$ extends the quotient map from
$\tpi^\th(\Omega^p \A_\th)$ onto $\Omega_\Dslash^p\,\A_\th$, which is
identified with a dense subspace of~$\H_p$. The possible ambiguity in
\eqref{eq:pre-action} due to the unfaithfulness of $\tpi^\th$
disappears if we define the functional action (noncommutative
Yang--Mills action) as:
\begin{equation}
YM(\a)
:= \frac{N!\,(2\pi)^N}{8g^2}\, \braket{P\tpi^\th(F)}{P\tpi^\th(F)}_2
\label{eq:Yang-Mills},
\end{equation}
where $\Omega^1_\Dslash\,\A_\th \ni \a = \tpi^\th(\eta)$ and
$F = \delta\eta + \eta^2$ is the curvature of the 1-form~$\eta$ and
$g$ is the coupling constant. It is shown in \cite{ConnesL1,Sirius}
that $YM(\a)$ is equal to the infimum of the preaction on all
$\eta \in \Omega^1\A_\th$ with the same image in
$\Omega^1_\Dslash\,\A_\th$:
$$
YM(\a)
= \frac{N!\,(2\pi)^N}{8g^2} \inf\set{I(\eta) : \tpi^\th(\eta) = \a}.
$$
This result justifies the notation $YM(\a)$, because this positive
quartic functional of $\eta$ depends only on its equivalence class in
$\Omega^1_\Dslash\A_\th$, namely~$\a$.

\begin{thm}
Let $\eta = -\eta^* \in \Omega^1\A_\th$. Then the Yang--Mills action
$YM(\a)$ of the universal connection $\delta + \eta$, with
$\a = \tpi^\th(\eta)$, is equal to
$$
YM(\a) = -\frac{1}{4g^2} \int F^{\mu\nu} \mop F_{\mu\nu}(x) \,d^{2N}x
= -\frac{1}{4g^2} \int F^{\mu\nu}(x) \,F_{\mu\nu}(x) \,d^{2N}x,
$$
where $F_{\mu\nu} :=
\half(\del_\mu A_\nu - \del_\nu A_\mu + [A_\mu,A_\nu]_\mop)$ and
$A_\mu$ is defined by $\a = L^\th(A_\mu) \ox \ga^\mu$.
\end{thm}

\begin{proof}
If $\eta = \sum_{j\in I} f_j\,\delta g_j$ for some $f_j,g_j \in \SS$
and a finite set $I$, then
$\a = \sum_{j\in I} L^\th_{f_j} L^\th_{\del_\mu g_j} \ox \ga^\mu =
\sum_{j\in I}L^\th(f_j \mop \del_\mu g_j) \ox \ga^\mu$. Thus
$A_\mu := \sum_{j\in I} f_j \mop \del_\mu g_j$ and, with a sum over
$j,k\in I$ understood,
\begin{align*}
\tpi^\th(\delta\eta + \eta^2)
&= \tpi^\th(\delta f_j\,\delta g_j
              + (f_j\,\delta g_j) (f_k\,\delta g_k))
= \tpi^\th(\delta f_j\,\delta g_j
              + f_j\,\delta(g_j \mop f_k) \delta g_k
              - (f_j \mop g_j)\,\delta f_k\,\delta g_k)
\\
&= L^\th(\del_\mu f_j \mop \del_\nu g_j
+ f_j \mop \del_\mu(g_j \mop f_k ) \mop \del_\nu g_k
- f_j \mop g_j \mop \del_\mu f_k \mop \del_\nu g_k) \ox \ga^\mu\ga^\nu
\\
&= L^\th(\del_\mu f_j \mop \del_\nu g_k
+ f_j \mop \del_\mu g_j \mop f_k \mop \del_\nu g_k) \ox \ga^\mu\ga^\nu
\\
&= L^\th(\del_\mu(f_j \mop \del_\nu g_j)
+ f_j \mop \del_\mu g_j \mop f_k \mop \del_\nu g_k) \ox \ga^\mu\ga^\nu
- L^\th(f_j \mop \del_\mu \del_\nu g_j) \ox \eta^{\mu\nu}\,1_{2^N}
\\
&= L^\th(\del_\mu A_\nu + A_\mu \mop A_\nu) \ox \half[\ga^\mu,\ga^\nu]
+ \eta^{\mu\nu} L^\th(\del_\mu A_\nu + A_\mu\mop A_\nu) \ox 1_{2^N}
\\
&\hspace{4em}
- \eta^{\mu\nu} L^\th(f \mop \del_\mu \del_\nu g) \ox 1_{2^N}.
\end{align*}
The two last terms are in $\tpi^\th(J^2)$. Thus,
\begin{align*}
P(\tpi^\th(F))
&= P\bigl( L^\th(\del_\mu A_\nu + A_\mu \mop A_\nu)
\ox \half[\ga^\mu,\ga^\nu] \bigr)
\\
&= P\bigl( L^\th(\half
(\del_\mu A_\nu - \del_\nu A_\mu + [A_\mu,A_\nu]_\mop)
\ox \ga^\mu \ga^\nu \bigr)
\\
&= P(L^\th(F_{\mu\nu}) \ox \ga^\mu \ga^\nu)
             = L^\th(F_{\mu\nu}) \ox \ga^\mu \ga^\nu,
\end{align*}
where the last equality follows because the junk affects only the
scalar part of $\tpi^\th(\Omega^\8\A_\th)$. To repeat: each
$\omega = \omega_{\mu\nu} \ox \ga^\mu \ga^\nu
\in \tpi^\th(\Omega^2\A_\th)$ can be uniquely decomposed as
$$
\omega = \omega_{\mu\nu} \ox \half(\ga^\mu \ga^\nu - \ga^\nu \ga^\mu)
+ \omega_{\mu\nu} \ox \half(\ga^\mu \ga^\nu + \ga^\nu \ga^\mu)
$$
in $\tpi^\th(\Omega^2\A_\th)_a \oplus \tpi^\th(\Omega^2\A_\th)_s
= \tpi^\th(\Omega^2\A_\th)_a \oplus \tpi^\th(J^2)$, the direct sum
of its alternating and symmetric parts.

Since $A_\mu = -A_\mu^*$, we also find $F^*_{\mu\nu} = -F_{\mu\nu}$
and therefore
$P(\tpi^\th(F))^* = L^\th(F_{\mu\nu}) \ox \ga^\mu \ga^\nu$. Then
$$\Tr^+ \bigl( L^\th(F_{\mu\nu} \mop F_{\rho\sigma})\,
(-\Delta+\eps^2)^{-N} \ox \ga^\mu \ga^\nu \ga^\rho \ga^\sigma \bigr)
= \Tr^+ \bigl( L^\th(F_{\mu\nu} \mop F_{\rho\sigma})\,
(-\Delta+\eps^2)^{-N} \bigr)\, \Tr(\ga^\mu\ga^\nu\ga^\rho\ga^\sigma).
$$
But $\Tr(\ga^\mu\ga^\nu\ga^\rho\ga^\sigma) =
2^N(\eta^{\mu\nu}\eta^{\rho\sigma} - \eta^{\mu\rho} \eta^{\nu\sigma} +
\eta^{\mu\sigma}\eta^{\nu\rho})$, so since $F_{\mu\nu} = -F_{\nu\mu}$
Proposition~\ref{pr:calcul}, computed with $-i\nb$ instead of
$\Dslash$, yields
$$
YM(\a) = - \frac{2\,N!\,(4\pi)^N}{8g^2}
\Tr^+\bigl( L^\th(F_{\mu\nu} \mop F^{\mu\nu})\,
(-\Delta + \eps^2)^{-N} \bigr)
= -\frac{1}{4g^2} \int (F_{\mu\nu} \mop F^{\mu\nu})(x) \,d^{2N}x,
$$
and according to Lemma~\ref{lm:propriete}(v) the pointwise product can
replace the Moyal product.
\end{proof}

\begin{rem}
The action as we have defined it is positive definite, since
$$
YM(\a) = \frac{1}{4g^2}
\int \sum_{\mu,\nu=1}^{2N} |F_{\mu\nu}(x)|^2 \,d^{2N}x.
$$
\end{rem}

\section{Conclusions and outlook}

We have shown in detail how to build noncompact noncommutative spin
geometries. As a consequence, the classical background of present-day
NCFTs is recast in the framework of the rigorous Connes formalism for
geometrical noncommutative spaces.

One can wonder about the uniqueness of the constructions presented
here. Our detailed scrutiny shows that appropriate algebras for the
spectral triples are to a large extent ``selected'' by the Dirac
operator itself. The choice of $\A = \SS$ for the original nonunital
algebra, made in the flat space cases, has much to recommend it, not
least Fourier invariance and the existence of a body of tempered
distribution analysis. However, an outcome of the study in this paper
is that, both in the commutative and the Moyal-algebra example, a more
canonical `arrival' point is the bigger algebra
$\A_1 := \D_{L^2}(\R^{2N})$; we found that nearly everything that works
for $\A$ works also for $\A_1$, with significant improvement of the
finiteness axiom; also, $\A_1$ yields the most advantageous framework
for quantization.

We can accommodate an $\A_1$-triple instead of an $\A$-triple, provided
we make a slight modification of the summability axiom. That is, we use
the data $(\A_1, \Aun, \H, D, J, \chi)$, and the suggested new version
of the noncompact noncommutative geometry postulates runs as follows.

\begin{enumerate}

\item\textit{Spectral dimension, 2nd version}:

There is a unique nonnegative integer $k$, the spectral or
``classical'' dimension of the geo\-me\-try, for which
$a(|D| + \eps)^{-1}$ belongs to the Schatten class $\L^p$ for
$p > k$ whenever $a \in \A_1$, for any $\eps > 0$; and moreover, for
$a$ in a dense ideal $\A$ of $\A_1$, $a(|D| + \eps)^{-1}$ lies in the
generalized Schatten class $\L^{k+}$ and the trace
$a \mapsto \Tr^+(a(|D| + \eps)^{-k})$ is finite and not identically
zero. This $k$ is even if and only if the spectral triple is even.

\addtocounter{enumi}{1}

\item\textit{Finiteness, 2nd version}:

The algebras $\A_1$ and its preferred unitization $\Aun$ are
pre-$C^*$-algebras. The space of smooth vectors $\H^\infty$ is the
$\A_1$-pullback of a finite projective $\Aun$-module. Moreover, an
$\A_1$-valued hermitian structure $\roundbraket{\.}{\.}$ is implicitly
defined on $\H^\infty$ with the noncommutative integral, as follows:
$$
\Tr^+\bigl( \roundbraket{a\xi}{\eta}(|D| + \eps)^{-k} \bigr)
= \braket{\eta}{a\xi},
$$
where $a \in \Aun$ and $\braket{\.}{\.}$ denotes the standard inner
product on~$\H$.

\end{enumerate}

In the other postulates $\A$ is replaced by $\A_1$; they are otherwise
unchanged. In our case $\A$ could be taken equal to $\SS$ or larger:
$\Tr^+(a(|D| + \eps)^{-k})<\infty$ is valid for $a$ belonging to a
larger ideal of $\D_{L^2}$.

Support for enrollment of $\A_1$ comes from physics, on one hand, and
abstract nonsense, on the other. Langmann and
Mickelsson~\cite{LangmannM} found existence of the quantum scattering
matrix for quantized fermions in external gauge potentials with
components precisely in the sibling $\NN$ of $\D_{L^2}$; this is a both
strong and significant result. Also, as exploited in
Section~\ref{sec:Moyal-Wick}, the more correct and general approach to
the construction of Wick monomials makes use precisely of the smooth
domain of the Dirac operator. The close relation of $\A_1$ to this
smooth domain points to generalizations of the pseudodifferential
calculus in the fully noncommutative
context~\cite{HigsonRes,LangmannM}.

\smallskip

The orientation condition and the required boundedness of the operators
$[D,a]$ give rather tight lower and upper bounds (so to speak) on what
the preferred compactification of~$\A_1$ should be. It would be good to
know whether these two conditions determine such a unitization
uniquely. The following conjecture is strengthened by the result
of~\cite{MeloM}.

\begin{conj}
\label{cj:parida}
$\Aun = \B(\R^{2N})$ is the largest Moyal multiplier algebra of
$\A_1 = \D_{L^2}(\R^{2N})$ such that $[\Dslash,a]$ is bounded for each
$a \in \Aun$.
\end{conj}

The clever argument in~\cite{ChakrabortyGS} leads one to ponder what
kind of boundary conditions one would impose on $\Dslash$ (without
presumably changing the leading term behaviour of its spectral density)
in order to obtain a \textit{compact} spectral triple canonically
associated to the given noncompact one. This should allow the anomaly
calculations in~\cite{Camilla,CarmeloJuggernaut} to be made more
rigorous. The subject of noncommutative manifolds with boundary is
still in its infancy, however, and we shall not elaborate the point.

\smallskip

Apart from eventually proving a reconstruction theorem (a rather
strenuous task), much remains to be done. There are probably already
enough examples of noncommutative spaces around for consideration of
the ``category'' of spectral triples to be promising. For instance, NC
tori are quotients of the spaces considered in this paper. A
mathematically important question is the computation of the Hochschild
cohomology of $\Aun_\th$. Another is the explicit lifting of (a central
extension of) the group of (nonlinear, in general) symplectomorphisms
(or at least, of those connected to the identity) to a group of inner
automorphisms of $M^\th$ (or of $A_\th$), which should be irreducibly
represented on $\H$. In this context, work on the geometry of the gauge
algebra in noncommutative Yang--Mills theories~\cite{LizziSZ} can be
pursued.

\section{Appendix: a few explicit formulas}

\subsection{On the oscillator basis functions}
\label{sec:expl-basis}

For $N = 1$ and $m,n \in \N$, the basic eigentransition
$f_{mn}(x_1,x_2)$ is explicitly given by
$$
2(-1)^{\min(m,n)} \frac{\sqrt{n!m!}}{\max(m,n)!}
(4H_1/\\th)^{|m-n|/2} e^{i(n-m)\arctan(x_2/x_1)}
\,\exp(-2H_1/\th) \,L_{\min(m,n)}^{|m-n|}(4H_1/\th),
$$
with $L_j^{r}$ being the generalized Laguerre polynomials of order~$j$
and $H_1 = \half(x_1^2 + x_2^2)$. In general,
$$
f_{mn}(x_1,\dots,x_N)
= f_{m_1n_1}(x_1,x_{1+N}) \dots f_{m_Nn_N}(x_N,x_{2N}).
$$

Also, using the coalgebra formula for the Laguerre polynomials
$$
L_n^{r+s+1}(u + v) = \sum_{j+l=n} L_j^{r}(u) L_l^{s}(v),
$$
one obtains~\cite{BayenFFLS2} eigenstates for $H = H_1 +\cdots+ H_N$:
$$
H \mop f_M = f_M \mop H = \th \biggl(M + \frac{N}{2}\biggr) f_M,
$$
where
$$
f_M(x_1,\dots,x_{2N})
:= \sum_{|m|=M} f_{m_1m_1}\dots f_{m_Nm_N}(x_1,\dots,x_{2N})
= 2^N (-)^M \exp(-2H/\th) \, L_M^{N-1}(4H/\th).
$$

\smallskip

It is known that
$\int |f_{nn}(x_1,x_2)|\,dx_1\,dx_2 \sim \sqrt{n} \as n \to \infty$.
{}From this, using the closed graph theorem, it easy to show that there
are non-absolutely integrable functions in~$\I_{00}$~\cite{Daubechies}.

\subsection{More junk}
\label{sec:more-junk}

\begin{lem}
\label{lm:basic-junk}
For $m,n,k,l \in \N^N$, let
$\omega_{mnkl} := f_{mk}\,\delta f_{kn} - f_{ml}\,\delta f_{ln} \in
\Omega^1\A_\th$ (no summation on~$k$ or~$l$). Then
$$
\tpi^\th(\omega_{mnkl}) = 0  \sepword{and}
\tpi^\th(\delta\omega_{mnkl})
= \tfrac{2}{\th}(|k| - |l|)\, L^\th(f_{mn}) \ox 1_{2^N}.
$$
\end{lem}

\begin{proof}
Using the creation and annihilation functions~\eqref{eq:crea-annl} we
may rewrite the Dirac operator as follows; we adopt the convention that
$j = 1,\dots,N$, and write $\del_{a_j} = \del/\del a_j$ and
$\del_{a_j^*} = \del/\del a_j^*$:
$$
\Dslash = -\frac{i}{\sqrt{2}} \sum_j \ga^j(\del_{a_j} + \del_{a_j^*})
+ i\ga^{j+N} (\del_{a_j}-\del_{a_j^*})
= -i \sum_j (\ga^{a_j}\,\del_{a_j} + \ga^{a_j^*}\,\del_{a_j^*}),
$$
where $\ga^{a_j} := \frac{1}{\sqrt{2}}(\ga^j + i\ga^{j+N})$ and
$\ga^{a_j^*} := \frac{1}{\sqrt{2}}(\ga^j - i\ga^{j+N})$.

Lemma~\ref{lm:propriete}(iv), applied to $a_j$ and $a_j^*$
respectively, yields
$$
\del_{a_j} = -\frac{1}{\th} \ad_\mop a_j^*
:= -\frac{1}{\th} [a_j^*,\.\,]_\mop,  \qquad
\del_{a_j^*} = \frac{1}{\th} \ad_\mop a_j
:= \frac{1}{\th} [a_j,\.\,]_\mop
$$
and hence
$$
\Dslash = - \frac{i}{\th} \sum_j
(\ga^{a_j^*} \ad_\mop a_j - \ga^{a_j} \ad_\mop a_j^*).
$$

Let $u_j := (0,0,\dots,1,\dots,0)$ be the $j$-th standard basis vector
of~$\R^N$. From the definition \eqref{eq:basis} of $f_{mn}$, we
directly compute:
\begin{align*}
a_j^* \mop f_{mn} &= \sqrt{\th(m_j+1)}\, f_{m+u_j,n},
& f_{mn} \mop  a_j^* &= \sqrt{\th n_j}\, f_{m,n-u_j},
\\
a_j \mop f_{mn} &= \sqrt{\th m_j}\, f_{m-u_j,n},
& f_{mn} \mop a_j &= \sqrt{\th(n_j+1)}\, f_{m,n+u_j}.
\end{align*}
Consequently,
\begin{align}
\Dslash(f_{mn})
&= -\frac{i}{\th} \sum_j \ga^{a_j} \bigl(\sqrt{\th n_j}\, f_{m,n-u_j}
- \sqrt{\th(m_j+1)}\, f_{m+u_j,n} \bigr)
\nonumber \\
&\hspace{4em} + \ga^{a_j^*} \bigl( \sqrt{\th m_j}\, f_{m-u_j,n}
- \sqrt{\th(n_j+1)}\, f_{m,n+u_j} \bigr).
\label{eq:plusun}
\end{align}

We are now able to compute $\tpi^\th(\omega_{mnkl})$ and
$\tpi^\th(\delta\omega_{mnkl})$. Firstly,
\begin{align*}
\tpi^\th(\omega_{mnkl})
&= \tpi^\th(f_{mk}\,\delta f_{kn} - f_{ml}\,\delta f_{ln})
= L^\th(f_{mk} \mop \del_\mu f_{kn} - f_{ml} \mop \del_\mu f_{ln})
\ox \ga^\mu
\\
&= \frac{1}{\th} \sum_j \bigl(
\sqrt{\th n_j}\, L^\th(f_{mk} \mop f_{k,n-u_j})
- \sqrt{\th(k_j+1)}\, L^\th(f_{mk} \mop f_{k+u_j,n})
\\
&\hspace{4em} - \sqrt{\th n_j}\, L^\th(f_{ml} \mop f_{l,n-u_j})
+ \sqrt{\th(l_j+1)}\,L^\th(f_{ml} \mop f_{l+u_j,n})\bigr) \ox \ga^{a_j}
\\
&\qquad + \bigl(
\sqrt{\th k_j}\, L^\th(f_{mk} \mop f_{k-u_j,n})
- \sqrt{\th(n_j+1)}\, L^\th(f_{mk} \mop f_{k,n+u_j})
\\
&\hspace{4em} - \sqrt{\th l_j}\, L^\th(f_{ml} \mop f_{l-u_j,n})
+ \sqrt{\th(n_j+1)}\, L^\th(f_{ml} \mop f_{l,n+u_j}) \bigr)
\ox \ga^{ a_j^*}
\\
&= 0.
\end{align*}

Secondly, we calculate that
$$
\tpi^\th(\delta\omega_{mnkl})
= \tpi^\th(\delta f_{mk}\,\delta f_{kn} - \delta f_{ml}\,\delta f_{ln})
= L^\th(\del_\mu f_{mk} \mop \del_\nu f_{kn} -
\del_\mu f_{ml} \mop \del_\nu f_{ln}) \ox \ga^\mu \ga^\nu
$$
equals
\begin{align*}
&\frac{1}{\th^2} \biggl\{ \sum_j \Bigl( \bigl(
\sqrt{\th k_j}\, L^\th(f_{m,k-u_j})
- \sqrt{\th(m_j+1)}\, L^\th(f_{m+u_j,k}) \bigr) \ox \ga^{a_j}
\\
&\hspace{5em} + \bigl( \sqrt{\th m_j}\, L^\th(f_{m-u_j,k})
- \sqrt{\th(k_j+1)}\, L^\th(f_{m,k+u_j}) \bigr) \ox \ga^{ a_j^*}
\Bigr)
\\
&\hspace{3em} \sum_p \Bigl( \bigl(
\sqrt{\th n_p}\, L^\th(f_{k,n-u_p})
- \sqrt{\th(k_p+1)}\, L^\th(f_{k+u_p,n}) \bigr) \ox \ga^{a_p}
\\
&\hspace{5em} + \bigl(
\sqrt{\th k_p}\, L^\th(f_{k-u_p,n})
- \sqrt{\th(n_p+1)}\, L^\th(f_{k,n+u_p}) \bigr) \ox \ga^{a_p^*}
\Bigr)
\\
&\qquad - \sum_j \Bigl( \bigl(
\sqrt{\th l_j}\, L^\th(f_{m,l-u_j})
- \sqrt{\th(m_j+1)}\, L^\th(f_{m+u_j,l}) \bigr) \ox \ga^{a_j}
\\
&\hspace{5em} + \bigl(
\sqrt{\th m_j}\, L^\th(f_{m-u_j,l})
- \sqrt{\th(l_j+1)}\, L^\th(f_{m,l+u_j}) \bigr) \ox \ga^{a_j^*}
\Bigr)
\\
&\hspace{3em} \sum_p \Bigl( \bigl(
\sqrt{\th n_p}\, L^\th(f_{l,n-u_p})
- \sqrt{\th(l_p+1)}\, L^\th(f_{l+u_p,n}) \bigr) \ox \ga^{a_p}
\\
&\hspace{5em} + \bigl(
\sqrt{\th l_p}\, L^\th(f_{l-u_p,n})
-\sqrt{\th(n_p+1)}\, L^\th(f_{l,n+u_p}) \bigr) \ox \ga^{a_p^*}
\Bigr) \biggr\}.
\end{align*}

Using the elementary properties of the $f_{mn}$ from
Lemma~\ref{lm:osc-basis}, this simplifies to
\begin{align}
\tpi^\th(\delta\omega_{mnkl})
&= \frac{1}{\th} \sum_j \bigl(
k_j L^\th(f_{mn}) \ox \ga^{a_j} \ga^{ a_j^*}
+ (k_j+1) L^\th(f_{mn}) \ox \ga^{a_j^*} \ga^{a_j}
\nonumber \\
&\hspace{3em} - l_j L^\th(f_{mn}) \ox \ga^{a_j} \ga^{ a_j^*}
- (l_j+1) L^\th(f_{mn}) \ox \ga^{ a_j^*} \ga^{a_j} \bigr)
\nonumber \\
&= \frac{1}{\th}\, L^\th(f_{mn}) \ox  \sum_j
(k_j - l_j)\,(\ga^{a_j} \ga^{ a_j^*} + \ga^{ a_j^*} \ga^{a_j})
\nonumber \\
&= \frac{2}{\th} \sum_j (k_j - l_j)\, L^\th(f_{mn}) \ox 1_{2^N}.
\tag*{\qed}
\end{align}
\hideqed
\end{proof}

\subsection*{Acknowledgments}

\hspace{\parindent}
We thank A.~Connes, K.~Fredenhagen, H.~Grosse, F.~Lizzi,
C.~P.~Mart\'{\i}n, M.~Puschnigg, M.~Rieffel, A.~Schwarz and
A.~Wassermann for suggestions and/or helpful discussions, and
G.~Rozenblum and T.~Weidl for correspondence on matters pertaining to
the subject of this paper.

The work of JMGB and JCV was supported by the Vicerrector\'{\i}a de
Investigaci\'on and the Facultad de Ciencias of the Universidad de
Costa Rica. VG and JCV are grateful to Vanderbilt University for
providing a splendid occasion and nice surroundings for discussions.
JMGB also thanks the Universit\'e de Provence, and JCV thanks the Abdus
Salam ICTP, for their customarily excellent hospitality during various
stages of this work.

\end{document}